\documentclass[prb,aps,twocolumn,showpacs,superscriptaddress,floatfix,10pt]{revtex4-1}
\usepackage{graphicx,amsfonts,amssymb,amsmath,xspace,dsfont}
\usepackage[colorlinks=true,citecolor=green,linkcolor=red,urlcolor=blue]{hyperref}
\usepackage{color}
\definecolor{g}{rgb}{.1,0.4,.1} 
\definecolor{b}{rgb}{0,0.2,1}
\definecolor{rouge}{rgb}{0.82,0.,0.}
\definecolor{vert}{rgb}{0.,0.82,0.}
\definecolor{orange}{rgb}{1,0.5,0.}
\definecolor{bleu}{rgb}{0.,0.,0.82}
\definecolor{m}{rgb}{0.82,0.,0.82}
\definecolor{vert2}{rgb}{0.,0.5,0.}
\definecolor{rougeclair}{rgb}{1.0,0.7,0.7}
\definecolor{marc}{rgb}{.27,.51,.70}
\definecolor{fiona}{rgb}{1.,0.,0.}

\newcommand{\mone}{\textcolor{black}{ \boldsymbol{ 1}}}
\newcommand{\msigma}{\textcolor{red}{ \boldsymbol{ \sigma}}}
\newcommand{\mpsi}{\textcolor{blue}{ \boldsymbol{ \psi}}}
\newcommand{\Jp}{J_{\rm{p}}}
\newcommand{\Jep}{J_{\rm e}^{\mpsi}}
\newcommand{\Jes}{J_{\rm e}^{\msigma}}
\newcommand{\me}{\textcolor{black}{ \boldsymbol{ e}}}
\newcommand{\mm}{\textcolor{black}{ \boldsymbol{ m}}}
\newcommand{\meps}{\textcolor{black}{ \boldsymbol{ \varepsilon}}}

\newcommand{\Isi}{Ising$\times \overline{\text{Ising}}$ }

\newcommand{\fb}{\it \color{fiona}}

\allowdisplaybreaks

\begin{document}

\title{Frustrated topological symmetry breaking: geometrical frustration and anyon condensation}

\author{Marc D. Schulz}
\email{mdschulz@umn.edu}
\affiliation{School of Physics and Astronomy, University of Minnesota, Minneapolis, MN 55455, USA}
\author{Fiona J. Burnell}
\email{fburnell@umn.edu}
\affiliation{School of Physics and Astronomy, University of Minnesota, Minneapolis, MN 55455, USA}%

\begin{abstract}
We study the phase diagram of a topological string-net type lattice model in the presence of geometrically frustrated interactions. These interactions drive several phase transitions that reduce the topological order, leading to a rich phase diagram including both Abelian ($\mathbb{Z}_2$) and non-Abelian (\Isi) topologically ordered phases, as well as phases with broken translational symmetry. Interestingly, one of these phases simultaneously exhibits (Abelian) topological order and long-ranged order due to translational symmetry breaking, with non-trivial interactions between excitations in the topological order and defects in the long-ranged order. We introduce a variety of effective models, valid along certain lines in the phase diagram, which can be used to characterize both topological and symmetry-breaking order in these phases, and in many cases allow us to characterize the phase transitions that separate them. We use exact diagonalization and high-order series expansion to study areas of the phase diagram where these models break down, and to approximate the location of the phase boundaries. 
\end{abstract}

\pacs{71.10.Pm, 75.10.Kt, 03.65.Vf, 05.30.Pr}

\maketitle
%
%
\section{Introduction}
In recent years, topological order has gained increasing interest, motivated in large part by potential applications in quantum computation\cite{nayak08,kitaev03,kitaev06,mong14}. These applications rely on the fact that the entanglement between certain states in a topologically ordered system is genuinely non-local, and thus cannot be disturbed by local perturbations\cite{kitaev03}, which constitute the main obstacle for a successful realization of a quantum computer.

This non-locality is also entrenched in the characteristics that identify phases as topologically ordered. These phases are characterized by intrinsically non-local properties such as the (finite) ground-state degeneracy on a torus, fractional excitations with non-trivial mutual statistics\cite{wen90b}, 
and patterns of long-ranged entanglement\cite{levin06,kitaev06b}. 
In particular, there is by definition no local order parameter that can be used to identify a topologically ordered phase. 
Among other things, this implies that the usual Landau-Ginzburg machinery for understanding phase diagrams and second-order critical points does not directly apply in these systems. 

It has been known for some time that transitions between phases without a local order parameter can exist\cite{wegner71}. 
In the case of $\mathbb{Z}_2$ topological order\cite{kitaev03}, the phase diagram has been extensively studied\cite{fradkin79,trebst07,hamma08,castelnovo08,tupitsyn10,dusuel11,wu12}. 
Additionally, Refs.~\onlinecite{bais02,bais03,bais09b} developed a mathematical framework identifying which topological orders can be related by condensing bosonic (albeit possibly non-Abelian) excitations. Examples of these more exotic transitions have been identified both in quantum Hall bilayers\cite{barkeshli10,barkeshli11,moeller14} and in a family of lattice models\cite{gils09,gils09ising,burnell11b,burnell12,schulz13,morampudi14}. 

One key difference between studying the phase diagrams of topological lattice models, relative to continuum systems, is the possibility of frustration. More specifically, beginning with an exactly solvable lattice Hamiltonian (see for example Refs.~\onlinecite{kitaev03,levin05,kitaev06}) that realizes a particular topological phase, with the appropriate lattice geometry one can typically add a perturbing Hamiltonian which on its own has an extensive ground state degeneracy.  This can lead to frustrated transitions in which the topological order is lost or reduced at a transition in which the system orders ``by disorder".  This intriguing possibility has been studied in the context of $\mathbb{Z}_2$ spin liquids\cite{schmidt13,roychowdhury15} and dimer models\cite{moessner00,moessner01,misguich02,poilblanc10}, but has received relatively little attention in the context of more complex topological orders. (See, however, Refs.~\onlinecite{schulz12,schulz14,schulz15}). 

The present work focuses on shedding light on this interplay of geometric frustration and topological order. Specifically, we introduce a model which contains both a phase with non-Abelian (Ising-like) anyons, and phases with $\mathbb{Z}_2$ or trivial topological order. We show that both $\mathbb{Z}_2$ and trivial topological orders can arise in conjunction with broken translational symmetry resulting from frustration. In the frustrated $\mathbb{Z}_2$ topologically ordered phase, we show that some excitations of the parent topological theory become confined and correspond to defects in the long-ranged translation-breaking order, while others remain deconfined and comprise the new topological quasi-particles. 

In addition to elucidating the mechanism allowing topological order and symmetry-breaking to coexist, we give a comprehensive description of the phase diagram of our model, for both frustrated and unfrustrated perturbations away from the non-Abelian regime. For each of the phases realized we provide an effective Hamiltonian whose ground state(s) can be determined exactly, allowing us to analytically identify the corresponding topological orders and symmetry breaking patterns. We complement this analysis with a numerical determination of the various phase boundaries, obtained through a combination of exact diagonalization and high-order perturbation theory.

The remainder of this work is structured as follows.
We present the details of the lattice model in Sec.~\ref{sec:model}. In Sec.~\ref{sec:phasediag}, we give an overview of the model's phase diagram, together with the methods used to obtain it. The details of the various phases are discussed in the remaining sections. First, in Sec.~\ref{sec:AIZ}, we describe in depth the frustrated $\mathbb{Z}_2$ topological phase. Our approach also describes a $\mathbb{Z}_2$ unfrustrated phase, and allows us to identify the transitions between both frustrated and unfrustrated $\mathbb{Z}_2$ phases and the parent doubled Ising topological order. In Sec.~\ref{sec:topaway}, we discuss the various non-topological phases, which can be obtained from the $\mathbb{Z}_2$ phases by tuning an additional parameter. A third $\mathbb{Z}_2$ topological phase, which arises through a fundamentally different mechanism than the other two, is presented in Sec.~\ref{ssec:uppercornerC}. We conclude with a discussion of the transitions not connected to the Ising-anyon phase in Sec.~\ref{sec:conclusion}.

\section{Model}\label{sec:model}
To explore the interplay between topological order and geometrical frustration, our starting point is an exactly solvable Levin-Wen type Hamiltonian\cite{levin05} $H_{\rm SN}$, which realizes the $\mathrm{Ising}\times\overline{\mathrm{Ising}}$ (or doubled Ising) topological order. 
This topological order describes a bilayer system, in which the two layers have topological orders with opposite chiralities. We describe the form of $H_{\rm SN}$ in Section~\ref{ssec:stringnet}.

To the solvable Hamiltonian $H_{\rm SN}$, we will add a second term which we call $V$. Terms in $V$ commute with each other, but not with $H_{\rm SN}$. Hence by adjusting the associated couplings, we can drive the system from the doubled Ising phase realized by $H_{\rm SN}$ into a variety of other phases with various combinations of symmetry-breaking and topological order. As we will see, studying a lattice model significantly enriches the phase diagram, producing a number of frustrated phases which are not natural in a continuum setting (as would be appropriate for the superconducting bilayer mentioned above).
We shall introduce the precise form of $V$ in Section~\ref{ssec:loc_pert}. 

The phase diagram of the resulting perturbed string-net Hamiltonian
\begin{align}
H=&H_{\rm SN}+V
\label{eq:fullham}
\end{align}
is discussed in Section~\ref{sec:phasediag}.

\subsection{The topological Hamiltonian}\label{ssec:topham}
\subsubsection{The Ising string-net \texorpdfstring{$H_{\rm SN}$}{HSN}}\label{ssec:stringnet}
We study a string-net model\cite{levin05} on the honeycomb lattice. The version studied here is based on the Ising CFT (see Ref.~\onlinecite{bonderson_thesis}), and as described below contains excitations which are either hardcore bosons or non-Abelian anyons. This model has been discussed at length in the literature (e.g.~in Refs.~\onlinecite{levin05,burnell12}), so we will constrain ourselves here to the facts relevant to this work, whereas technical details can be found in Appendix \ref{app:stringnetdetails}.

\begin{figure}[htp]%
\includegraphics[width=1cm]{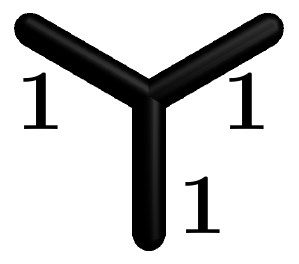}\quad%
\includegraphics[width=1cm]{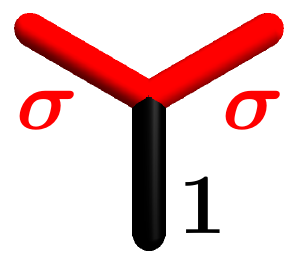}\quad%
\includegraphics[width=1cm]{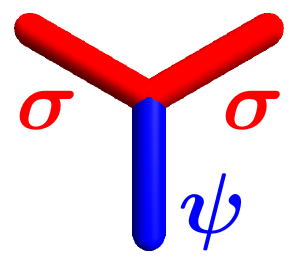}\quad%
\includegraphics[width=1cm]{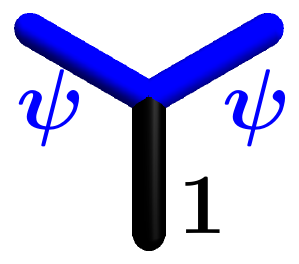}%
\caption{Vertex constraints for the Hilbert space. States in the constrained Hilbert space must be in one of the configurations shown here (up to rotations) at each vertex.}%
\label{fig:lattice_constraints}%
\end{figure}

The Hilbert space in our model consists of three possible states for each edge of the honeycomb lattice, which we label $\mone$, $\msigma$, and $\mpsi$. We impose the constraint that at each vertex of the lattice we have one of the configurations depicted in Fig.~\ref{fig:lattice_constraints}.
In particular, edges with the label $\msigma$ always form closed loops, and chains of $\mpsi$-labeled edges must either form closed loops or terminate at a vertex with two $\msigma$-edges.

Imposing the above constraints differs from the original construction of Levin and Wen\cite{levin05}, which allows violations of these constraints at finite energy cost $\epsilon_v$. This introduces additional type of quasi-particles not present in our model. However, this technical difference will not affect the spectrum of our model at energies below $\epsilon_v$ anywhere in the phase diagram, and does not affect our conclusions about the phase diagram or criticality.

In the constrained Hilbert space, the Levin-Wen (or string net) Hamiltonian $H_{\rm SN}$ is given by
\begin{align}
\!\!H_{\rm{SN}} =\! -\Jp\sum\limits_p\! B_p = -\frac{\Jp}{4}\sum\limits_p \!\left(\! B_p^{\mone} + \sqrt{2}B_p^{\msigma} + B_p^{\mpsi}\! \right),
\label{eq:HSN_def}
\end{align}
where the operators $B_p^{\boldsymbol{s}}$ induce fluctuations between different string-net states by ``raising'' the labels of the links around the plaquette $p$ by the label $\boldsymbol{s}$. 
More specifically, $B_p^{\boldsymbol{s}}$ acts via
\begin{align}
B_p^{\boldsymbol{s}} = \prod\limits_{v \in p} \phi(v) \prod\limits_{e\in p} S_e^{\boldsymbol{s}},
\label{eq:bps_def}
\end{align}
where the coefficients $\phi(v)$ (given in App.~\ref{app:details_bps}) depend on the configuration at the vertex $v$ of the initial and final state. 
The operators $S_e^{\boldsymbol{s}}$ acting on the label of edge $e$ are given in the basis $\left\lbrace \left|\mone\right\rangle_e,\left|\msigma\right\rangle_e,\left|\mpsi\right\rangle_e \right\rbrace$ by
\begin{align}
S_e^{\mone}=\!\!\left(\!\!{\scriptstyle\begin{array}{c c c} 1 & 0 & 0 \\ 0 & 1 & 0 \\ 0 & 0 & 1 \end{array}}\!\!\right),\ S_e^{\msigma}\!\!=\!\!\left(\!\!{\scriptstyle\begin{array}{c c c} 0 & 1 & 0 \\ 1 & 0 & 1 \\ 0 & 1 & 0 \end{array}}\!\!\right),\ S_e^{\mpsi}\!\!=\!\!\left(\!\!{\scriptstyle\begin{array}{c c c} 0 & 0 & 1 \\ 0 & 1 & 0 \\ 1 & 0 & 0 \end{array}}\!\!\right).
\label{eq:S_operators}
\end{align}
The coefficients $\phi(v)$ are chosen such that the operators $B_p^{\boldsymbol{s}}$ annihilate states not fulfilling the constraints shown in Fig.~\ref{fig:lattice_constraints} and commute among themselves, which ensures the exact solvability of the model.

The coefficients of the $B_p^{\boldsymbol{s}}$ in Eq.~(\ref{eq:HSN_def}) are chosen such that $B_p$ is a projector. For $J_{\rm{p}}>0$, which shall be assumed throughout this work, ground states $\left|\Psi_0\right\rangle$ of $H_{\rm{SN}}$ fulfill $B_p\left|\Psi_0\right\rangle=+1\left|\Psi_0\right\rangle$ for all plaquettes $p$. 

The topological order of the resulting gapped phase is characterized by two physical properties: (1) the topological ground state degeneracy, and (2) the mutual statistics of its low-energy point-like excitations. The topological ground state degeneracy results from the fact that if there are non-contractible loops in the space in which the lattice is embedded in, it is possible to construct loop operators that commute with the Hamiltonian and measure additional conserved quantum numbers, leading to multiple physically distinct ground states. A closely related set of {\it open string} operators, which commute with the Hamiltonian everywhere except at their endpoints, can be used to generate quasi-particles and determine their mutual statistics\cite{kitaev03,levin05}. 

To understand the topological ground-state degeneracy of our string-net Hamiltonian, let us detail the loop operators $W_{\mathcal{C}_i}^{(\alpha,\beta)}$. We will restrict our discussion to the torus (i.e. to lattices with periodic boundary conditions), which is the simplest spatial topology with non-contractible loops. On the torus there are two inequivalent non-contractible closed loops, $\mathcal{C}_1$ and $\mathcal{C}_2$. The loop operators are defined similarly to the operators $B_p^{\boldsymbol{s}}$, by
\begin{align}
W_{\mathcal{C}_i}^{(\alpha,\beta)} = \prod\limits_{v \in \mathcal{C}_i} \omega(v) \prod\limits_{e\in \mathcal{C}_i} S_e^{\alpha}S_e^{\beta},
\label{eq:wcab_def}
\end{align}
with $\alpha,\beta\in\left\lbrace\mone,\msigma,\mpsi\right\rbrace$.
The coefficients $\omega(v)$, given in App.~\ref{app:stringoperatorsdetails}, depend on the initial and final configuration of the edge labels of the vertices $v$ crossed by $\mathcal{C}_i$. 
From the full set of loop operators, one can choose the mutually commuting set $\lbrace W_{\mathcal{C}_1}^{(\alpha,\beta)} \rbrace$ with $\alpha,\beta\in\left\lbrace \mone,\msigma,\mpsi \right\rbrace$ to characterize the nine distinct ground states through their possible eigenvalues. The operators $\lbrace W_{\mathcal{C}_2}^{(\alpha,\beta)} \rbrace$, which commute with $H_{\rm{SN}}$ but not with $\lbrace W_{\mathcal{C}_1}^{(\alpha,\beta)} \rbrace$, alter these eigenvalues and thus map between the different ground states. Details are given in App.~\ref{app:stringoperatorsdetails}. 

The elementary excitations of $H_{\rm SN}$ correspond to plaquettes on which $B_p$ has eigenvalue $0$. 
Because the Hamiltonian is comprised of commuting projectors, the eigenvalue of $B_p$ on each plaquette is conserved, and these excitations are static and non-interacting. As described in Refs.~\onlinecite{levin05,burnell12}, in the absence of violations of the vertex constraints there are two types of plaquette excitations. The first, which we call a $\mpsi$-flux, is a hardcore boson.
The other excitation, which we call a $\msigma$-flux, is a non-Abelian boson. 

In terms of the Ising$ \times \overline{\text{Ising}}$ topological order, these excitations can be understood as follows.  Topologically, the Ising CFT is very similar to a chiral $p_x + i p_y$ topological superconductor: It contains two types of anyons, a fermion $\mpsi$, and a non-Abelian anyon $\msigma$.  Analogous to the vortices of the $p_x + ip_y$ superconductor, each pair of $\msigma$ anyons can have even or odd fermion parity.  In the bilayer Ising$ \times \overline{\text{Ising}}$ system, the $\mpsi$-flux corresponds to a bound state of one fermion excitation in each layer, and the $\msigma$-flux corresponds to a bound state of one $\msigma$-anyon in each layer.  The $\msigma$-fluxes are non-Abelian anyons in the sense that braiding can change the internal (fermion parity) state of each bound pair.  
Chiral excitations, which live on only one layer of the bilayer system, are not present in our model due to the Hilbert space constraint.

In the lattice model, these excitations are pair-created by open string operators defined on curves $\mathcal{C}_{1,2}$ connecting plaquettes $p_1$ and $p_2$. Away from their endpoints, open string operators are defined in the same way as the loop operators (\ref{eq:wcab_def}), and can be chosen to commute with the vertex constraint (but not $B_p)$ at their endpoints. Specifically, $\mpsi$-fluxes are created in pairs by open strings $W_{\mathcal{C}_{1,2}}^{(\mpsi,\mpsi)}$, which obey $\left( W_{\mathcal{C}_{1,2}}^{(\mpsi,\mpsi)}\right)^2 =1$. 
$\msigma$-fluxes are pair-created by open strings $W_{\mathcal{C}_{1,2}}^{(\msigma,\msigma)}$. These excitations are non-Abelian, as evidenced by the fact that $W_{\mathcal{C}_{1,2}}^{(\msigma,\msigma)} = \mathbf{1} + W_{\mathcal{C}_{1,2}}^{(\mpsi,\mpsi)}$: creating $\msigma$-fluxes twice on the same pair of plaquettes can lead to no flux, or to a pair of $\mpsi$-fluxes. 
 
For our purposes, it is important that both excitations are bosons in the sense of Ref.~\onlinecite{bais09b}, i.e.~both types of excitations can be condensed, leading to various (possibly second order) phase transitions out of the doubled Ising phase.

\subsubsection{Topological properties of \texorpdfstring{$\mathbb{Z}_2$}{Z(2)}-phases}
The phase diagram studied here also includes phases with topological order that is distinct from that of the string net discussed in Sec.~\ref{ssec:topham}. As we show below, all of these phases have $\mathbb{Z}_2$ topological order -- i.e that of an Ising gauge theory, or equivalently of the Toric code\cite{kitaev03,moessner01c}. In these $\mathbb{Z}_2$-phases, there are three non-trivial loop-operators for each non-contractible curve, which we will call $W^{\me}$, $W^{\mm}$, and $W^{\meps} = W^{\me} W^{\mm}$, which can be shown to imply a four-fold ground state degeneracy on the torus. The corresponding open strings produce quasi-particle pairs in the unconstrained Hilbert space which are hard-core bosons ($\me$ and $\mm$) or spinless fermions ($\meps$).

\subsection{The non-topological Hamiltonian \texorpdfstring{$V$}{V}}\label{ssec:loc_pert}
The second key element of our model is the term $V$, which will allow us to tune the system out of the doubled Ising phase by condensing appropriate combinations of the two flux excitations described above. We take
\begin{align}
V=-\frac{\Jes}{2}\sum\limits_e \left(n_e^{\mone}-n_e^{\mpsi}\right)+\frac{J_{\rm{e}}^{\mpsi}}{2}\sum\limits_e n_e^{\msigma},
\label{eq:pert_def}
\end{align}
where the operators $n_e^{\bf{\alpha}}$ measures whether the edge $e$ carries the label $\bf{\alpha}$, i.e. $n_e^{\bf{\alpha}}\left|\beta\right\rangle=\delta_{\alpha,\beta}\left|\beta\right\rangle$. The combination of operators $n_e^{\bf{\alpha}}$ in Eq.~\ref{eq:pert_def} is chosen such that the term proportional to $J_{\rm{e}}^{\bf{\alpha}}$ introduces dynamics and pair-creation/annihilation of excitations of $\alpha$-fluxes\cite{burnell12}. The Hamiltonian $V$ consists of commuting projectors, which in combination with the vertex constraints gives rise to either polarized or frustrated phases\cite{burnell12,schulz14}; for this reason we will sometimes refer to $V$ as the non-topological Hamiltonian. Special cases of this form have been studied e.g.~in Refs.~\onlinecite{gils09ising,burnell12,schulz14}.
We will describe these different polarized and frustrated phases, which are relevant to understanding the phase diagram of the full Hamiltonian (\ref{eq:fullham}), in the following sections. 

\section{The phase diagram: overview and methods}\label{sec:phasediag}
In this section, we give an overview of the various gapped phases of the perturbed Ising string-net model (\ref{eq:fullham}), together with the numerical methods used to determine the phase boundaries. 
\begin{figure}[htp]%
\centering
\includegraphics[width=\columnwidth]{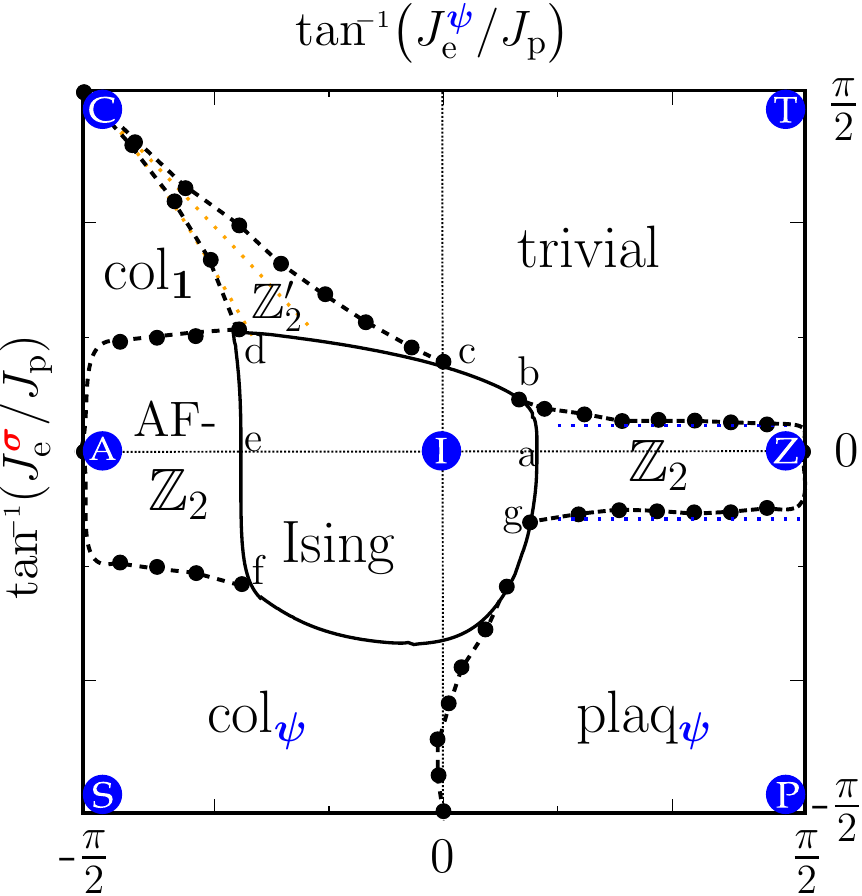}%
\caption{The phase diagram of the perturbed Ising string-net as a function of the ratios of the non-topological couplings $\Jes$, $\Jep$ over the topological coupling $\Jp$. We find topologically-ordered phases like the Ising- and the various $\mathbb{Z}_2$-phases, a trivial phase as well as phases described in terms of effective dimer models (either columnar ($\rm col$) or plaquette-type ($\rm plaq$) of order\cite{moessner01}). The phase boundary of the Ising string-net phase (solid line) has been obtained by series expansion techniques. The location of the other phase transitions have been extracted from exact diagonalization (dots). The dashed lines are guide for the eye. The different limits, denoted by the blue dots, are described in the main text. The dotted blue (orange) lines show the location of the phase transitions out of the $\mathbb{Z}_2$-($\mathbb{Z}_2^{\prime}$-)phase for the effective model derived in Sec.~\ref{ssec:pzt} (Sec.~\ref{ssec:uppercornerC}).}%
\label{fig:phase_diag}%
\end{figure}

\subsection{The phase diagram}
The phase diagram is shown in Fig.~\ref{fig:phase_diag}. 
It consists of eight phases, each distinguished from its neighbor by differing topological orders and/or patterns of translational symmetry breaking. At the center of the phase diagram (for $|\Jes|$ and $|\Jep|$ small, i.e.~at point $\rm I$), we find the doubled Ising phase described in Sec.~\ref{ssec:topham}. 
For small $|\Jes|$ but large $|\Jep|$ we find two phases with $\mathbb{Z}_2$ topological order: around point $\rm A$ the frustrated or ``antiferromagnetic" $\mathbb{Z}_2$ phase denoted by ${\rm AF}-\mathbb{Z}_2$, which also exhibits three-sublattice translation breaking, and around point $\rm Z$ the unfrustrated or ferromagnetic phase denoted by $\mathbb{Z}_2$, which does not. In both phases the topological order is due to fluctuating $\mpsi$-edge labels, whereas $\msigma$-labels are largely absent in the ferromagnetic phase, and are numerous but ordered in the antiferromagnetic phase. 
 
Increasing $|\Jes|$ in either $\mathbb{Z}_2$ topological phase destroys the topological order, by either favoring or disfavoring $\mpsi$-edges. 
For $\Jep>0$ this leads to either a ``trivial" phase, whose ground state is adiabatically connected to a product state with $n_e^{\msigma} = n_e^{\mpsi} =0$ (realized at point $\rm T$ in the phase diagram), or a frustrated phase labeled by ${\rm plaq}_{\mpsi}$ with three-fold translational symmetry breaking due to long-ranged order in the pattern of $\mpsi$-labels. This order is described by the ``plaquette" phase of an appropriate quantum dimer model\cite{moessner01,schlittler15}, realized at point $\rm P$. 
For $\Jep<0$ the disappearance of topological order in the antiferromagnetic phase coincides with a change in the long-ranged order, from a ``plaquette"-type order of $\msigma$ edges in the ${\rm AF}-\mathbb{Z}_2$ phase, to a ``columnar" order in phases ${\rm col}_{\mone}$ and ${\rm col}_{\mpsi}$. (We elaborate on the details of these two ordering patterns in Secs.~\ref{sec:AIZ} and \ref{sec:topaway} below). 
Finally for $- \Jep \approx \Jes \gg \Jp$, we find a second phase with $\mathbb{Z}_2$ topological order and no translational symmetry breaking. We label this phase by $\mathbb{Z}_2^{\prime}$. Unlike in the ferromagnetic $\mathbb{Z}_2$ topological phase, where $\msigma$-labels are sparse and the topological order can be attributed to extended $\mpsi$-loops, in this region $\mpsi$-labels are sparse and the $\mathbb{Z}_2^{\prime}$ topological order arises due to extended $\msigma$-loops.

Also indicated in the Figure are certain phase boundaries across which we can identify the universality class of the phase transition. The lines $\rm d-e-f$ ($\rm b-a-g$), separating the (un-) frustrated topological phases respectively from the doubled Ising phase, can be shown to be exactly in the universality class of the 2D triangular lattice quantum Ising model with (un-) frustrated interactions. These transitions are therefore of the 3D-XY (3D Ising) type. As is well known, the same applies to the two transitions out of the ferromagnetic $\mathbb{Z}_2$ topological phase: the line $\rm Z-b$ separating the trivial and the $\mathbb{Z}_2$ phase represents transitions in the 3D Ising universality class\cite{fradkin79,trebst07,hamma08,vidal_par,wu12}, while the line $\rm Z-g$ separating the ${\rm plaq}_{\mpsi}$ is 3D-XY\cite{blankschtein84,isakov03}. We speculate on the nature of some of the remaining transitions in Sec.~\ref{sec:conclusion}. 

In the following sections, for each phase identified in Fig.~\ref{fig:phase_diag} we derive an effective Hamiltonian, valid along a certain line within this phase, which we can show explicitly has the symmetry breaking and/or topological order described here. These effective models also allow us to identify the phase transitions described in the preceding paragraph.

\subsection{Methods}
To support our theoretical analysis we also present numerical results, which we use both to estimate the location of the phase boundaries shown in Fig.~\ref{fig:phase_diag}, and to verify that the phases described above are the complete set of phases of our model. 
We employ two complementary approaches: high-order series expansion and exact diagonalization. 
The phase boundary of the \Isi-phase as well as for the $\mathbb{Z}_2$ phase in the effective model (\ref{eq:hameffz}) is obtained by determining the approximate parameter value at which the low-energy gap closes. The gap for the different excitations has been obtained by means of perturbative continuous transformations (pCUT)\cite{knetter00} and extrapolated by dlog-Pad\'e approximants\cite{domb89} in the same fashion as used e.g.~in Refs.~\onlinecite{schulz12,schulz13}. The leading-order expressions can be found in App.~\ref{app:seriesising}. We complement these perturbative results with exact diagonalization (ED) distinguishing for the relevant translation-symmetry and topological sectors on systems with up to $\approx 2\cdot10^8$ states. We determine the location of a phase boundary by identifying divergences developing in the second derivative of the ground-state energy. Since for finite systems $\partial^2e_0$ will show a developing divergence for either first or second order phase transitions, this allows us to infer the location, but not necessarily the order, of the transition.

The two approaches are both limited in the degree to which they can accurately describe our system. These limitations are, in a sense, complementary: whereas the perturbative expansions are valid in the thermodynamic limit, but limited by the finite order of the expansion, exact diagonalization is non-perturbative but limited by finite-size effects. Here our ED results treat only systems up to $13$ plaquettes for the full model (\ref{eq:fullham}), so that the finite-size effects can be substantial, and can result in significant differences between the actual phase boundaries and those obtained here.  This is particularly true in the frustrated phases, as we discuss in App.~\ref{app:fss_z2prime}.  As a rough benchmark for the accuracy of these two methods, we present results for the PZT line, in which our model is simply the Toric code in an appropriate magnetic field, in App.~\ref{app:z2}.

\section{The topological line A-I-Z}\label{sec:AIZ}
We begin by considering the line A-I-Z in Fig.~\ref{fig:phase_diag}, where $\Jes=0$. We call this the topological line, as all phases arising here have topological order of the \Isi- or $\mathbb{Z}_2$-type. 

Along this line, the Hamiltonian has the form
\begin{align}
H^{\rm AIZ}=H_{\rm SN}+\frac{\Jep}{2}\sum\limits_en_e^{\msigma}.
\label{eq:heff_aiz}
\end{align}
As we will show, the model (\ref{eq:heff_aiz}) has three gapped phases. For $\Jep=0$ we recover the (gapped) Ising string-net Hamiltonian; hence for small $\left|\Jep\right|$ the system realizes a phase with doubled Ising topological order. For large positive (negative) $\Jep$, $\msigma$-labels on the edges are energetically disfavored (favored) compared to the other two labels. 
This leads to two additional gapped phases, both of which have $\mathbb{Z}_2$ topological order resulting from the fluctuating $\mpsi$-edge labels. However, these phases are fundamentally different: for $\Jep\gg\Jp$ the low-energy properties are captured by a standard $\mathbb{Z}_2$ (or Toric-code type) lattice model for $\Jep\ll \Jp$ is frustrated, and spontaneously breaks lattice translation symmetry. This phase, in which topological order and spontaneous symmetry breaking coexist, is one of our model's most striking features.

Our objective here is to clarify the nature of the phases for large, negative $\Jep$. However, for pedagogical reasons, we also review the case of positive $\Jep$ to highlight similarities and differences between the two regimes. 
This review largely follows the treatment of Refs.~\onlinecite{burnell11b,burnell12}, which studied the phase diagram of (\ref{eq:heff_aiz}) for $\Jep>0$ in detail. 

\subsection{Effective low-energy model for the A-I-Z line}\label{ssec:efftfim}
To obtain a more quantitative description of the two transitions along the line $\Jes=0$, we follow Ref.~\onlinecite{burnell12} and introduce an effective Hamiltonian $H_{\rm{eff}}^{\rm{AIZ}}$ which faithfully reproduces $H^{\text{AIZ}}$ when acting on states with no $\msigma$ fluxes. Since $\msigma$-fluxes are gapped and conserved under the Hamiltonian (\ref{eq:heff_aiz}), this effective model allows us to confirm the presence and nature of the phase transitions. We will then re-introduce the $\msigma$-fluxes in order to study the resulting gapped phases, labeled by $\mathbb{Z}_2$ and ${\rm AF}-\mathbb{Z}_2$ in Fig.~\ref{fig:phase_diag}, in the limit of $\left|\Jep\right| \gg \Jp$.

The effective model of Ref.~\onlinecite{burnell12} follows from the fact that,
if we neglect the static and gapped $\msigma$-excitations, the only remaining degrees of freedom are the $\mpsi$-excitations, which are (hardcore) bosons. We can therefore introduce a dual pseudo spin-$\frac{1}{2}$ variable on each plaquette $p$, where $\left|\uparrow\right\rangle_p$ ($\left|\downarrow\right\rangle_p$) denotes the absence (presence) of a $\mpsi$-excitation. In the absence of $\msigma$-fluxes, the Hamiltonian (\ref{eq:heff_aiz}) in the dual pseudo-spin basis is exactly the transverse field Ising model on the dual triangular lattice\cite{fradkin79}:
\begin{align}
H_{\rm{eff}}^{\rm{AIZ}}=-\frac{\Jp}{2}\sum\limits_p\left(\mathds{1}+\tau_p^z\right)-\frac{J_e^{\mpsi}}{4}\sum\limits_{\left\langle p^{\phantom{\prime}}\!\!,p^{\prime}\right\rangle}\left(\mathds{1}+\tau_{p^{\phantom{\prime}}}^x\!\tau_{p^{\prime}}^x\right),
\label{eq:hameffaiz}
\end{align}
where $\tau^z_p,\tau^x_p$ are Pauli matrices acting on the plaquette pseudo-spins, and the second term is the representation of $n_e^{\msigma}$ in the $\msigma$-flux free Hilbert space. We emphasize that the mapping of Ref.~\onlinecite{burnell12} is valid independent of the sign of $\Jep$.

$H_{\rm{eff}}^{\rm{AIZ}}$ has been extensively studied\cite{coppersmith85,kim90,blankschtein84,moessner01,moessner01c}, and is known to have three distinct gapped phases: a paramagnetic phase for $\left|\Jep\right|\ll\Jp$, a ferromagnetic phase for $\Jep\gg\Jp$, and an anti-ferromagnetic phase for $-\Jep\gg\Jp$. Since the $\msigma$-fluxes are non-dynamical throughout, re-introducing them cannot lead to additional phase transitions. It follows that the perturbed string-net model also undergoes two phase transitions (one for $\Jep>0$ located at point $\rm a$ in the phase diagram Fig.~\ref{fig:phase_diag} and one for $\Jep<0$ located at point $\rm e$ in Fig.~\ref{fig:phase_diag}) out of the \Isi-phase arising at $\Jep=0$.

\subsection{The ferromagnetic topological \texorpdfstring{$\mathbb{Z}_2$}{Z(2)}-phase} \label{Sec:z2p}
The effective Hamiltonian (\ref{eq:hameffaiz}) allows us to identify the location and universality classes of the phase transitions in our system, but does not fully describe the corresponding gapped phases in the original model (\ref{eq:fullham}). To understand these gapped phases we must re-introduce the $\msigma$-fluxes, which we will now do for the two phases with large $|\Jep|$. 

We begin at large positive $\Jep$, where the $\mpsi$-fluxes have condensed (i.e. deep in the ferromagnetic phase of the effective model (\ref{eq:hameffaiz})). In the limit $\Jep \rightarrow \infty$ (denoted by $\rm Z$ in Fig.~\ref{fig:phase_diag}), where $\msigma$-edges are effectively absent from the ground state, the low-energy effective Hamiltonian is: 
\begin{align}
H_{\rm eff}^{Z}=PHP=-\frac{\Jp}{4}\sum\limits_p\left(\mathds{1}+B_p^{\mpsi}\right)=\frac{1}{2} H^{\mathbb{Z}_2}_{\rm SN},
\label{eq:hameffz}
\end{align}
where $P$ is a projector onto the low-energy Hilbert space -- which in this case is the set of states with no $\msigma$-edges. The relation to the $\mathbb{Z}_2$ string-net Hamiltonian $H^{\mathbb{Z}_2}_{\rm SN}$ results from identifying the edge-label $\mone$ ($\mpsi$) with the label $\mone$ ($-\mone$)\footnote{The Hamiltonian $H^{\mathbb{Z}_2}_{\rm SN}$ described here differs from the $\mathbb{Z}_2$ string-net Hamiltonian in that the matter (i.e.~the electric charge $\boldsymbol{e}$) is fermionic and not bosonic. However, in 2D this does not affect the topological order.} of the $\mathbb{Z}_2$-algebra with the particle content $\mone$ (trivial), $\me$ (electric), $\mm$ (magnetic), $\meps$ (fermion)\cite{kitaev03}. 

For finite $\Jep$, the projector $P$ (and the corresponding Hamiltonian) must be modified to include fluctuations generating short $\msigma$-loops. However, since the topological order cannot change unless the system undergoes a phase transition, the analysis above suffices to characterize the entire phase. 

The $\mathbb{Z}_2$-topological order can also be deduced at more general values of $\Jep$ from the set of string-operators that commute with $P$. Specifically, $PW_{\mathcal{C}_i}^{(\alpha,\beta)}P=0$ for either $\alpha={\msigma}\neq\beta$ or $\alpha\neq{\msigma}=\beta$, since these strings create extended $\msigma$-loops, whereas $P$ projects onto states with only short $\msigma$-loops. For the remaining operators, we have
\begin{align}
PW_{\mathcal{C}_i}^{({\mone},{\mone})}P=PW_{\mathcal{C}_i}^{({\mpsi},{\mpsi})}P\equiv W_{\mathcal{C}_i,\rm{FM}}^{\mone},\label{eq:stringoperatorsz2_1a}\\
PW_{\mathcal{C}_i}^{({\mpsi},{\mone})}P=PW_{\mathcal{C}_i}^{({\mone},{\mpsi})}P\equiv W_{\mathcal{C}_i,\rm{FM}}^{\meps}.
\label{eq:stringoperatorsz2_1}
\end{align}
The first identity follows from the fact that the only operator that can distinguish between $W_{\mathcal{C}_i}^{({\mone},{\mone})}$ and $W_{\mathcal{C}_i}^{({\mpsi},{\mpsi})}$ is an extended (non-contractible) $\msigma$-string; hence $P$ eliminates the distinction between these two states. The second line is a consequence of the first, since $W_{\mathcal{C}_i}^{({\mone},{\mpsi})} = W_{\mathcal{C}_i}^{({\mpsi},{\mone})} \times W_{\mathcal{C}_i}^{({\mpsi},{\mpsi})}$. 

Additionally we have the non-trivial relation\cite{burnell12}
\begin{align}
PW_{\mathcal{C}_i}^{({\msigma},{\msigma})}P\equiv W_{\mathcal{C}_i,\rm{FM}}^{\me} + W_{\mathcal{C}_i,\rm{FM}}^{\mm},
\label{eq:stringoperatorsz2_2}
\end{align}
The form of the operators $W_{\mathcal{C}_i,\rm{FM}}^{{\me}}$ and $W_{\mathcal{C}_i,\rm{FM}}^{{\mm}}$ is given in App.~\ref{app:stringoperatorsz2details}. 
Essentially, however, Eq.~(\ref{eq:stringoperatorsz2_2}) follows from the fact that $W_{\mathcal{C}_i}^{({\msigma},{\msigma})}$ strings come in two ``flavors". These are mixed in the presence of extended (or open, in the original Levin-Wen formulation) $\msigma$-strings\cite{kitaev12}, but become physically distinct excitations when these extended strings are confined. 

\subsection{The anti-ferromagnetic \texorpdfstring{$\mathbb{Z}_2$}{Z(2)}-phase}\label{ssec:afmz2}
We now turn to the phase at $- \Jep \gg \Jp$. The effective model (\ref{eq:hameffaiz}) dictates that there is a single phase transition for negative $\Jep$, separating the paramagnetic phase (which, in the full model, corresponds to a phase with $\mathrm{Ising}\times\overline{\mathrm{Ising}}$ topological order) from a phase with partial anti-ferromagnetic order which breaks three-sublattice translation symmetry. 
As for $\Jep>0$, to characterize this phase in the full topological model (\ref{eq:heff_aiz}), we must re-introduce the $\msigma$-fluxes and deduce the resulting topological order. 

To do so, we will again construct an effective Hamiltonian $H_{\rm eff}^A$, valid in the limit $\Jep\rightarrow - \infty$, by projecting onto the corresponding low-energy Hilbert space. 
\begin{figure}[htp]%
\begin{minipage}{\columnwidth}
\begin{minipage}{.4\columnwidth}
\includegraphics[width=\columnwidth]{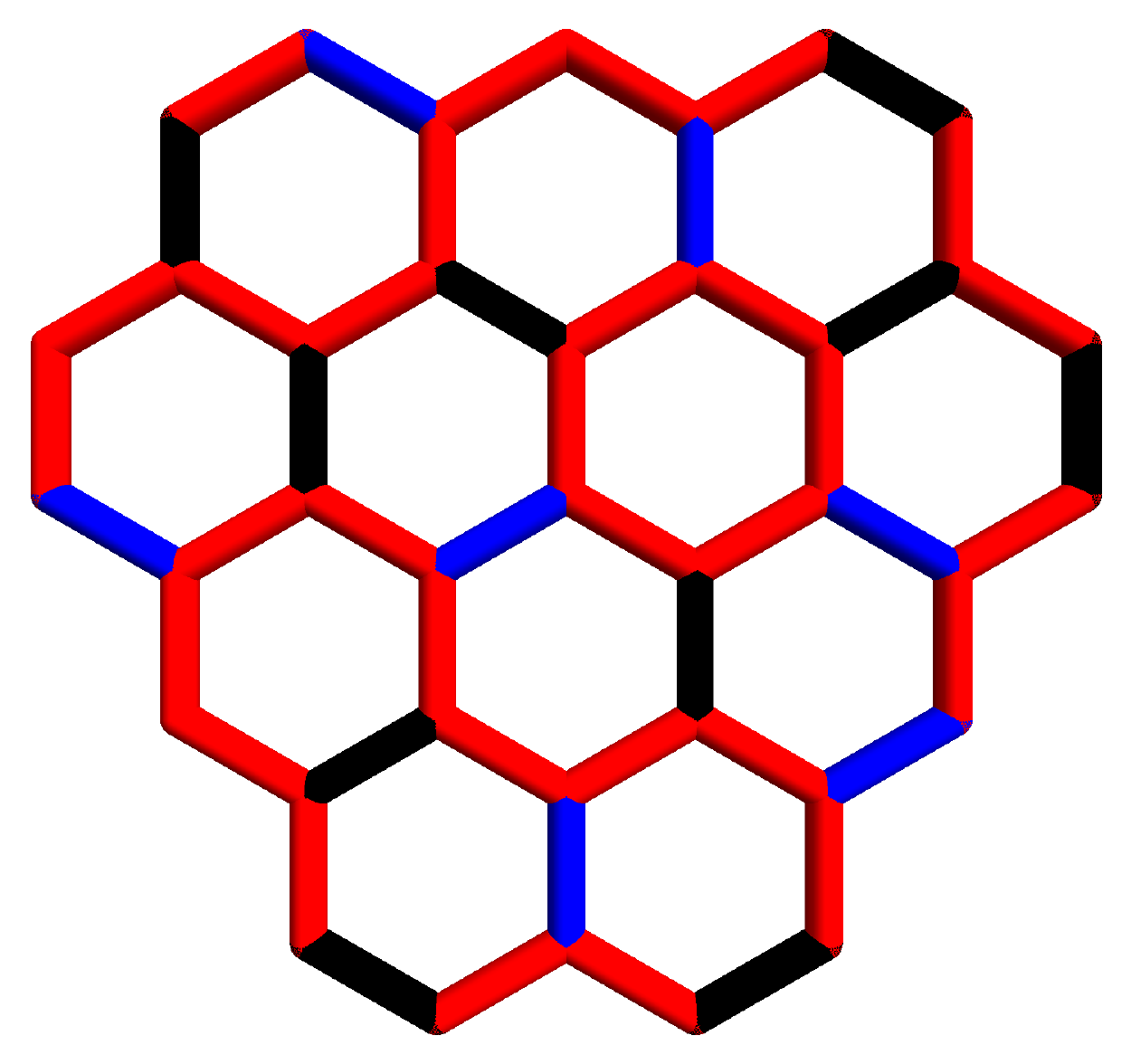}%
\end{minipage}
\begin{minipage}{.1\columnwidth}
\Large$\Rightarrow$
\end{minipage}
\begin{minipage}{.4\columnwidth}
\includegraphics[width=\columnwidth]{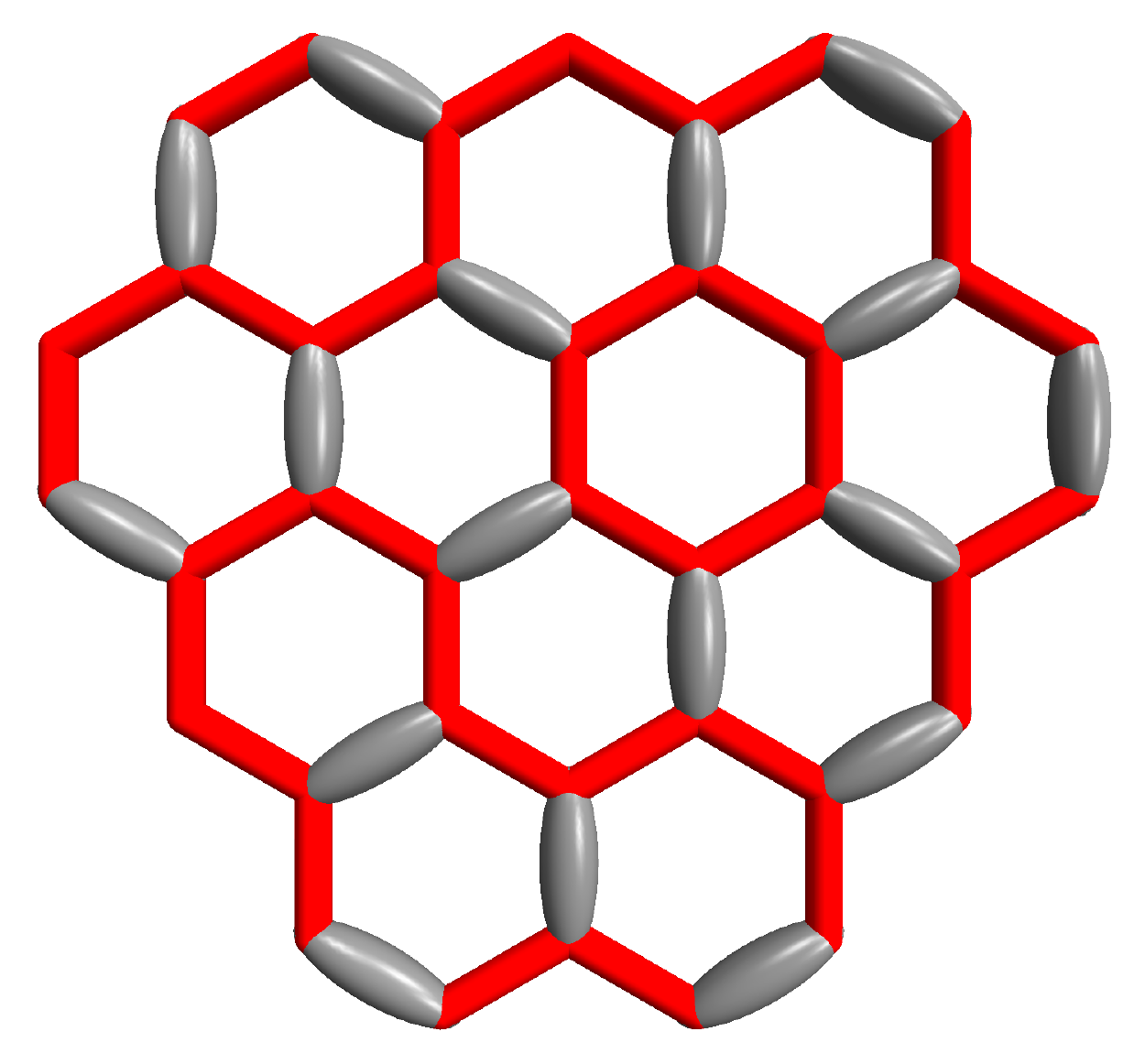}%
\end{minipage}
\end{minipage}
\caption{In the limit of large negative $\Jep$, the number of of $\msigma$-edges (depicted in red) is maximized leading to low-energy states, which can be described as dimer coverings of the hexagonal lattice. Dimers are depicted as thickened gray objects. The dimers have an 'internal' degree of freedom, corresponding to whether the corresponding edge is in the $\mone$ (black) or $\mpsi$ (blue) state.}%
\label{fig:dimeridentification}%
\end{figure}
In this low-energy Hilbert space the number of $\msigma$-labels on the edges is maximized. The corresponding projector $\bar{P}$ onto this Hilbert space therefore selects dimer coverings of the honeycomb lattice, as noted by Ref.~\onlinecite{schulz14}, with dimers representing edges that do not carry the $\msigma$-label. 
However, with this definition there are two vertex configurations in Fig.~\ref{fig:lattice_constraints} (up to rotations) with a single dimer. To account for this, 
each dimer carries an ``internal'' degree of freedom, indicating whether the corresponding edge is labeled $\mone$ (black) or $\mpsi$ (blue). We will use gray dimers to represent edges for which the label may be either $\mone$ or $\mpsi$. One example of this identification is shown in Fig.~\ref{fig:dimeridentification}. 

The effective low-energy Hamilton $H_{\rm eff}^A$ in this limit reads
\begin{align}
H_{\rm eff}^A=&\bar{P}H\bar{P}\\
=&-\frac{\Jp\sqrt{2}}{4}\sum\limits_p\bar{P}B_p^{\msigma}\bar{P}-\frac{\Jp}{4}\sum\limits_p\bar{P}B_p^{\mpsi}\bar{P}\nonumber\\&+\left(-\frac{\Jp}{4}+\Jep\right)N_{\rm p}\bar{P}\\
=&-\frac{\Jp}{8}\sum\limits_{p,i,f} \left(\beta(i,f) \left|\begin{array}{c}\includegraphics[width=.75cm]{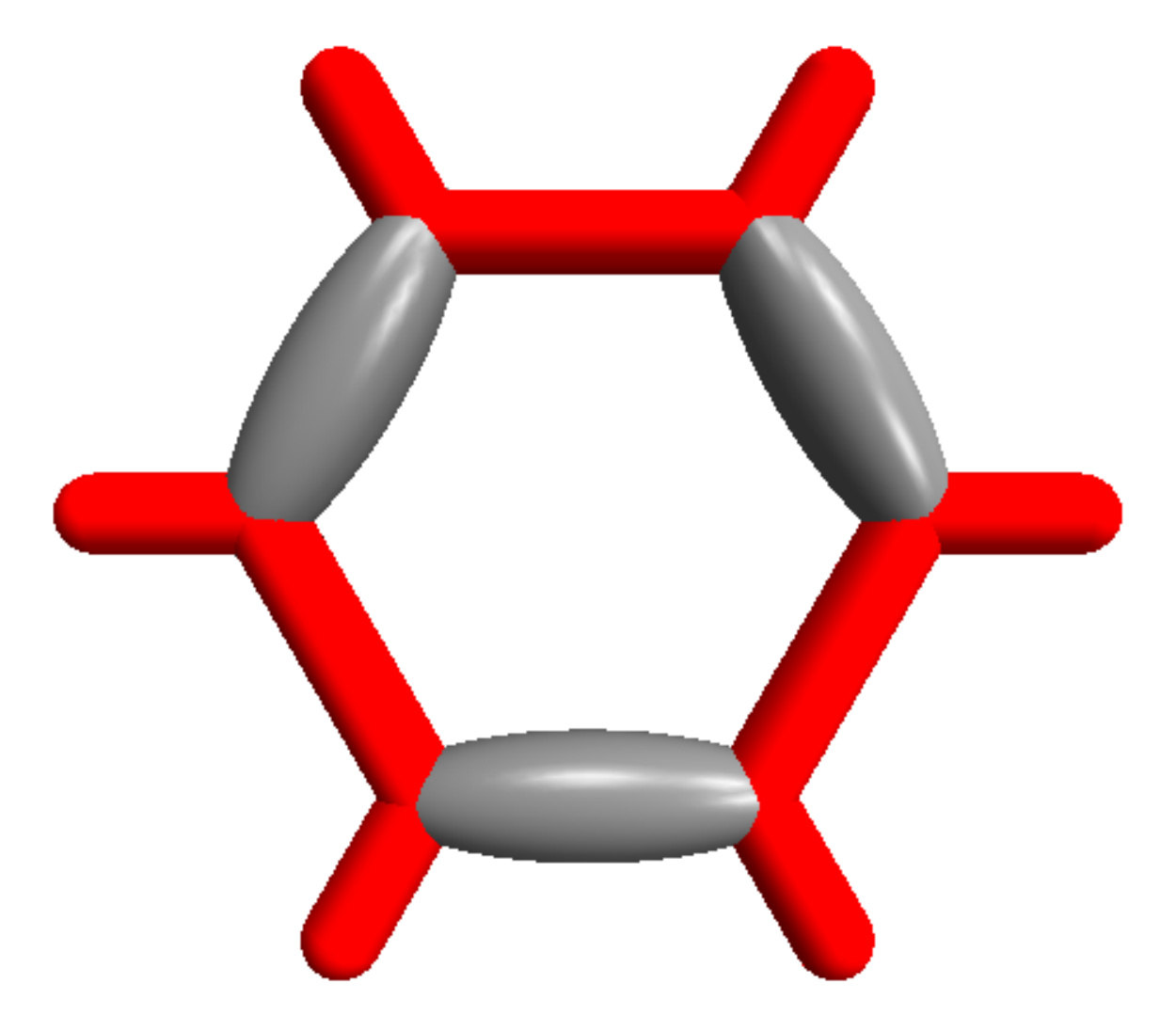}\end{array} \right\rangle\left\langle \begin{array}{c}\includegraphics[width=.75cm]{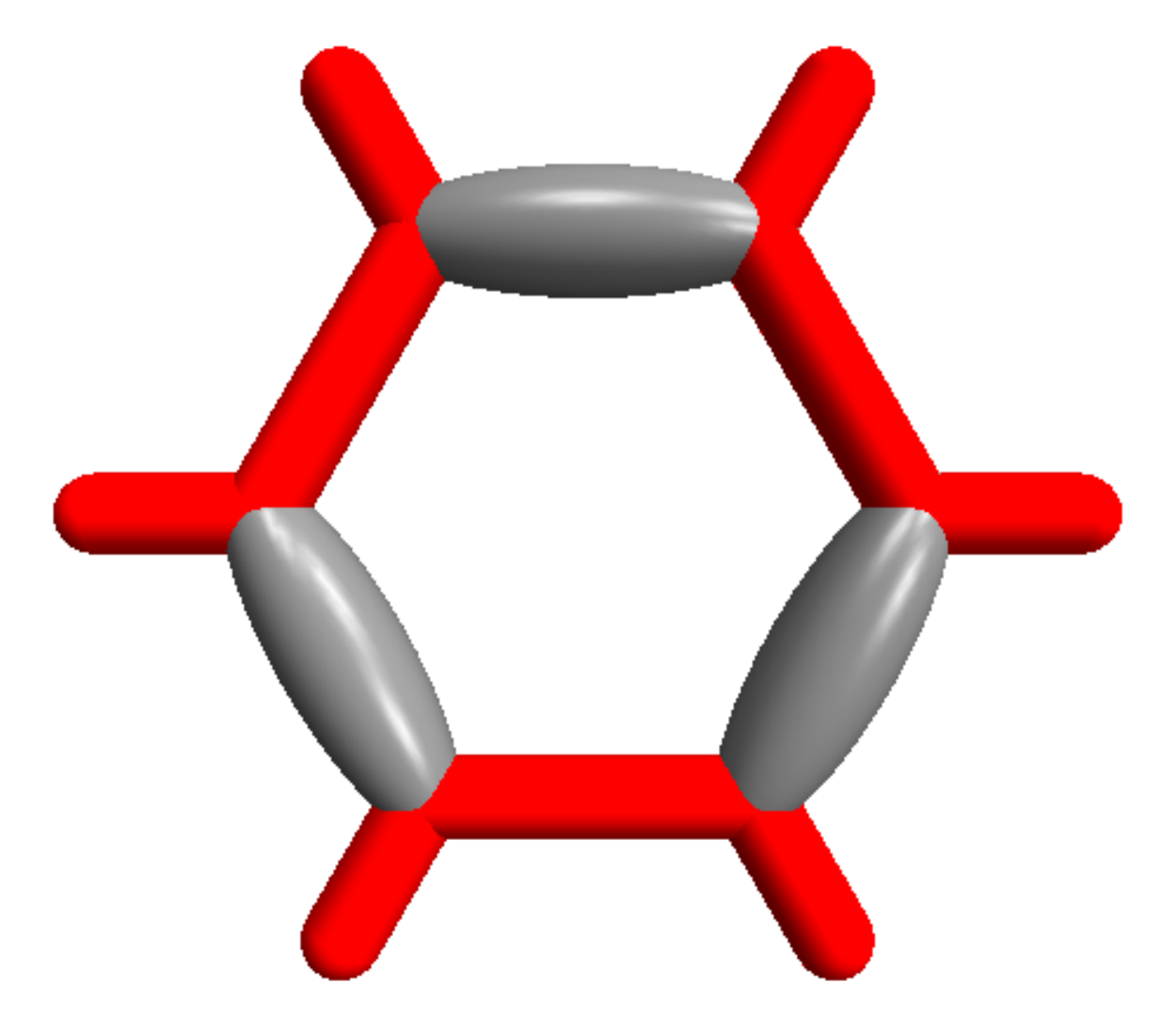}\end{array} \right| + {\rm h.c.}\right) \nonumber\\
&-\frac{\Jp}{4}\sum\limits_{p,i,f} \left(\gamma(i,f) \left|\begin{array}{c}\includegraphics[width=.75cm]{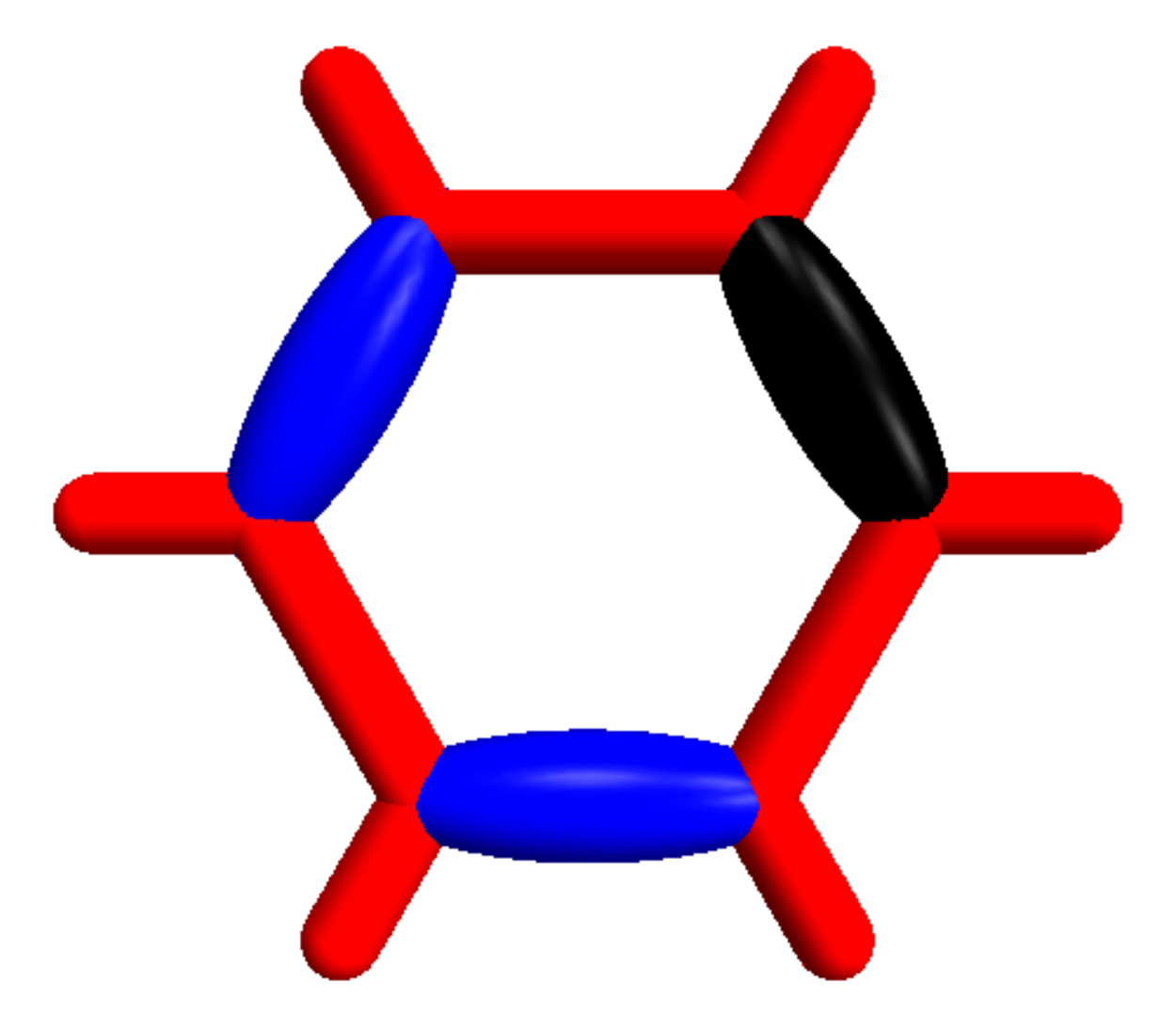}\end{array} \right\rangle\left\langle \begin{array}{c}\includegraphics[width=.75cm]{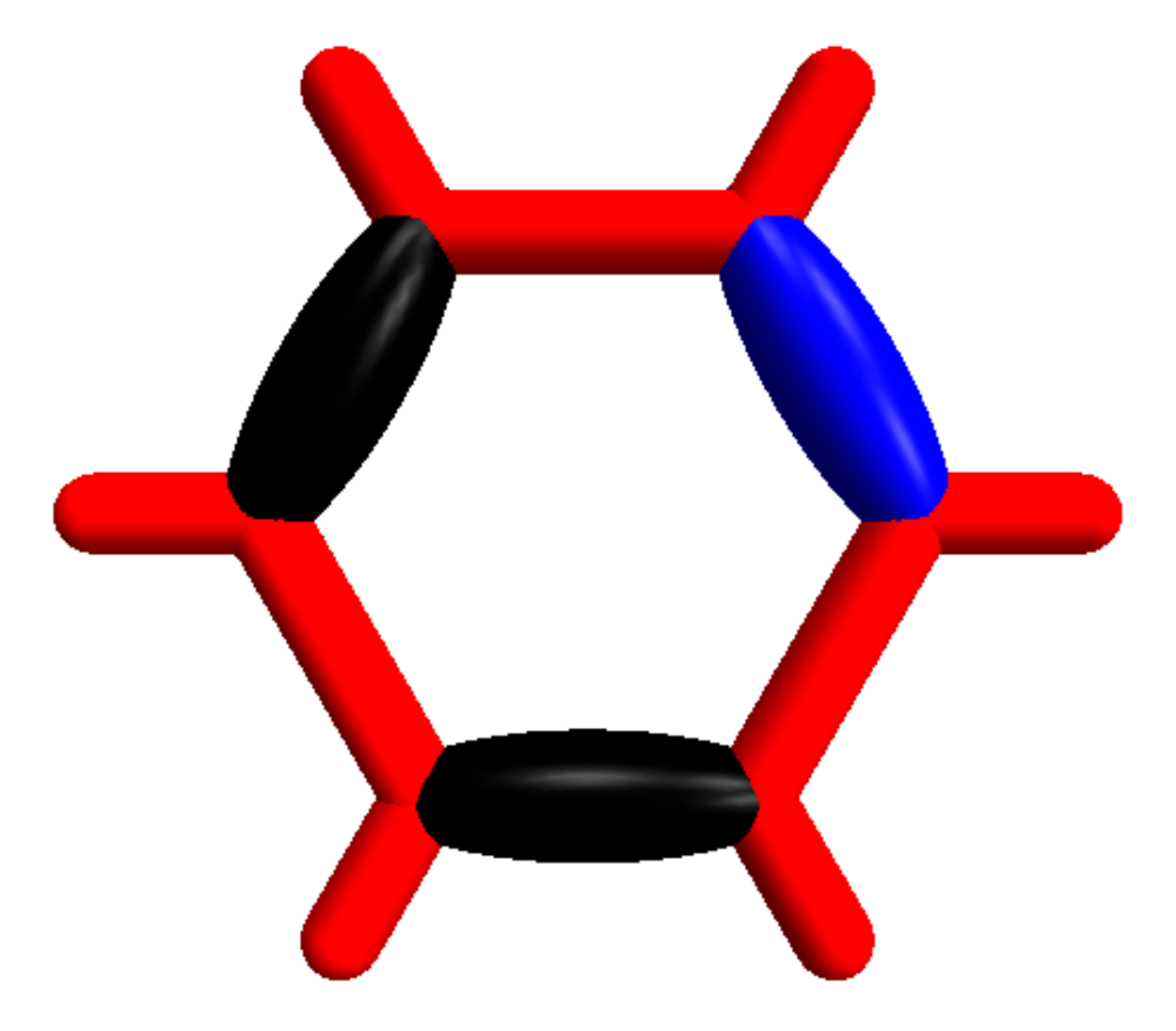}\end{array} \right| + \ldots\right)\nonumber\\&+\left(-\frac{\Jp}{4}+\Jep\right)N_{\rm p}\bar{P}
\label{eq:hameffa}
\end{align}
The first line of Eq.~(\ref{eq:hameffa}) describes the action of $\bar{B}_p^{\msigma}\equiv\bar{P}B_p^{\msigma}\bar{P}$. This term annihilates plaquettes with less than three dimers, and interchanges dimer and non-dimer edges on ``flippable" plaquettes (with exactly three dimer edges). Here we have left the action on the internal dimer labels ambiguous; however the amplitudes $\beta(i,f) \in \lbrace \pm 1\rbrace$, given in App.~\ref{app:dimer2dimer}, depend on both the internal dimer states and the dimer locations in the initial ($i$) and final ($f$) state. This is also the case for the coefficients $\gamma(i,f)$ appearing in the second line of Eq.~(\ref{eq:hameffa}), which gives the action of $\bar{B}_p^{\mpsi}\equiv\bar{P}B_p^{\mpsi}\bar{P}$. The operator $B_p^{\mpsi}$ flips the internal dimer labels and therefore commutes with $\bar{P}$. Consequently, the restriction to the dimer model has non-zero matrix elements when acting on any dimer configuration and does not favor any particular state. The ``$\ldots$" therefore represents a sum of such terms for all possible states allowed by the dimer and vertex constraints. Further, since $[B_p^{\mpsi}, \bar{P}] = [B_p^{\mpsi}, B_p^{\msigma}] =0$, $\bar{B}_p^{\mpsi}$ and $\bar{B}_p^{\msigma}$ commute with each other and therefore their effects can be considered separately.

\begin{figure}[htp]%
\begin{minipage}{.5\columnwidth}
\includegraphics[width=\columnwidth]{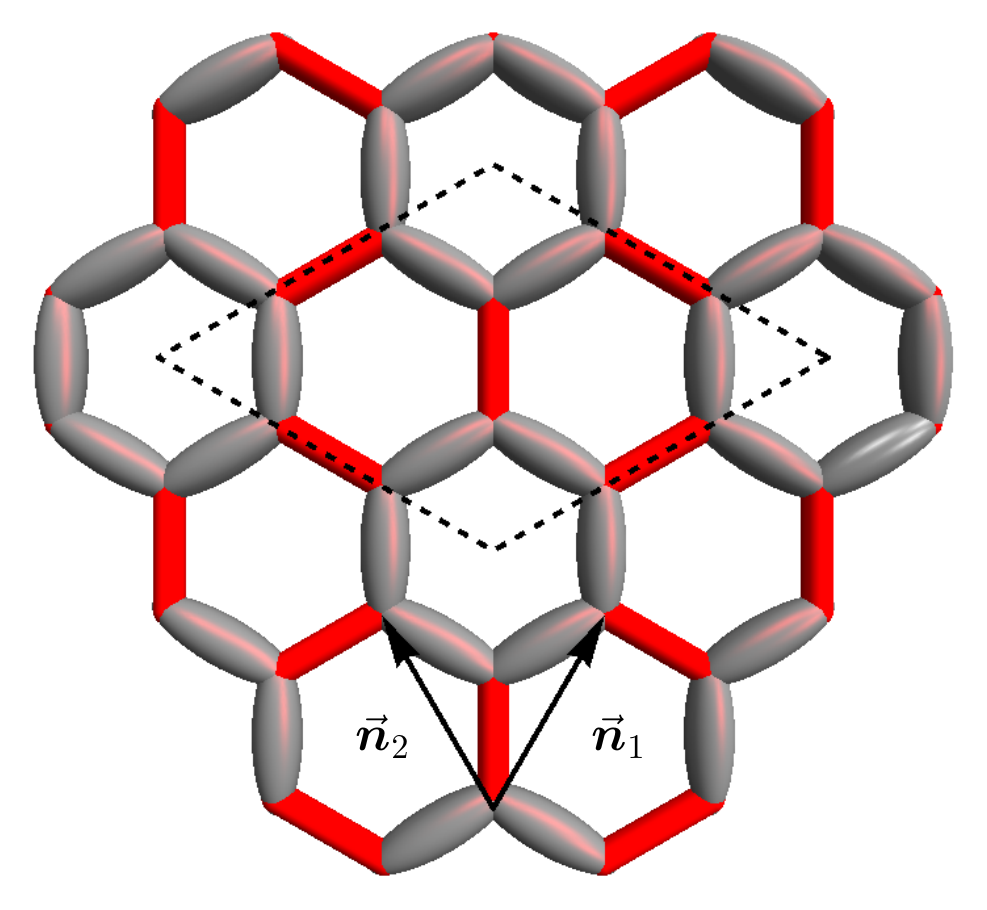}%
\end{minipage}
\begin{minipage}{.1\columnwidth}
\end{minipage}
\begin{minipage}{.35\columnwidth}
\vspace{1cm}
$\begin{array}{c}
\includegraphics[width=.75cm]{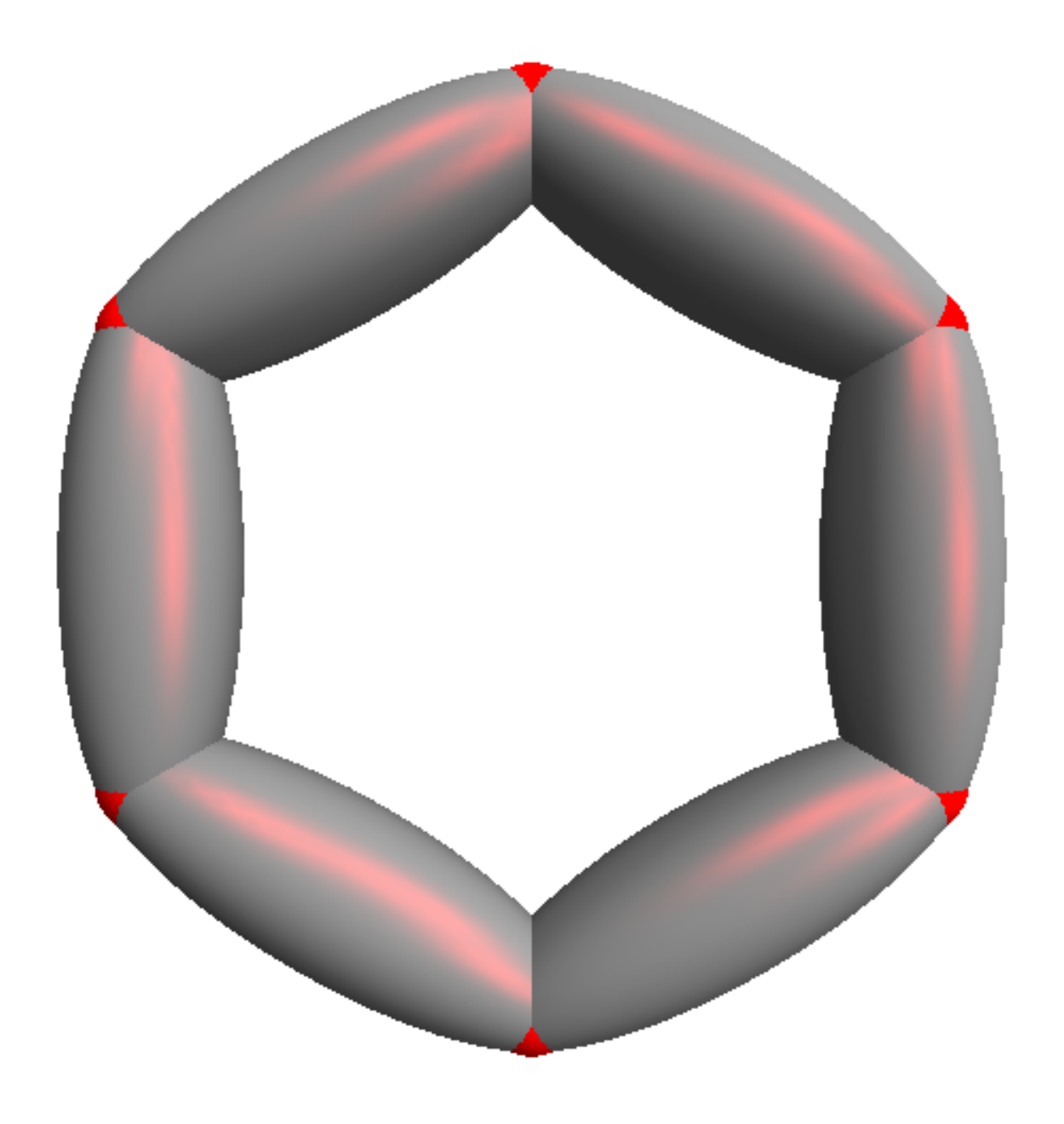}
\end{array}
\!\!\!=\!\!\!
\begin{array}{c}
\includegraphics[width=.75cm]{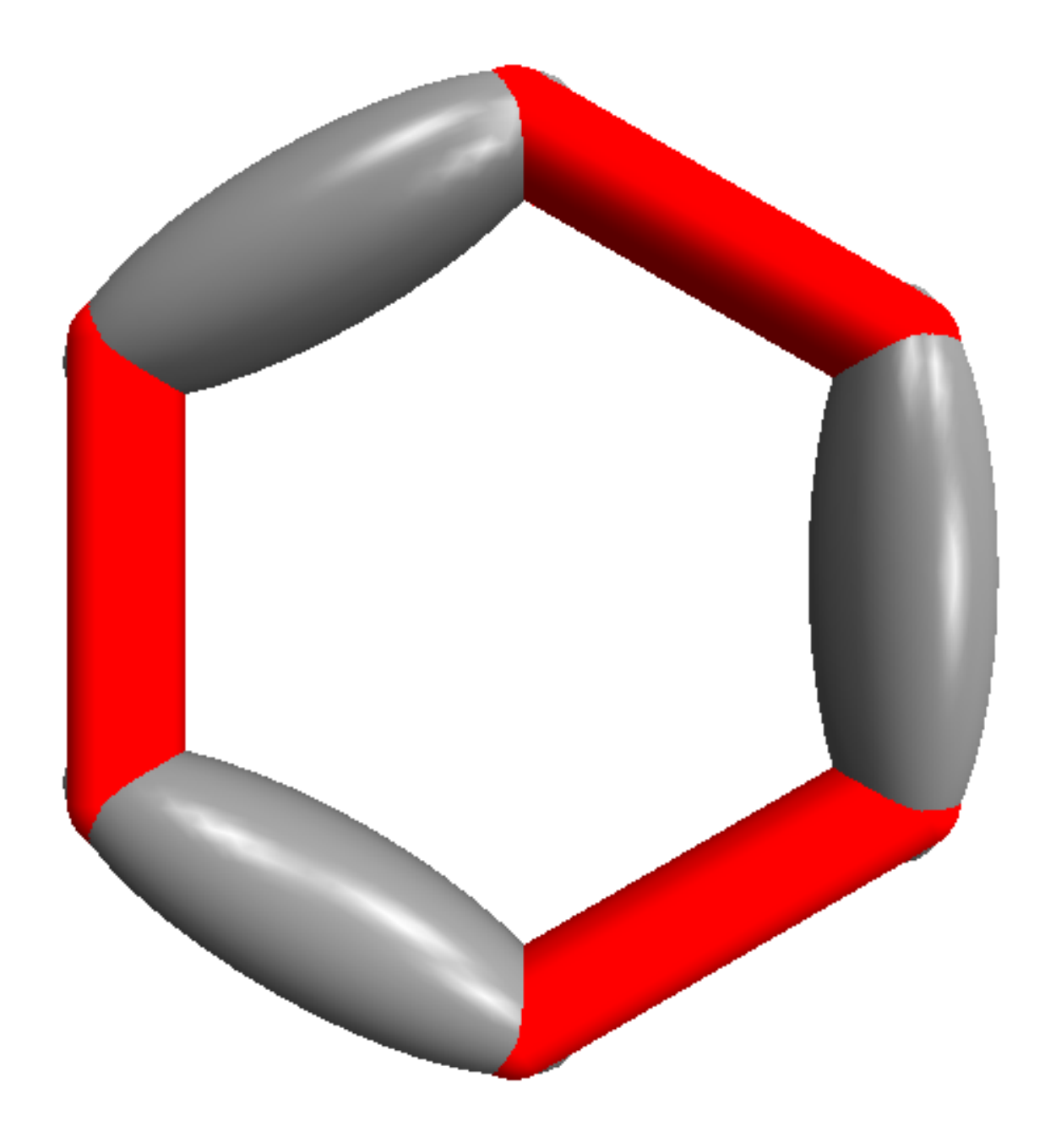}
\end{array}\!\!\!
+\!\!\!
\begin{array}{c}
\includegraphics[width=.75cm]{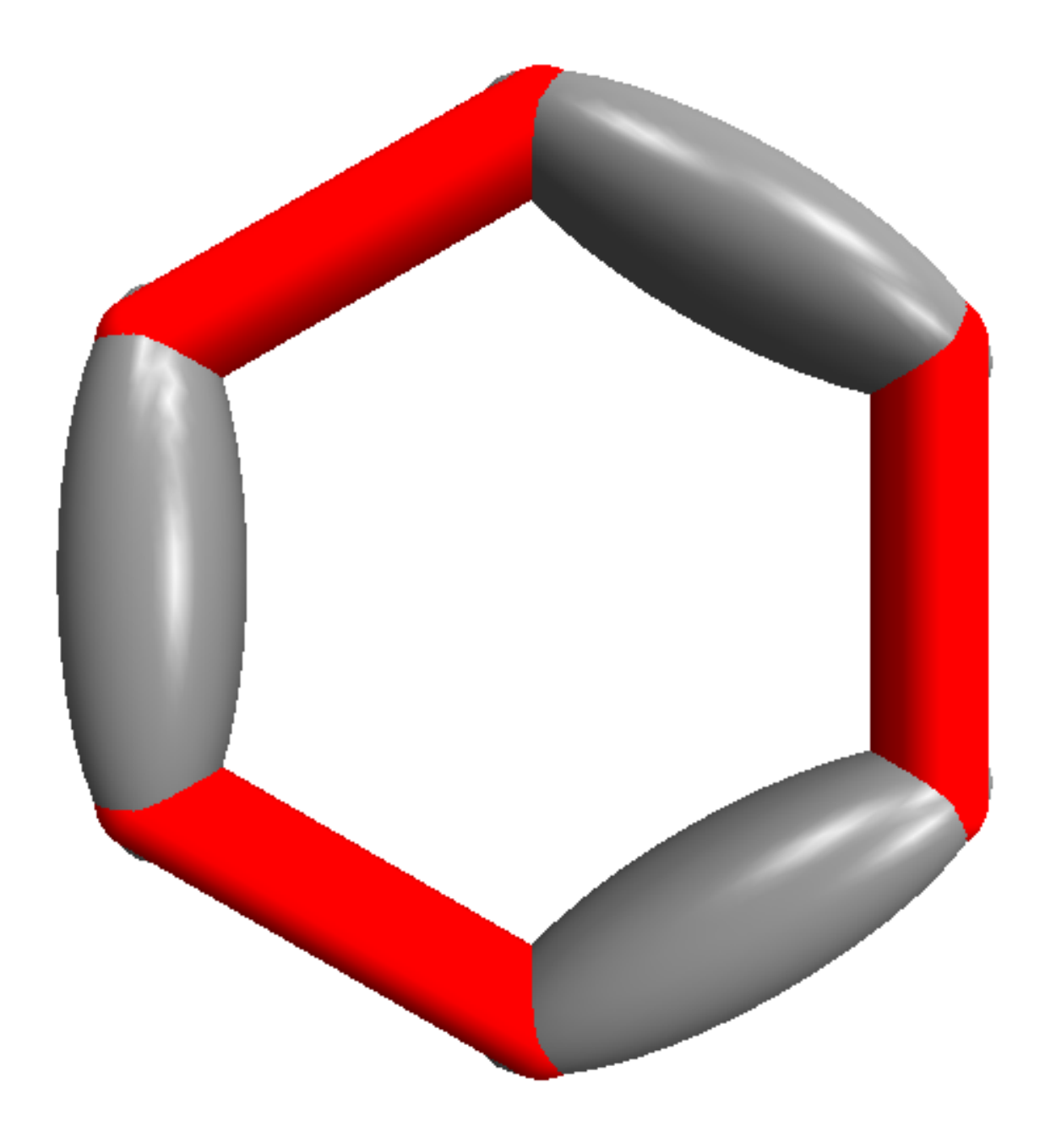}
\end{array}
$
\end{minipage}
\caption{Plaquette (positional) order of the dimers. The dimers fluctuate around $1/3$ of the plaquettes as shown on the right. The dashed lines depict the unit cell. The internal state of the dimers (depicted in gray) is either $\mone$ or $\mpsi$ and results in $\mathbb{Z}_2$-topological order on top of the translational symmetry broken background. }%
\label{fig:plaquette_order}%
\end{figure}

Because the second line of Eq.~(\ref{eq:hameffa}) affects only the internal dimer labels, the positional dimer order is completely determined by $\bar{B}_p^{\msigma}$. This operator corresponds to the so-called resonance-term of the quantum-dimer model on the honeycomb lattice\cite{moessner01}; it favors configurations in which the number of fluctuating plaquettes is maximized. 
This leads to a three-sublattice (so-called plaquette) order, which is adiabatically connected to the state depicted in Fig.~\ref{fig:plaquette_order}. In this phase, one third of the plaquettes are resonating (i.e.~in an eigenstate of $\bar{B}_p^{\msigma}$ with maximal eigenvalue $+1$ as depicted in the right side of Fig.~\ref{fig:plaquette_order}), whereas the other plaquettes remain frustrated (i.e.~non resonating).

This three-sublattice order is identical to that of the dual Ising model (\ref{eq:hameffaiz}), which for $-\Jep \gg \Jp$ is also described by an effective dimer Hamiltonian consisting only of this resonance term.\cite{moessner01b} 
In terms of the pseudo-spins introduced in Eq.~(\ref{eq:hameffaiz}), the resulting plaquette phase corresponds to a magnetization pattern {$\left\langle\tau^x\right\rangle=(0,m,-m)$} for the three different sublattices. Sites with $\left\langle\tau^x\right\rangle=\pm m$ correspond to non-resonating plaquettes, on which the $\mpsi$-flux eigenvalue (which is measured by the eigenvalue of $B_P^{\msigma}$) is not fixed.
Sites with magnetization $\left\langle\tau^x\right\rangle=0$ correspond to resonating plaquettes, which carry zero $\mpsi$-flux since $\bar{B}_p^{\msigma}$ has eigenvalue $1$. The corresponding pseudo-spins are therefore polarized along the $+\hat{z}$ axis.
Excited states with $\left|\downarrow\right\rangle_p$ (i.e.~a $\mpsi$-flux located on a resonating plaquette $p$), for which $\bar{B}_p^{\msigma}$ has eigenvalue $-1$, correspond to visons in the dimer model. These non-topological excitations of the dimer model are gapped with a gap renormalized from the bare value of $\Jp/8$ by about $25\%$\cite{schlittler15}. 

We note that generically, a quantum dimer model\cite{moessner01b} consists of two terms: a resonance term, which flips the position of the dimers along a plaquette of the underlying lattice and a so-called potential term, which assigns a relative energy cost to different dimer configurations. As discussed above, the former favors configurations that maximize fluctuations, and in our case leads to the plaquette order. The latter stabilizes other, less fluctuating, orders. Though the effective model (\ref{eq:hameffa}) contains only the resonance term, later we will see that for $\Jes \neq 0$ both terms are generated, leading to different translation-breaking orders. 

So far, we have ignored the internal degrees of freedom of the dimers, which describe (in a dual basis) precisely the degrees of freedom that are not captured by the effective spin Hamiltonian (\ref{eq:hameffaiz}). 
The unusual properties of our three-sublattice ordered phase become apparent when we consider their dynamics, given by the second line of Eq.~(\ref{eq:hameffa}). Since $\left(\bar{B}_p^{\mpsi}\right)^2=\bar{P}$, $\bar{B}_p^{\mpsi}$ has possible eigenvalues $\pm 1$ in the dimer subspace. Because this term does not compete with the dimer fluctuations, the ground state(s) $\left|\Psi_0\right\rangle$ obey $\bar{B}_p^{\mpsi}\left|\Psi_0\right\rangle=+\left|\Psi_0\right\rangle$. We will show that the resulting ground state superposition of different internal dimer configurations leads to $\mathbb{Z}_2$-topological order.

To demonstrate topological order, we will construct a set of loop operators, valid in the dimer limit, which lead to a four-fold ground state degeneracy on the torus. We also discuss (with more details given in App.~\ref{app:stringoperatorsz2details}) the low-energy excitations in this phase, proving that the quasi-particle types are isomorphic to those of the Toric code\cite{kitaev03}.

In contrast to the ferromagnetic case, in the antiferromagnetic phase there are dimer configurations in which we can interchange $\msigma$ and non-$\msigma$ edges along extended non-contractible loops without leaving the dimer Hilbert space. 
Hence $\overline{P}$ does not annihilate operators such as $W_{\mathcal{C}_i}^{(\msigma,\mone)}$. 
Instead, operators $W_{\mathcal{C}_i}^{(\alpha,\beta)}$ which change $\msigma$ to $\mone$ or $\mpsi$ either map a given state outside of the dimer Hilbert space, or to a state with a defect in the long-ranged three-sublattice plaquette order along the length of the string $\mathcal{C}_i$. Thus extended $\msigma$-strings have finite energy cost, and are not associated with topological order.
In contrast, string-operators which only act on the internal states of the dimers do not disrupt the long-range order, and act within the dimer Hilbert space.  

Because they also commute with $H_{\rm{SN}}$ before projecting to the dimer subspace, these loop operators map the system between (topologically) distinct ground states.  In analogy to Eqs.~(\ref{eq:stringoperatorsz2_1a}-\ref{eq:stringoperatorsz2_2}),  we have:
\begin{align}
\bar{P}_pW_{\mathcal{C}_i}^{({\mone},{\mone})}\bar{P}_p=\bar{P}_pW_{\mathcal{C}_i}^{({\mpsi},{\mpsi})}\bar{P}_p\equiv W_{\mathcal{C}_i,\rm{AFM}}^{\mone},\label{eq:afmops1}\\
\bar{P}_pW_{\mathcal{C}_i}^{({\mpsi},{\mone})}\bar{P}_p=\bar{P}_pW_{\mathcal{C}_i}^{({\mone},{\mpsi})}\bar{P}_p\equiv W_{\mathcal{C}_i,\rm{AFM}}^{\meps},\label{eq:afmops2}\\
\bar{P}_pW_{\mathcal{C}_i}^{({\msigma},{\msigma})}\bar{P}_p\equiv W_{\mathcal{C}_i,\rm{AFM}}^{b_1} + W_{\mathcal{C}_i,\rm{AFM}}^{b_2}.\label{eq:afmops3}
\end{align}
The explicit construction of those operators is detailed in App.~\ref{app:stringoperatorsz2details}. As in the ferromagnetic case, these relations are justified by the absence of extended $\msigma$-strings, such as $W_{\mathcal{C}_i}^{(\msigma,\mone)}$, in the low-energy Hilbert space. Since $\msigma$-strings are confined at long length scales by the three-sublattice order, the relations remain valid everywhere in the translation-breaking phase. However, the reduced topological order is tied to the translational symmetry breaking, such that it is not possible to separate the topological and symmetry-breaking phase transitions.

It is worth elaborating on the nature of the loop operators in this case, to clarify why low-energy extended $\msigma$-loops are required to alter the topological order. 
In both Eqs.~(\ref{eq:afmops1}) and (\ref{eq:afmops2}), the matrix elements given in App.~\ref{app:stringoperatorsz2details} for the two loop operators of the Ising string net differ by phase factors of $-1$ for each $\msigma$-edge crossed by the non-contractible curve $\mathcal{C}_i$. However, in the ordered phase $\mathcal{C}_i$ will cross an even number of $\msigma$-edges in any low-energy state, such that each pair of operators have identical matrix elements in the low-energy Hilbert space.

Eq.~(\ref{eq:afmops3}) is slightly more involved. In the absence of non-contractible $\msigma$-loops, our lattice admits a bipartition into ``black" and ``white" regions, separated by $\msigma$-loops. Within each domain type, the operator $W_{\mathcal{C}_i}^{(\msigma,\msigma)}$ splits into two operators, one of which raises the edges in $\mathcal{C}$ by $\mpsi$, analogous to $W_{\mathcal{C}_i,\rm{AFM}}^{\me}$ in the limit $\Jep \gg \Jp$, and the other of which measures the number of $\mpsi$-labeled edges crossed by $\mathcal{C}$, analogous to $W_{\mathcal{C}_i,\rm{AFM}}^{\mm}$. (See App.~\ref{app:stringoperatorsz2details} for details). Upon crossing from a ``white" to a ``black" region (i.e. upon crossing a $\msigma$-edge), the two types are interchanged. In other words, one of the two string-operators arising from $W_{\mathcal{C}_i}^{(\msigma,\msigma)}$ raises the edges by $\mpsi$ in the black partition, and measures crossed $\mpsi$-edges in the white partition; the other operator measures the $\mpsi$-labels in the black partition, and raises them by $\mpsi$ in the white partition. The crucial point is that though $\msigma$-loops are densely packed in the ordered state, an ambiguity between these two operators can arise only when the distinction between black and white regions is lost -- in other words, only when $\mathcal{C}$ crosses an odd number of $\msigma$-edges. We refer to the two distinct loop operators as $W^{b_1}$ and $W^{b_2}$. 

In App.~\ref{app:stringoperatorsz2details}, we also describe how to construct open string operators in the dimer limit. Interestingly, in contrast to the $\mathbb{Z}_2$ phase at $\Jep \gg \Jp$, in the dimer limit all three quasi-particle types can be realized, even in the presence of vertex constraints, and we explicitly give open string operators for two mutually semionic bosons, and one fermion. The corresponding string endpoints create a minimum of either one (for $W^{b_i}$) or two (for $W^{\meps})$ plaquettes on which $B_p^{\mpsi}$ has eigenvalue $-1$.  In the latter case, one defect is necessarily in the ``black" region, and the other in the ``white" region.  This means that the two bosons $b_1$ and $b_2$ are in fact distinguished by which of these regions they occupy, as the difference between a boson in the black region and a boson in the white region is the fermion.  (As discussed in App.~\ref{app:stringoperatorsz2details}, open string operators necessarily create either pairs of $b_1$ or pairs of $b_2$ excitations).

It is interesting to note that the energy of such a defect depends on which sublattice(s) the flux defect(s) occupy. 
As operators, we have $B_{p}^{\msigma} = B_{p}^{\msigma} B_{p}^{\mpsi}$, from which it follows that if $B_p^{\mpsi}\left|\bar{\Psi}\right\rangle = - \left|\bar{\Psi}\right\rangle$, 
\begin{align}
B_p^{\msigma}\left|\bar{\Psi}\right\rangle=& B_p^{\msigma} \frac{1}{2}\left(\mathds{1}+B_p^{\mpsi}\right)\left|\bar{\Psi}\right\rangle = B_p^{\msigma}\frac{1}{2}(1-1)\left|\bar{\Psi}\right\rangle=0.
\label{eq:nofluctuationonflux}
\end{align}
Consequently, plaquettes that are eigenstates of $B_p^{\mpsi}$ with eigenvalue $-1$ cannot resonate. Therefore in the plaquette-ordered ground state the gap of these excitations is $\Delta\approx\frac{\Jp}{2}$ for non-resonating plaquettes, but $\Delta\approx\frac{5 \Jp}{8}$ for a resonating plaquette.

Because the bosonic string operators are ``$\me$-like'' in one region, and ``$\mm$-like" in the other, we will call the two bosons $b_1$ and $b_2$, rather than $\me$ and $\mm$. The reason that this does not conflict with $\mathbb{Z}_2$ topological order is that this last is invariant under interchange of the two bosons $\me$ and $\mm$: such an exchange preserves all topological data, including the particles' mutual statistics. Thus in our model the mutual statistics, which are determined by the commutation relations of the string operators far from their endpoints, are well-defined independent of whether the two defects are in the same region. 

From the above discussion, it is apparent that any topological distinction between $b_1$ and $b_2$ (or $W^{b_1}$ and $W^{b_2}$) disappears in the presence of extended $\msigma$-lines (or of open $\msigma$-lines, if we were to allow these in our Hilbert space), since in this case we can bring a $b_1$ excitation around a non-contractible curve on the lattice and have it return to the same point as a $b_2$. This is anticipated by Refs.~\onlinecite{kitaev12,barkeshli13,barkeshli13b,barkeshli13c}, who showed in the Toric code that crossing such a defect line interchanges the excitations $\me\leftrightarrow\mm$. 
As a consequence, once the linear confining energy for extended $\msigma$-loops (or open $\msigma$-strings) disappears, $b_1$ and $b_2$ are no longer physically distinct excitations, but rather the two internal states of the non-Abelian $\msigma$-flux defect found in the \Isi phase. 

Thus, we have established that in the regime $\Jep\rightarrow-\infty$ (point $\rm A$ in Fig.~\ref{fig:phase_diag}) we have on the one hand the long-ranged order and translational symmetry breaking due to the dimer locations, and on the other hand topologically-ordered internal states of the dimers. 
Further, we have shown that disintegration of the long-ranged order (for $\Jes=0$) necessarily restores the full topological order of the \Isi phase. 

\subsection{Away from \texorpdfstring{$\Jes=0$}{Js=0}: numerical results}
We conclude this section with a discussion of the fate of the phase transitions described in Sec.~\ref{Sec:z2p} for finite $\Jes$. Because the nature of the condensing excitation remains the same at finite $\Jes$, on general grounds we expect that both of these transitions remain in the universality class of the $\Jep=0$ line throughout the region separating the $\mathbb{Z}_2$ topologically ordered phases from the \Isi phase. Here we present numerical results supporting this expectation.

We begin with the phase transition between the \Isi- and the $\mathbb{Z}_2$-phase at positive $\Jep$, which is in the $3D$ classical Ising universality class for $\Jes=0$. 
In that case the critical value is known from Monte Carlo simulations to be at $\left.\Jep/\Jp\right|_c=0.419$\cite{bloete02} (point $\rm a$ in Fig.~\ref{fig:phase_diag}), with an exponent for the gap closure $z\nu=0.63$\cite{hasenbusch10}. Our series expansion at this point gives a transition at $\left.\Jep/\Jp\right|_c=0.415$ and an exponent of $z\nu=0.637$, in good agreement with the exact results. 

\begin{figure}[htp]%
\includegraphics[width=.8\columnwidth]{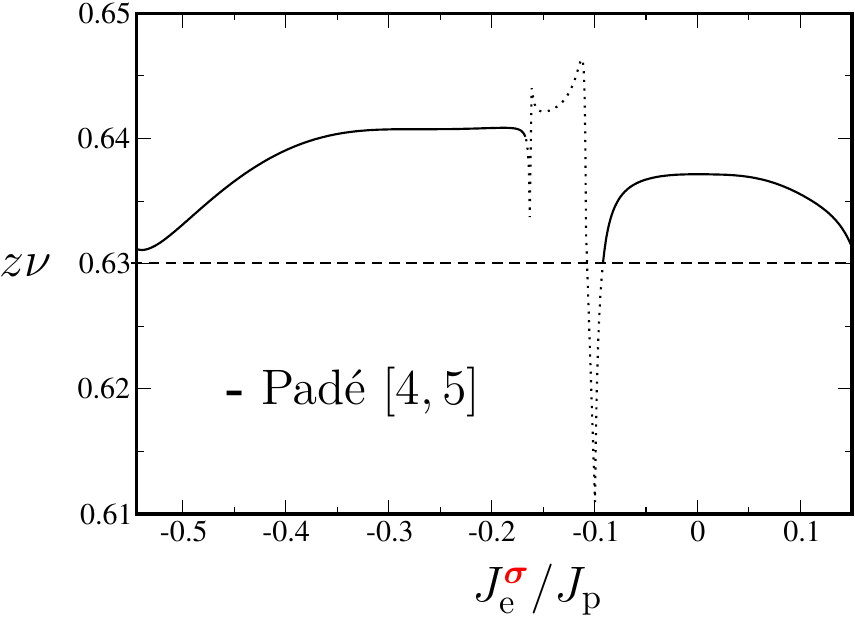}%
\caption{The critical exponent $z\nu$ obtained from the gap closure via a dlog-Pad\'e $[4,5]$ extrapolation for the transition between the Ising (I) and the ferromagnetic topological phase (Z). The dotted line indicates values obtained from defective extrapolants\cite{domb89}. The deviations from the Monte Carlo value of $0.63$\cite{hasenbusch10}, depicted as dashed line, appear to be within the precision of the series expansions, which we estimate from the discrepancy between series expansion and Monte Carlo results at  $J_e^{\msigma }=0$.\cite{schulz12}}%
\label{fig:z2fm_exponent}%
\end{figure}

Given the good agreement between series expansion at $J_e^{\msigma }=0$ and the best-known results for the 3D Ising critical point, what does series expansion predict about the nature of the phase transition along the critical line ($\rm b-a-g$)?  For $J_e^{\msigma}\neq 0$ the $\msigma$-excitations become mobile, and strictly speaking, the dual mapping to the effective spin Hamiltonian $H_{\rm{eff}}^{\rm{AIZ}}$ (\ref{eq:hameffaiz}) is not valid anymore. We give the leading orders of the corresponding series in App.~\ref{app:seriesising}. The main result is that these predict a value of the critical exponent $z\nu$ that remains constant (within the uncertainties of the method, which can be estimated from its error at $J_e^{\msigma }=0$) up to the phase boundaries of the $\mathbb{Z}_2$ phase (points $\rm b$, $\rm g$ in Fig.~\ref{fig:phase_diag}), as shown in Fig.~\ref{fig:z2fm_exponent}. 

\begin{figure}[htp]%
\includegraphics[width=\columnwidth]{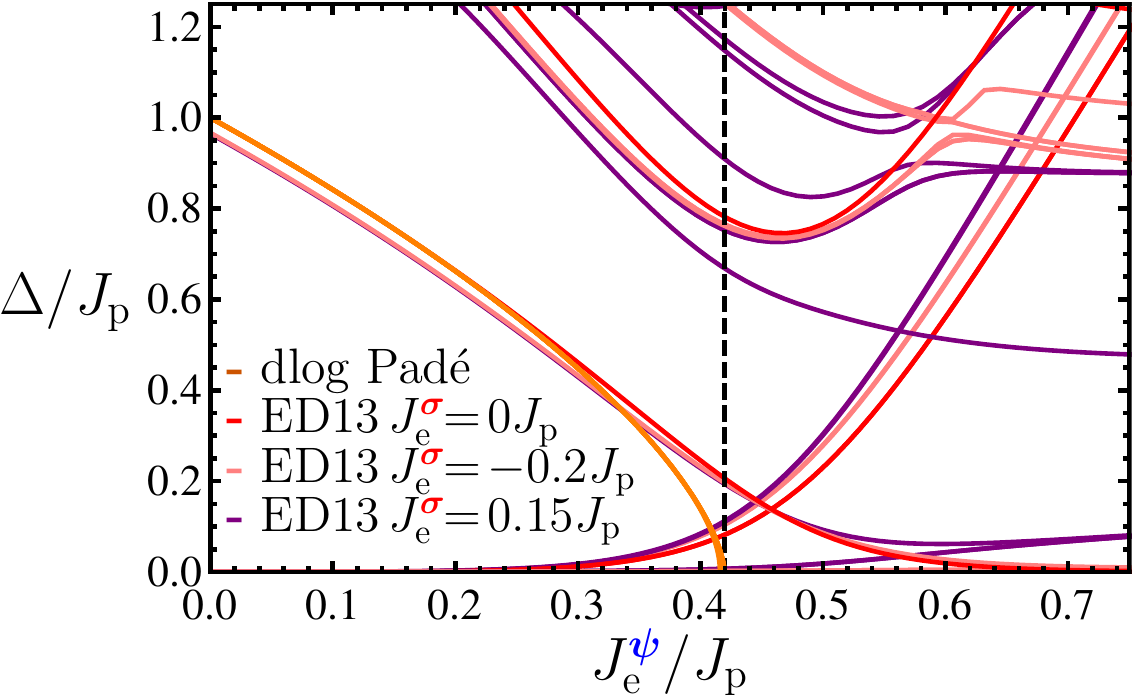}%
\caption{Low-energy gap in the regime of the transition between the Ising ($\Jep=0$) and the topological phase ($\Jep \gg \Jp$). The exact diagonalization results for the $\vec{k}=\vec{0}$ sector are shown for the largest system ($N_{\rm p}=13$) in red for $\Jes=0\Jp$. Note that the Hilbert space of the Hamiltonians (\ref{eq:fullham},\ref{eq:heff_aiz}) contains the single-particle mode condensing at the phase transition $\left.\frac{\Jep}{\Jp}\right|_c=0.415$ (dashed line). The orange lines denote the dlog-Pad\'e extrapolants obtained from series expansion for the single-$\mpsi$-flux mode in the Ising phase up to order $10$. For $\Jes=0.15\Jp$ ($\Jes=-0.2\Jp$), the ED-spectrum is shown in purple (pink).  The low-energy features are similar to those at $\Jes =0$.}%
\label{fig:z2fm_spectrum}%
\end{figure}

Fig.~\ref{fig:z2fm_spectrum} compares the gap to the first excited state obtained from series expansion with the low-energy spectrum of the full topological Hamiltonian Eq.~(\ref{eq:fullham}) obtained from exact diagonalization. The ED-data indicates a transition around $\Jep/\Jp\approx 0.45$ overestimating this value by roughly $10\%$. 
Additionally, the ED-spectrum shown in Fig.~\ref{fig:z2fm_spectrum} for values of $\Jes$ close to the boundary of the resulting $\mathbb{Z}_2$-phase is very similar to that at $\Jes=0$ (point $\rm a$ in Fig.~\ref{fig:phase_diag}), suggesting that there are no qualitative changes to the critical behavior along this line. This is in agreement with the (low-order) perturbative arguments (for small $\Jes/\Jp$) in Ref.~\onlinecite{burnell11b}, but also extends also all the way up to the phase boundary of the $\mathbb{Z}_2$ topological phase. 
Therefore our numerics support our expectation that the perturbations introduced by (gapped) $\msigma$-fluxes at the critical point do not change the universality class of the transition. 


\begin{figure}[htp]%
\includegraphics[width=.8\columnwidth]{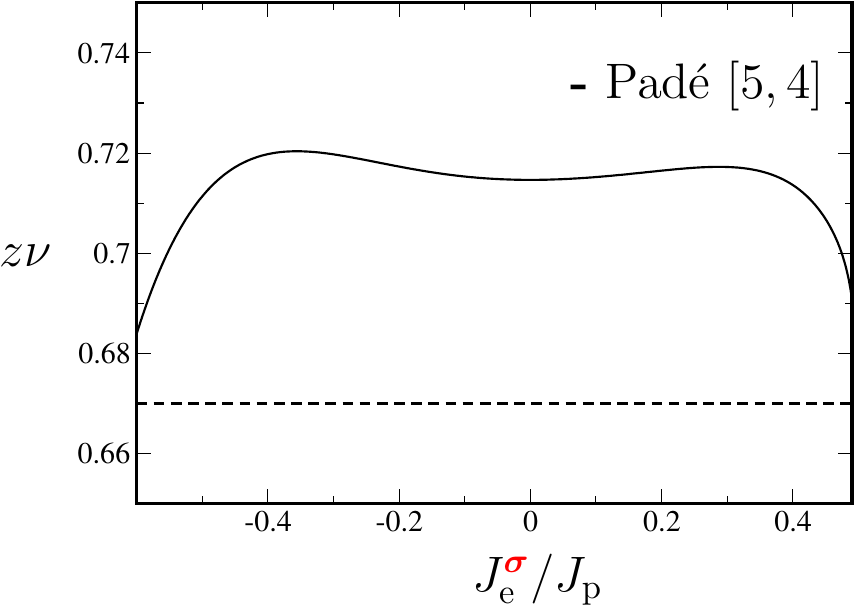}%
\caption{The critical exponent $z\nu$ obtained from the gap closure via a dlog-Pad\'e extrapolation $[5,4]$ for the transition between the Ising and the anti-ferromagnetic topological phase. The deviations from the Monte Carlo value $z\nu=0.67$\cite{gottlob94}, depicted as dashed line, are typical for the series expansion results\cite{schulz12}.}%
\label{fig:expsafm}%
\end{figure}

\begin{figure}[htp]%
\includegraphics[width=\columnwidth]{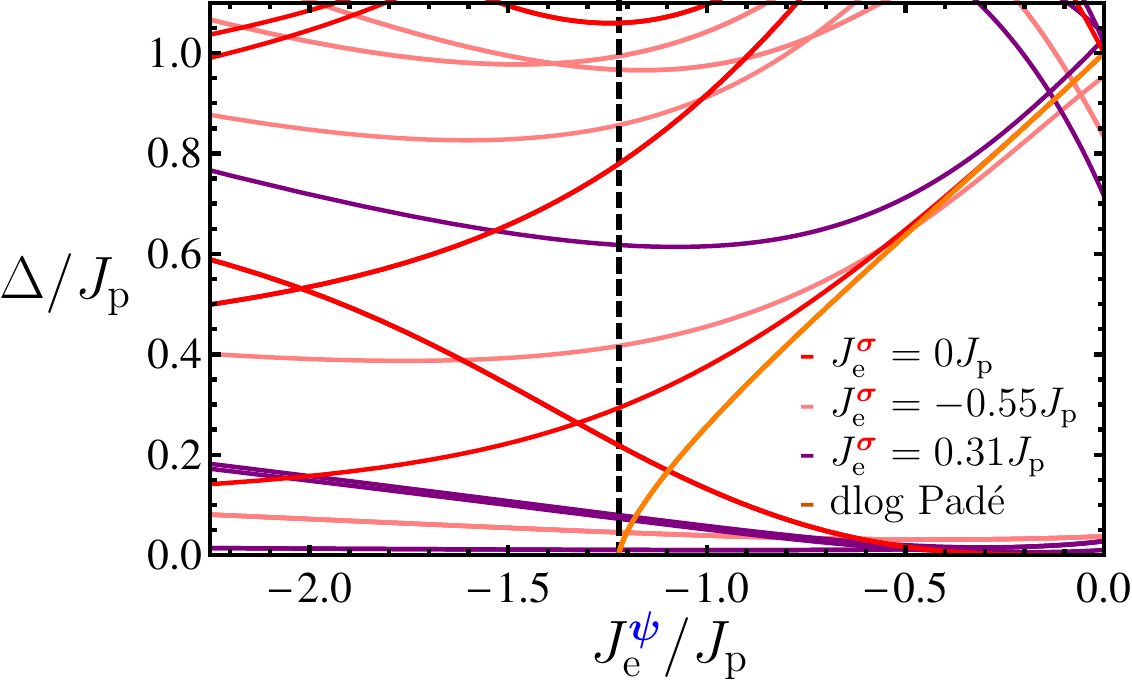}%
\caption{Low-energy gap in the regime of the transition between the Ising ($\Jep=0$) and the topological phase ($-\Jep \gg \Jp$). The exact diagonalization results for the $\vec{k}=\vec{0}$- and $\vec{k}=(\pm \frac{2\pi}{3},\mp \frac{2\pi}{3})$-sector are shown in red for the largest system commensurate with the three-sublattice order ($N_{\rm p}=12$). Note that the Hilbert space of the Hamiltonians (\ref{eq:fullham},\ref{eq:heff_aiz}) contains the single-particle mode condensing at the phase transition $\left.\frac{\Jep}{\Jp}\right|_c=-1.222$ (dashed line). The orange lines denote the dlog-Pad\'e extrapolants obtained from series expansion for the single-$\mpsi$-flux mode in the Ising phase up to order $10$. The low-energy features are similar for non-zero $\Jes$ indicating the phase transition to be of the same universality class. Note however the enhanced finite-size splittings to the vicinity of the topological phase for $\Jes>0$.}%
\label{fig:aiffm}%
\end{figure}

For $\Jep<0$, the universality class of the transition between the symmetry-breaking topological and the \Isi-phase is that of the classical 3D XY-model\cite{blankschtein84,isakov03,powalski13}. High-order series expansion of the Hamiltonian (\ref{eq:heff_aiz}) (see Fig.\ref{fig:aiffm}) pinpoints the phase transition at $\left.\Jep/\Jp\right|_c=-1.222$ (point $\rm e$ in Fig.~\ref{fig:phase_diag}), and the critical exponent (Fig.~\ref{fig:expsafm}) $z\nu=0.714$. This is in reasonable agreement with the literature for the 3D- model: Quantum Monte Carlo studies give a transition at $\left.\Jep/\Jp\right|_c=-1.212$\cite{isakov03}, whereas both series expansions studies\cite{powalski13} and Monte Carlo\cite{gottlob94} give $z \nu = 0.67$. 
In Fig.~\ref{fig:expsafm}, we show our results for the transition between these two phases for $\Jes\neq 0$, i.e. along the line $\rm d-e-f$. As for $\Jep>0$, the exponent remains constant (within the uncertainties of the method). Again, this is consistent with our expectation that introducing gapped $\msigma$-flux excitations should not change the universality class.

In Fig.~\ref{fig:aiffm}, we show the low-energy spectrum for $\Jes=0$, $\Jep<0$.  Again, the ED-data overestimates the location of the transition by roughly $10\%$ yielding $\Jep/\Jp\approx-1.37$.
We also display results for $\Jes\neq 0$, for which the spectra are qualitatively similar (though quantitatively different), in that we do not find evidence for intermediate phases or phase transitions as $\Jep$ increases. For $\Jes =0$ this follows from the known behavior of the effective model (\ref{eq:hameffaiz}), and our numerics suggest that this remains the case for over a range of $|\Jes|>0$.  


In summary, we have established that there are three distinct topologically ordered phases on the line $\Jes=0$. In particular, in addition to the \Isi and $\mathbb{Z}_2$-topological phase already discussed in the literature, there exists a phase in the regime of large negative $\Jep$ which simultaneously exhibits topological and long-ranged order. 
Additionally, we have identified the phase transitions separating these phases, and presented numerical evidence that they remain in the universality class appropriate to the transverse-field triangular lattice Ising model for all values of $\Jes$ for which these phases persist.

\section{Away from the topological line}\label{sec:topaway}
In this section, we describe the physics arising in the limits of large $\left|\Jep\right|$ for $\Jes\neq 0$. Our objective is to understand the regime where $\Jes$ is large enough to drive the system out of the gapped $\mathbb{Z}_2$ phases described in Sec.~\ref{sec:AIZ}. For $\Jep>0$, this problem is well studied\cite{fradkin79,hamma08,trebst07,tupitsyn10,wu12} and results in distinct transitions out of the $\mathbb{Z}_2$-topologically ordered phase whose nature depends on the sign of $\Jes$. We will review these results in the context of our model in Sec.~\ref{ssec:pzt}, since they provide a useful context for our discussion of the regime $\Jep<0$ presented in Sec.~\ref{ssec:sac}.

\subsection{The standard line P-Z-T}\label{ssec:pzt}
In this section, we briefly discuss the phases and transitions arising in the limit $\Jep\gg \Jp$ for $\Jes\neq 0$.
In this limit, $\msigma$-edges are absent from the low-energy Hilbert space, leading to the effective Hamiltonian
\begin{align}
H_ {\rm eff}^{\rm PZT}=-\frac{\Jp}{4}\sum\limits_p\left(\mathds{1}+B_p^{\mpsi}\right) -\frac{\Jes}{2}\sum\limits_e\left(n_e^{\mone}-n_e^{\mpsi}\right)\\
=\frac{1}{2}\left(H^{\mathbb{Z}_2}_{\rm SN}-\Jes\sum\limits_e\left(n_e^{\mone}-n_e^{\mpsi}\right)\right).
\label{eq:hameffpzt}
\end{align}
This is the $\mathbb{Z}_2$ string-net model\cite{levin05} (or equivalently the Toric code\cite{kitaev03} on the honeycomb lattice) with a perturbation that either favors or disfavors the non-trivial edge label $\mpsi$. This type of model has been studied extensively\cite{fradkin79,hamma08,trebst07,tupitsyn10,wu12}; here we review the features that will be germane to our discussion of the analogous model in the frustrated $\mathbb{Z}_2$ phase.

For $\Jes=0$, we showed in Sec.~\ref{sec:AIZ} that the ground state has $\mathbb{Z}_2$-topological order. 
In the constrained Hilbert space the only deconfined excitation here is a $\mathbb{Z}_2$-flux. Hence in the limit $\Jep\rightarrow\infty$, the effective model (\ref{eq:hameffpzt}) can be mapped exactly onto the transverse field Ising model on a triangular lattice\cite{wegner71,fradkin79}, where in this case the ferromagnetic Ising coupling is given by $\Jes$ (instead of $\Jep$ in Eq.~(\ref{eq:hameffaiz})). Thus, as discussed in the previous section, we find for $\Jes \gg \Jp$ an unfrustrated
phase, in which $\mpsi$-loops are confined. This results in a trivial ground state ($\rm T$), which is adiabatically connected to the (polarized) product state in which all edges carry the $\mone$-label.
For $-\Jes \gg \Jp$ we find an anti-ferromagnetic phase, with broken translational symmetry and a three-sublattice magnetic order.\cite{coppersmith85,blankschtein84} In terms of the edge variables, this phase is described by a dimer model of the form (\ref{eq:hameffa}), but with only the resonance term,\cite{moessner01} which now acts on $\mpsi$- instead $\msigma$-edges. Consequently, this antiferromagnetic phase is adiabatically connected to the plaquette phase of a quantum dimer model, where the dimers are now given by $\mone$-dimers in the background of $\mpsi$-edges. This phase is labeled as ${\rm plaq}_{\mpsi}$ in Fig.~\ref{fig:phase_diag}.

In contrast to the case discussed in Sec.~\ref{sec:AIZ}, here the dual Ising model (\ref{eq:hameffaiz}) describes all of the system's degrees of freedom. Therefore there is no additional ground state degeneracy in the limit of large $\left|\Jes\right|$, and these phases have trivial topological order\cite{wu12,vidal_par}. 
This can also be inferred from the loop operators. Specifically, none of the loop operators $W^{\alpha}_{{\mathcal{C}}_i,\rm FM}$ (\ref{eq:stringoperatorsz2_1a}-\ref{eq:stringoperatorsz2_2}) commute with the term $\propto \Jes$. It follows that the topological degeneracy is lifted completely in the limit of large $\left|\Jes\right|$. Our numerical and series expansion results for this line can be found in App.~\ref{app:z2}.

\subsection{The frustrated line S-A-C}\label{ssec:sac}
Having described the limit of large positive $\Jep$, in which the $\msigma$-links are absent, let us now turn to the limit of large negative $\Jep$, where the number of $\msigma$-links is maximized. As we have observed for $\Jes=0$, projecting onto states with maximal numbers of $\msigma$-edges leads to an effective dimer Hamiltonian in which dimers carry an additional internal label ($\mone$ or $\mpsi$). The gapped phases of these dimer models necessarily break the translational symmetry of the underlying lattice. 
Here we will extend this dimer description to the entire region $\Jp\ll-\Jep$, $\Jes<\Jep$. In Sec.~\ref{ssec:uppercornerC} we will discuss the behavior for $\Jes \approx -\Jep$, where the dimer projection is no longer valid and a competing order arises. Projecting onto the dimer Hilbert space we obtain the effective Hamiltonian:
\begin{align}
H^{\rm SAC}_{\rm eff}=& \bar{P}H\bar{P}\\
=&-\frac{\Jp\sqrt{2}}{4}\sum\limits_p\bar{P}B_p^{\msigma}\bar{P}-\frac{\Jp}{4}\sum\limits_p\bar{P}B_p^{\mpsi}\bar{P}\nonumber\\
&-\frac{\Jes}{2}\sum_e\left(n_e^{\mone} - n_e^{\mpsi}\right)+ \left( -\frac{\Jp}{4}+\Jep \right)N_p\bar{P}\\
=&-\frac{\Jp}{8}\sum\limits_{p,i,f} \left(\beta(i,f) \left|\begin{array}{c}\includegraphics[width=.75cm]{./figures/plaqterm1gray_mf.pdf}\end{array} \right\rangle\left\langle \begin{array}{c}\includegraphics[width=.75cm]{./figures/plaqterm2gray_mf.pdf}\end{array} \right| + {\rm h.c.}\ \right) \nonumber\\
&-\frac{\Jp}{4}\sum\limits_{p,i,f} \left(\gamma(i,f) \left|\begin{array}{c}\includegraphics[width=.75cm]{./figures//plaqterm1flavorflip_mf.pdf}\end{array} \right\rangle\left\langle \begin{array}{c}\includegraphics[width=.75cm]{./figures//plaqterm2flavorflip_mf.pdf}\end{array} \right| + \ldots\right)\nonumber\\
&-\frac{\Jes}{2}\sum_e\left(n_e^{\mone} - n_e^{\mpsi}\right)+ \left( -\frac{\Jp}{4}+\Jep \right)N_p\bar{P},
\label{eq:effdimer}
\end{align}
where again $i$ denotes the initial and $f$ the final states after action of the corresponding operator.

For $\Jes=0$, we recover $H_{\rm eff}^A$ (\ref{eq:hameffa}). For finite $\Jes$, one of the internal states of the dimers is disfavored with respect to the other. This effect competes with the second line of Eq.~(\ref{eq:effdimer}), which flips the internal states of all dimers. As $\left|\Jes\right|$ increases, this produces a transition in which the $\mathbb{Z}_2$ topological order associated with the internal dimer labels disappears. 

To understand the effect of varying $\Jes$, it is useful to consider the limiting cases of the effective dimer model (\ref{eq:effdimer}).
For $-\Jes \gg \Jp$, 
configurations with internal $\mone$-dimer states are energetically costly, and the low-energy states consist entirely of configurations involving the third vertex configuration of Fig.~\ref{fig:lattice_constraints}. The Hamiltonian (\ref{eq:effdimer}) reduces to
\begin{align}
H_{\rm SC}=&
-\frac{\Jp}{8}\sum\limits_p\left( \left| \begin{array}{c}\includegraphics[width=.75cm]{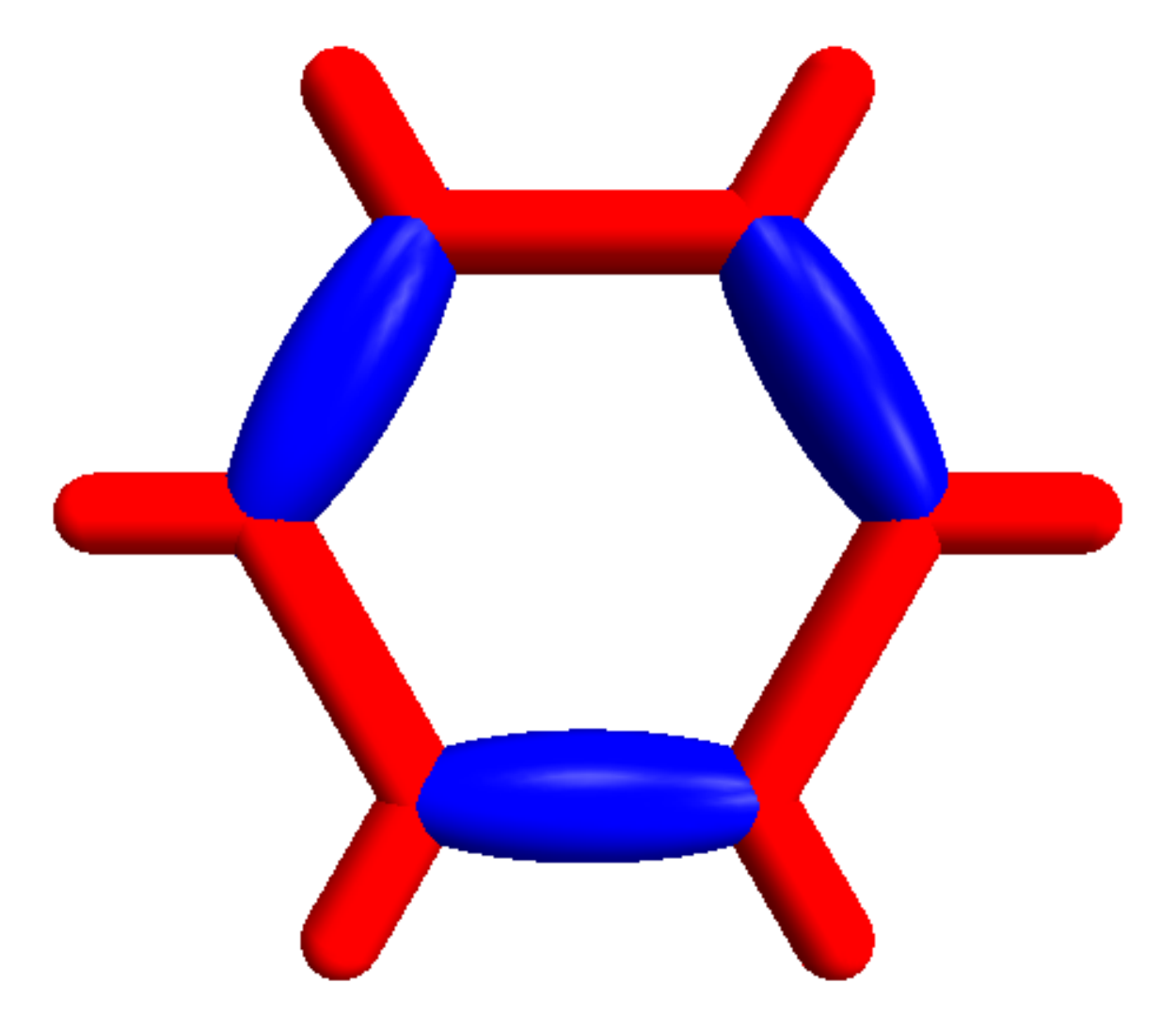}\end{array}\right\rangle \left\langle \begin{array}{c}\includegraphics[width=.75cm]{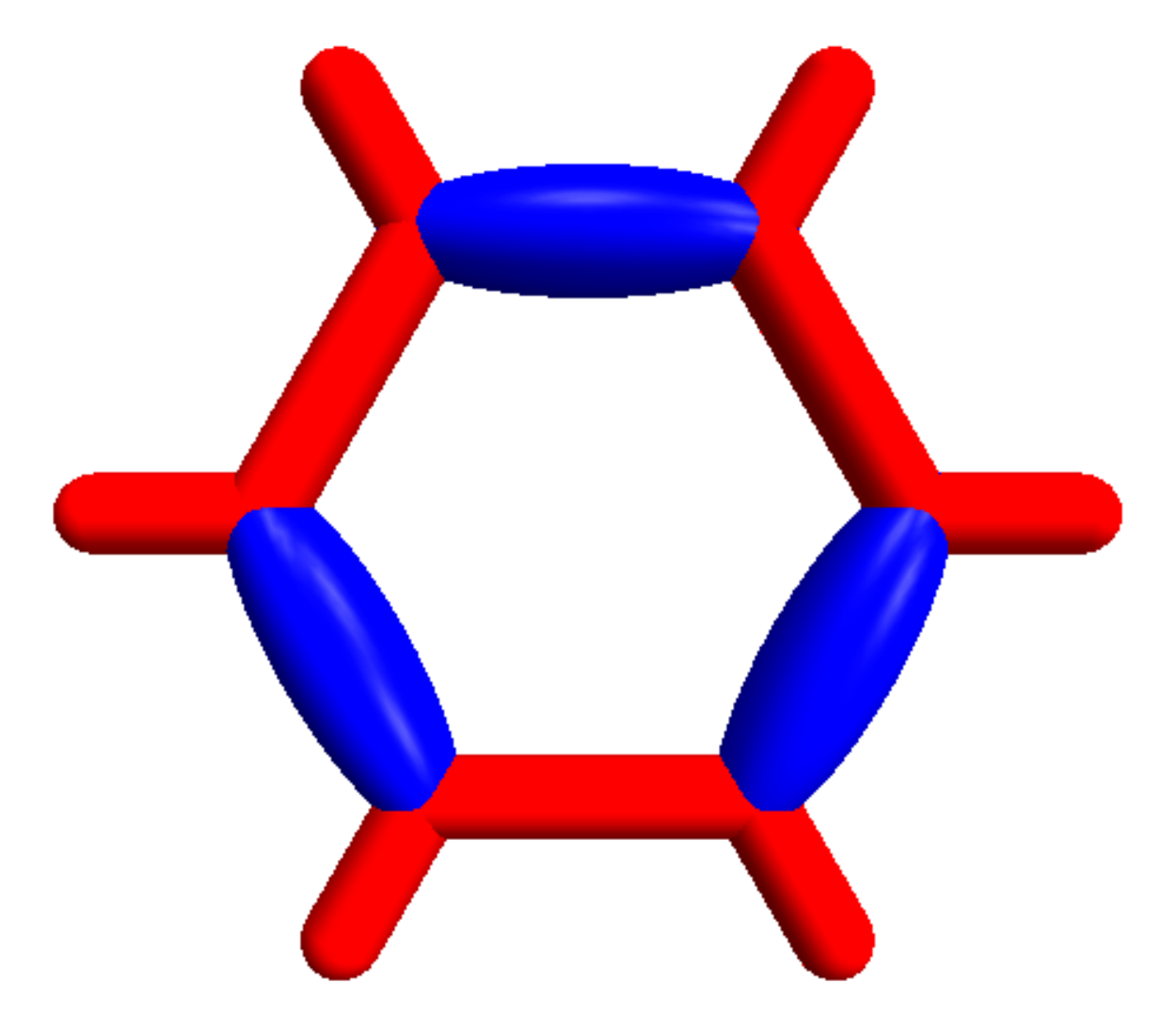}\end{array}\right| +{\rm h.c.} \right)\nonumber\\
&-\frac{\Jp}{4}\sum\limits_p \left| \begin{array}{c}\includegraphics[width=.75cm]{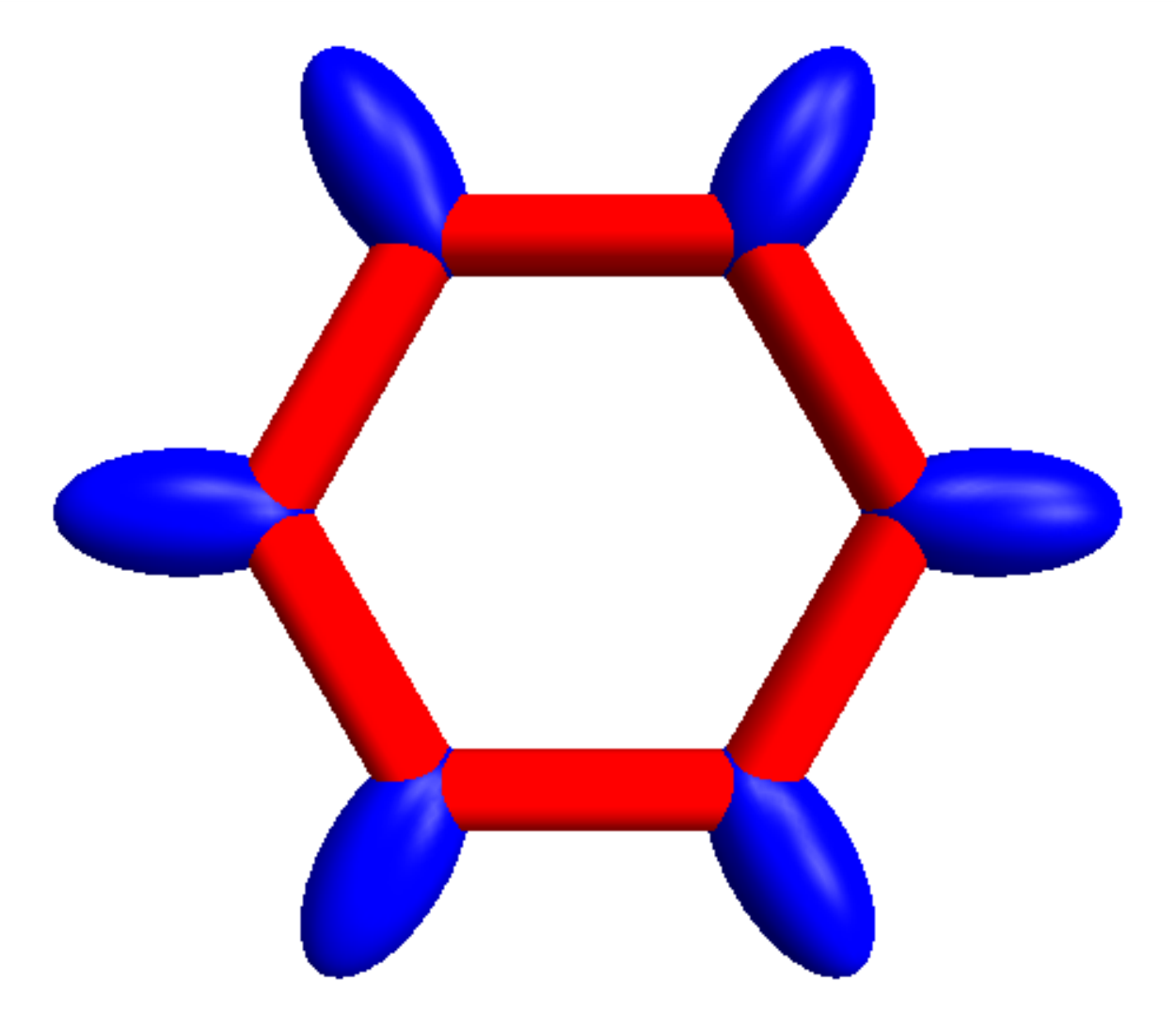}\end{array}\right\rangle \left\langle \begin{array}{c}\includegraphics[width=.75cm]{./figures/colterm_mf.pdf}\end{array}\right|\hfill
+{\rm const}.%
\label{eq:lowdimdimer}
\end{align}
Importantly, projecting out the $\mone$-labels annihilates all of the terms on the second line of Eq.~(\ref{eq:effdimer}) except for one. This introduces a new type of interaction, known as a potential term, in the dimer model, which explicitly favors certain dimer configurations and can help stabilize particular ordered states.  Comparing the spectra of the effective model (\ref{eq:lowdimdimer}) and the original model (\ref{eq:fullham}) shows that the latter accurately describes the low-energy spectrum of the full model for sufficiently large values of $\Jp$.\cite{schulz14}

\begin{figure}[htp]%
\includegraphics[width=.5\columnwidth]{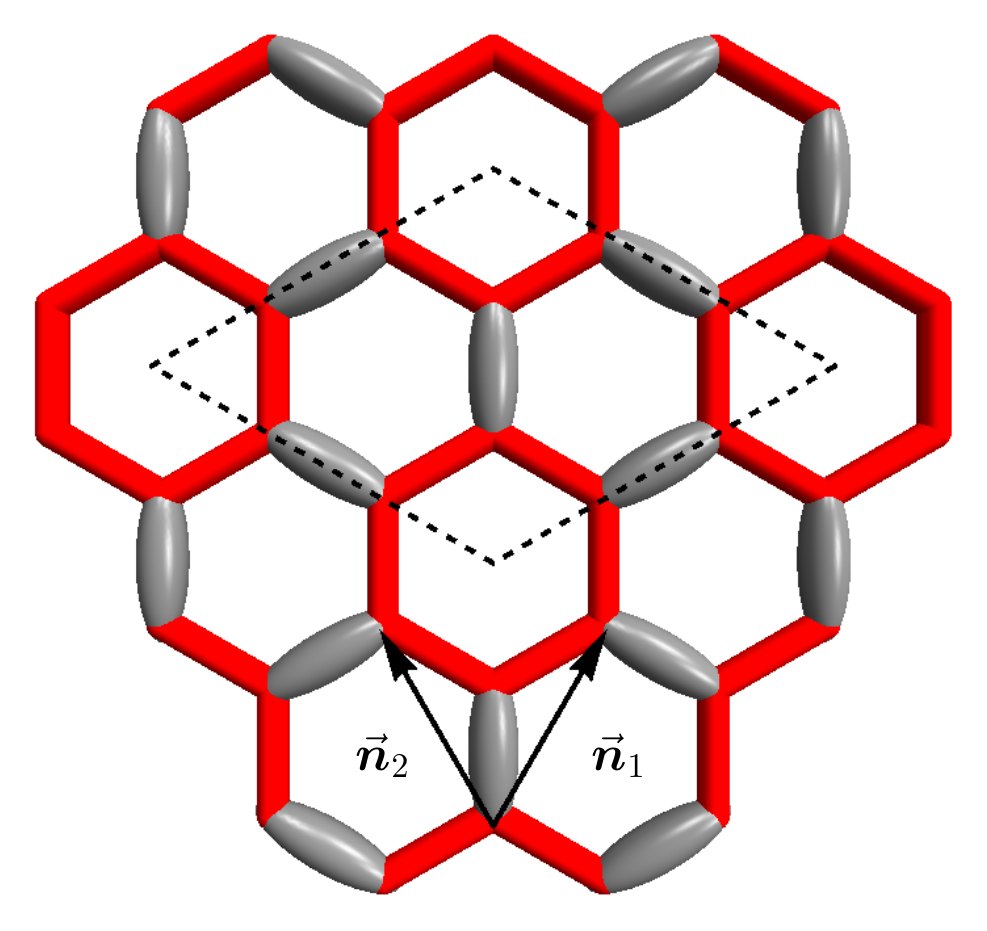}%
\caption{Columnar or star-crystal (positional) order of the dimers. The dashed lines depict the unit cell. The dimers remain static. The internal state of the dimers (depicted in gray) is either $\mone$ for positive $\Jes$ or $\mpsi$ for negative $\Jes$. }%
\label{fig:columnarorder}%
\end{figure}

The effective model (\ref{eq:lowdimdimer}) is now an undecorated quantum dimer model on the honeycomb lattice, whose phase diagram was established by Refs.~
\onlinecite{schulz14,schlittler15b}. As the potential term increases, the dimer model undergoes a transition from the plaquette phase to a phase with the ``columnar order" shown in Fig.~\ref{fig:columnarorder}. This order consists of static dimers, which are arranged such that one third of the plaquettes have dimers located only on their outgoing edges. This region is labeled by ${\rm col}_{\mpsi}$ in Fig.~\ref{fig:phase_diag}.

Excitations within the dimer Hilbert space consist of dimer configurations not maximizing the number of perfect (i.e.~$\msigma$-) hexagons. As detailed in Ref.~\onlinecite{schulz14}, the (non-topological) low-energy excitations of this phase eliminate three $\msigma$-hexagons; these are created by the action of the resonance term in (\ref{eq:lowdimdimer}) and thus exist on the sublattices formed by the non-perfect hexagons. We will return to this point later, when we discuss transitions between the topological and non-topological three-sublattice ordered phases.

An analogous picture holds in the limit $\Jes \gg \Jp$, where the roles of the states $\left|\mone\right\rangle_e$ and $\left|\mpsi\right\rangle_e$ are interchanged. As this does not affect the effective Hamiltonian, the same reasoning applied above implies that the resulting phase is a columnar phase, labeled by ${\rm col}_{\mone}$ in Fig.~\ref{fig:phase_diag}, where the internal state of the dimers is fixed to be $\left|\mone\right\rangle$. 

\subsubsection{Transitions between the frustrated phases}
Having discussed the limiting columnar orders for large $\left|\Jes\right|$ in the dimer limit, let us now discuss the effect of varying $\Jes$: for $\Jes=0$, as discussed above, the gapped phase with two qualitatively different types of excitations. The non-topological excitations correspond to a deviation from the ordering pattern of the dimers from the ground states. As the operator $\propto\Jes$ only acts on the internal states of the dimers, this operator does not impact the positional order of the dimers and therefore does not interact with these non-topological excitations. 

However, the topological excitations of the $\mathbb{Z}_2$ phase, which are static for $\Jes=0$, become mobile for finite $ \Jes$. As detailed in Sec.~\ref{ssec:afmz2}, these excitations also prevent excited plaquettes from resonating, such that it is energetically unfavorable for these excitations to occupy the resonating sublattice. 
Therefore the lowest-energy topological excitations are located on non-resonating plaquettes (which correspond to the sublattices with finite magnetizations in terms of the pseudo-spin $\tau$). Further, the dynamics that result from finite $\Jes$ cannot hop these defects between the two distinct non-resonating sublattices.  This is because the inequivalent non-resonating sublattices are always separated by domain walls formed by $\msigma$-edges, 
whereas the term $\propto \Jes$ annihilates states in which the edge $e$ does not contain a dimer. Therefore finite $\Jes$ endows the lowest-energy quasi-particles with dynamics such that they hop on one of two disjunct triangular lattices. {\fb }

The hopping between these sublattices occurs via virtual states on the resonating sublattice. This has two notable consequences: first,
the effective hopping matrix elements between sites in a given non-resonating sublattice are even in $-\Jes$. Therefore, in the dimer limit, to leading order the sign of $\Jes$ does not have an impact on the dispersion of the topological excitations, leading to an effective symmetry $\Jes\leftrightarrow-\Jes$ which can e.g.~be seen in the ED spectrum (see Fig.~\ref{fig:phi_plot}).

\begin{figure}[htp]%
\includegraphics[width=.5\columnwidth]{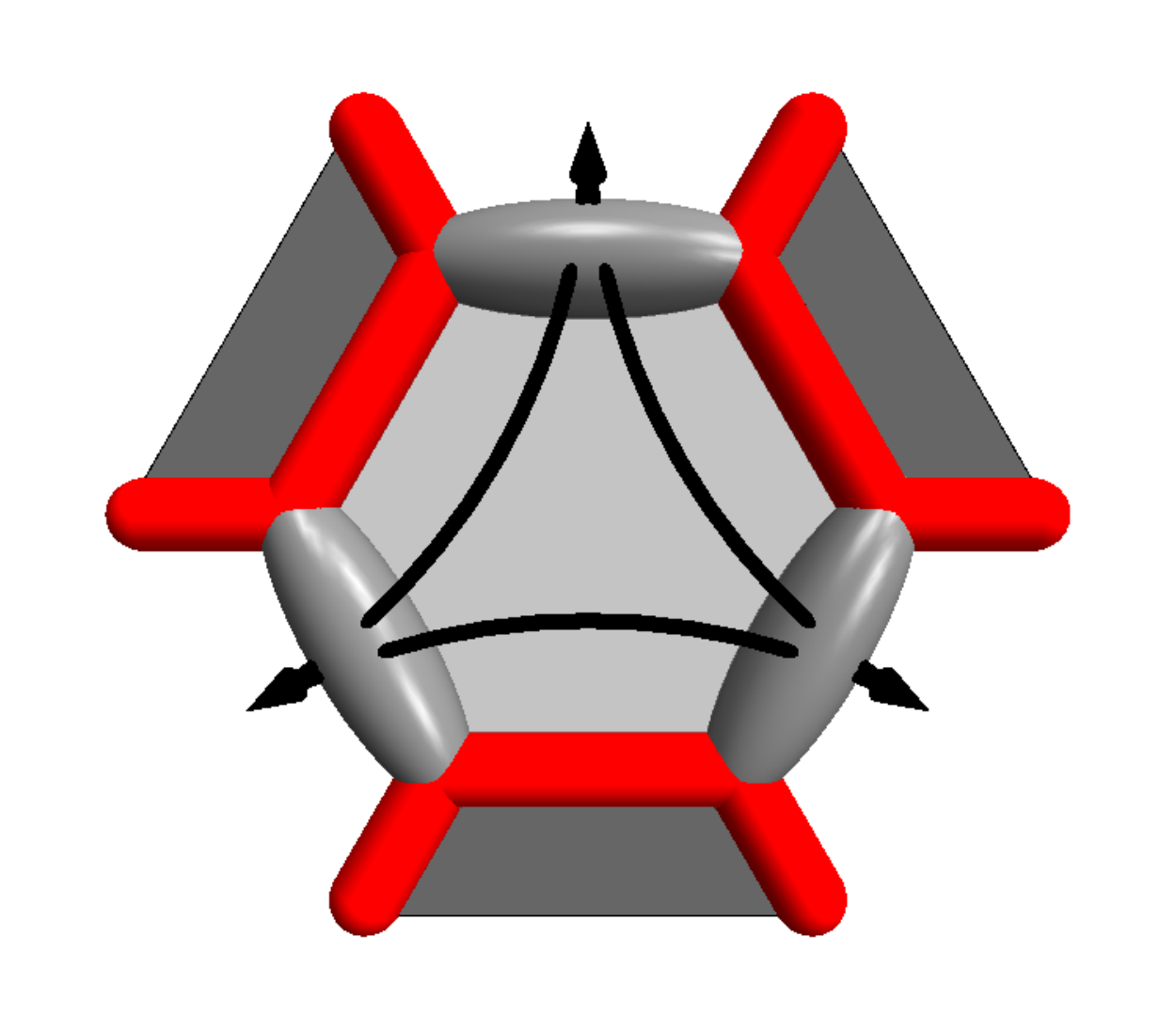}%
\includegraphics[width=.5\columnwidth]{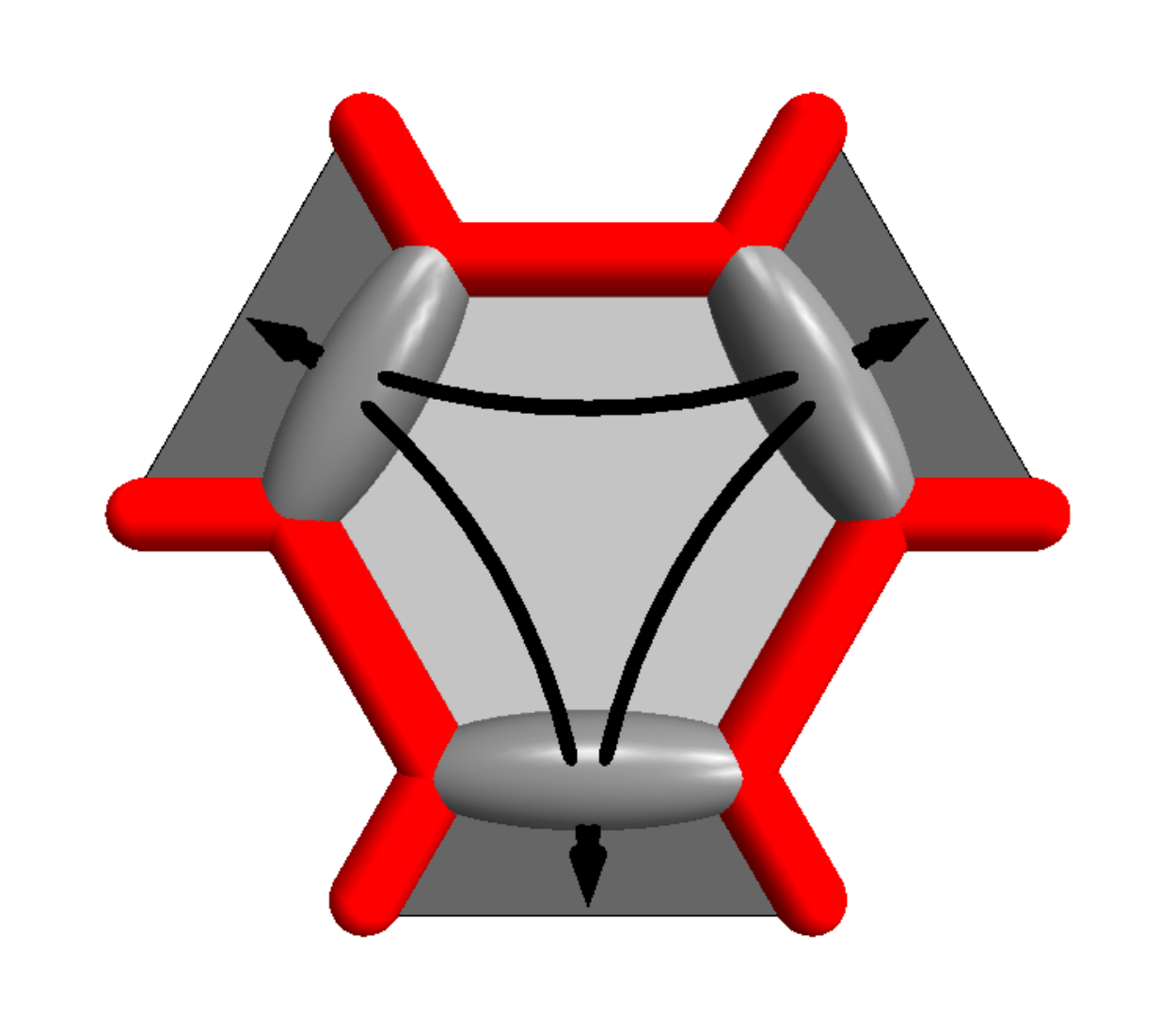}%
\caption{The lowest-energy excitations in the anti-ferromagnetic $\mathbb{Z}_2$-phase are located on the two non-resonating sublattices (depicted here in black and white). Their dynamics occurs via the intermediate state on the resonating (gray) plaquette, shown as black arrows. Thus each of the configurations depicted here allows only excitations on one sublattice to gain kinetic energy. Therefore the two non-resonating sublattices compete for the resonating sublattice. }%
\label{fig:sigma_plaq_dyn}%
\end{figure}

Second, since hopping between sites on a given sublattice can only occur for one of the two possible configurations of the resonating plaquette, the two non-resonating sublattices are competing for the resonating sublattice in order to gain kinetic energy. An example of this is depicted in Fig.~\ref{fig:sigma_plaq_dyn}; it can be viewed as a result of the fact that the model's dynamics cannot transform $b_1$-type excitations into $b_2$-type excitations, such that hopping of the defects can occur only within the same region (black or white). Thus the two inter-penetrating sublattices are mutually frustrating. At the transition the symmetry between these two sublattices is spontaneously broken and the resonating plaquette is largely pinned in one of its two configurations, leading to the star-crystal order shown in Fig.~\ref{fig:columnarorder}.

This correlation between the change in the three-sublattice order and proliferation of plaquette defects on one of the three sublattices suggests that the transition in the dimer order generically coincides with the loss of $\mathbb{Z}_2$ topological order (which disappears when these defects condense, confining fluctuations of the internal dimer labels). This is supported by our numerics, which are consistent with a single phase transition between the $\mathbb{Z}_2$ plaquette and star-crystal phases as $\Jes$ varies (e.g.~in Fig.~\ref{fig:phi_plot}), where the long-range and topological orders change simultaneously. 

Though our analysis is not sufficient to resolve the order of this transition, in the pure quantum dimer model, the transition between the plaquette and columnar phases is first order\cite{blankschtein84,moessner01,schlittler15,schlittler15b}. Hence for our model we expect, in analogy to the dimer model, first-order transitions for both signs of $\Jes$. 

The conclusions we have just drawn from the effective dimer model (\ref{eq:lowdimdimer}) must be applied to the full Hamiltonian with care, as there are important differences. First, the effect of decreasing $\left|\Jep\right|$ from infinity allows fluctuations out of the dimer Hilbert space, such that our effective Hamiltonian no longer applies. However, our numerical analysis shows no signature of intermediate phases between the columnar phase and plaquette phase (P) for finite $\Jep$.

Second the symmetry $\Jes \leftrightarrow -\Jes$ of the effective dimer model (\ref{eq:effdimer}) is approximate at best, and is strongly broken in the original model (\ref{eq:fullham}) for $\left|\Jes\right|\approx -\Jep$ (i.e. after the transition into the star-crystal phase). This turns out to have important implications for the phase diagram. 
For $\Jes= \Jep <0$, the non-topological term is {$V=-\frac{J_{\rm e}^{\msigma}}{2}\sum_e\left(n_e^{\mone}-n_e^{\mpsi}- n_e^{\msigma} \right)=|J_{\rm e}^{\msigma}|\sum_e\left(n_e^{\mone}-\frac{1}{2}\right)$}. Due to the vertex constraints, minimizing the number of $\mone$-edges yields a low-energy manifold of states formed by dimer coverings, and the effective description given above remains valid, as evidenced by the extent of the ${\rm col}_{\mpsi}$ phase in Fig.~\ref{fig:phase_diag}. However, for $\Jes=-\Jep>0$, {$V=-\frac{J_{\rm e}^{\msigma}}{2}\sum_e\left(n_e^{\mone}-n_e^{\mpsi}+n_e^{\msigma} \right)=|J_{\rm e}^{\msigma}|\sum_e\left(n_e^{\mpsi}-\frac{1}{2}\right)$}. Minimizing the number of $\mpsi$-edges does not restrict the Hilbert space to dimer coverings, since it is possible for three $\mone$-edges to meet at a vertex. Hence in this limit, as (to leading order) $\msigma$-edges and $\mone$-edges have the same energy, the low-energy Hilbert space contains superpositions of $\msigma$-loops of arbitrary length. This leads to a breakdown of long-ranged order, and a new $\mathbb{Z}_2$ topologically ordered phase in the upper left corner (C) of the phase diagram Fig.~\ref{fig:phase_diag}, which we discuss in the following section.

\section{An emergent \texorpdfstring{$\mathbb{Z}_2$}{Z(2)} topological phase}\label{ssec:uppercornerC}
For $\Jes \approx - \Jep$, the effective dimer model of the previous section breaks down, and a new $\mathbb{Z}_2$ topological phase emerges, which we describe here. We will keep the ratio $\Jes/\Jp$ large and positive, such that $\mpsi$-edges are very energetically costly, and thus essentially absent from the low-energy Hilbert space. However, we will consider the limit $-\Jep \approx \Jes$, such that for $\msigma$-edges (which are favored by $-\Jep$, but disfavored by $\Jes$), the potential and kinetic terms are of the same order of magnitude.
  
To study the region $J_{\rm e}^{\msigma} \approx -J_{\rm e}^{\mpsi} \gg J_p$, it is useful to use $\mathds{1} = n_e^{\msigma}+ n_e^{\mpsi}+n_e^{\mone}$ to write
\begin{align}
 V=J_{\rm e}^-\sum\limits_e\left(n_e^{\mpsi}-\frac{1}{2}\right)-\frac{J_e^+}{2}\left(n_e^{\mone}-n_e^{\msigma}-n_e^{\mpsi}\right),
\end{align} 
where $J_{\rm e}^{\pm}=\frac{1}{2}\left(\Jes\pm\Jep\right)$. For $\Jes\approx-\Jep$, we have $J_e^-\gg J_e^+,\Jp$. 

We next project the Hamiltonian (\ref{eq:fullham}) into the low-energy Hilbert space, where there are no $\mpsi$-edges in the system. The resulting effective Hamiltonian reads
\begin{align}
H^{C}_{\rm eff}= &-\frac{\Jp}{4} \sum\limits_p \left| 
\begin{array}{c}\includegraphics[width=.75cm]{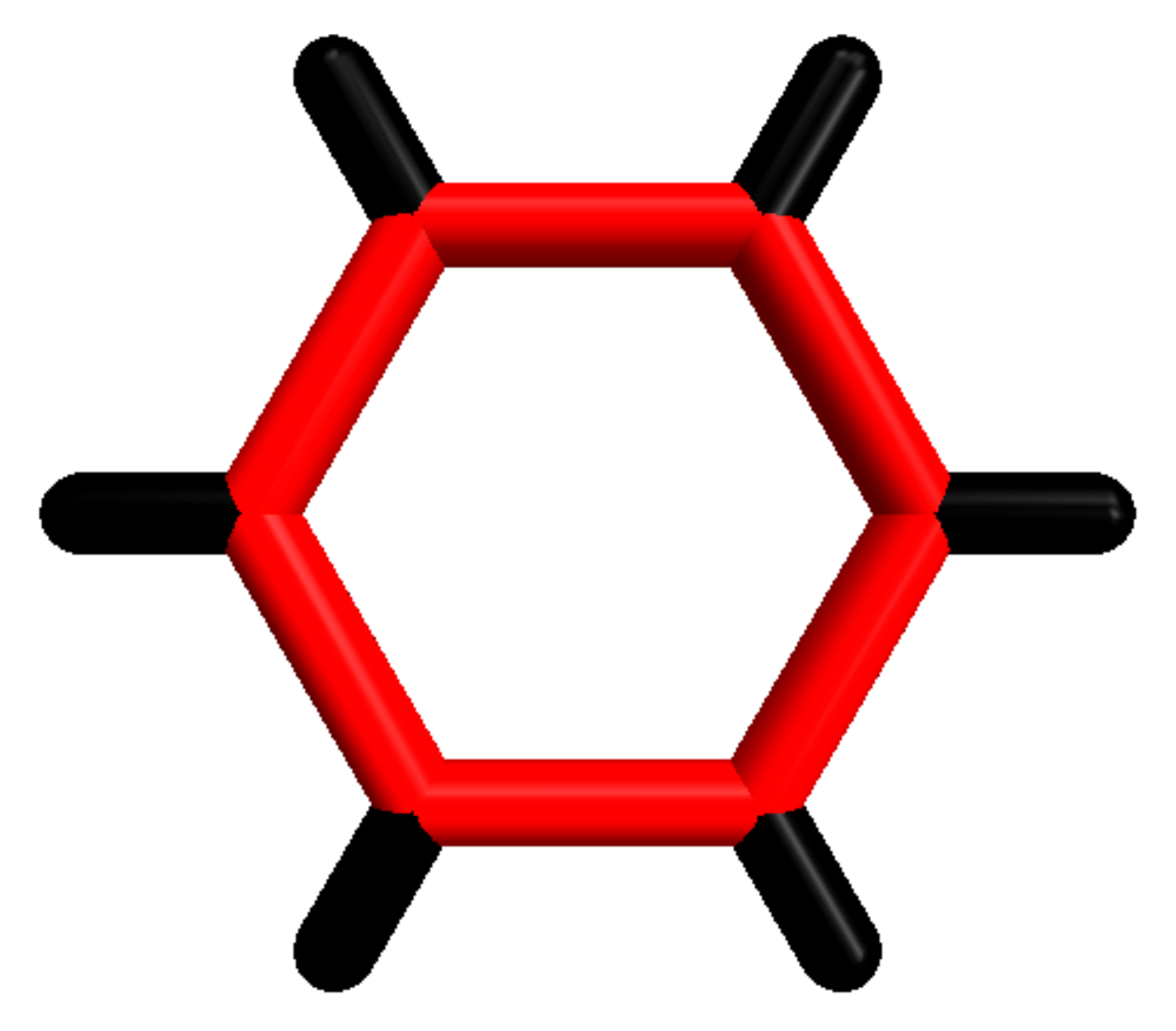}\end{array} 
\right\rangle\left\langle 
\begin{array}{c}\includegraphics[width=.75cm]{./figures/hexa_mf.pdf}\end{array} 
\right|\nonumber\\
&-\frac{\Jp}{4} \sum\limits_p \left( \left(\sqrt{2}\left| 
\begin{array}{c}\includegraphics[width=.75cm]{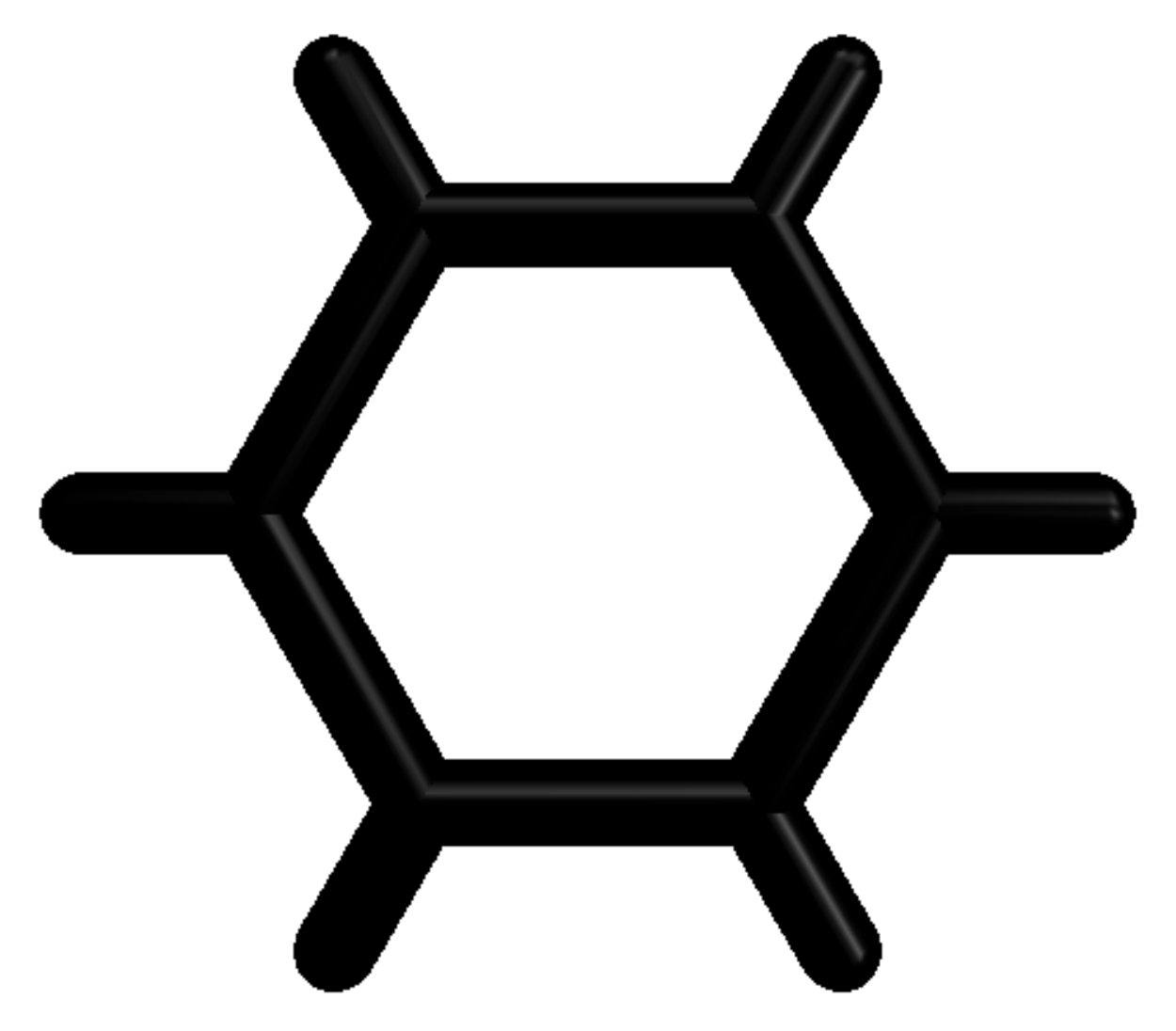}\end{array} 
\right\rangle\left\langle 
\begin{array}{c}\includegraphics[width=.75cm]{./figures/hexa_mf.pdf}\end{array} 
\right| +{\rm h.c.}\right)\right.\nonumber\\
&\phantom{-\Jp\sum\limits_p\ \ \ }\left.
+\left( \left|
\begin{array}{c}\includegraphics[width=.75cm]{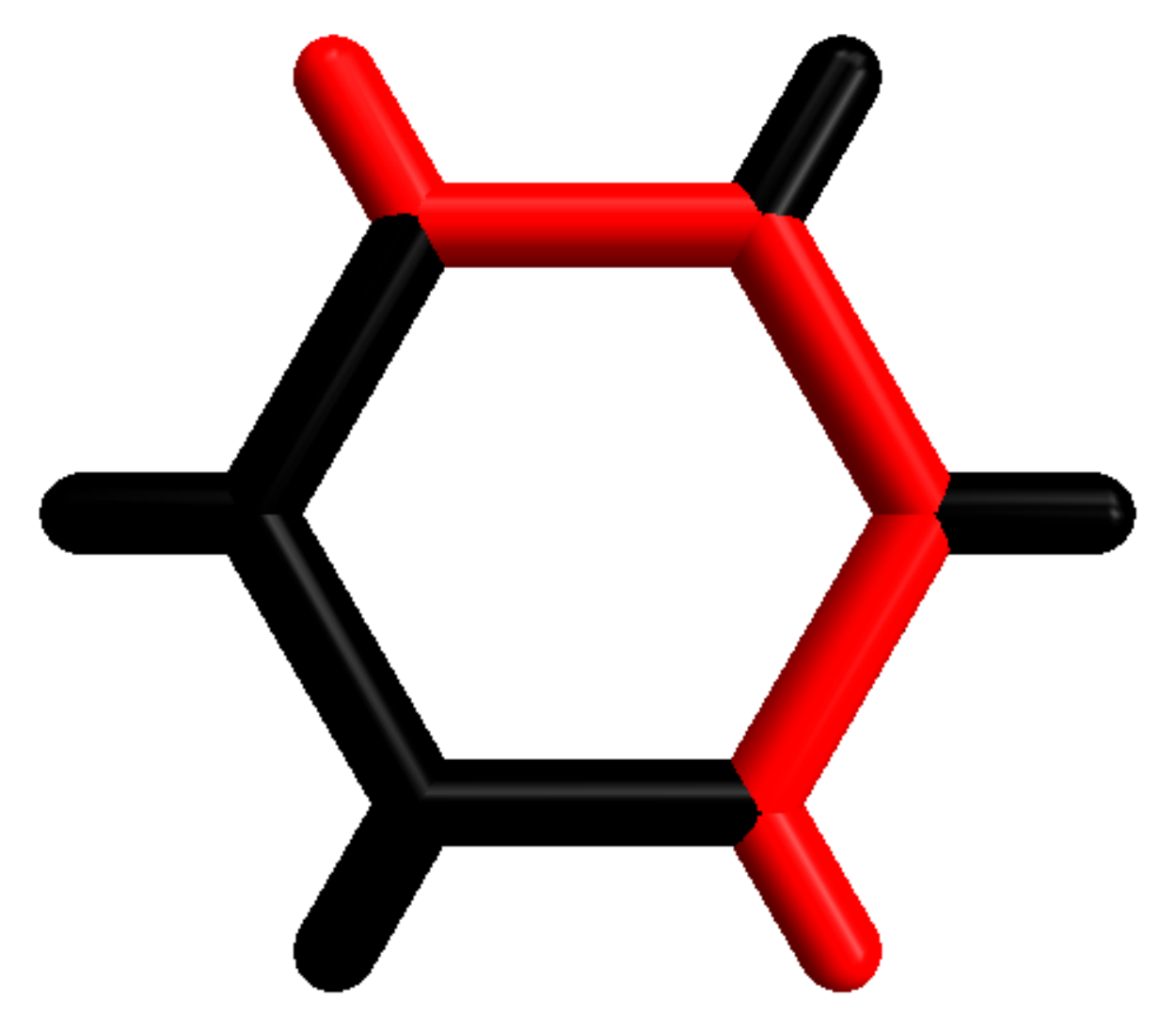}\end{array} 
\right\rangle\left\langle 
\begin{array}{c}\includegraphics[width=.75cm]{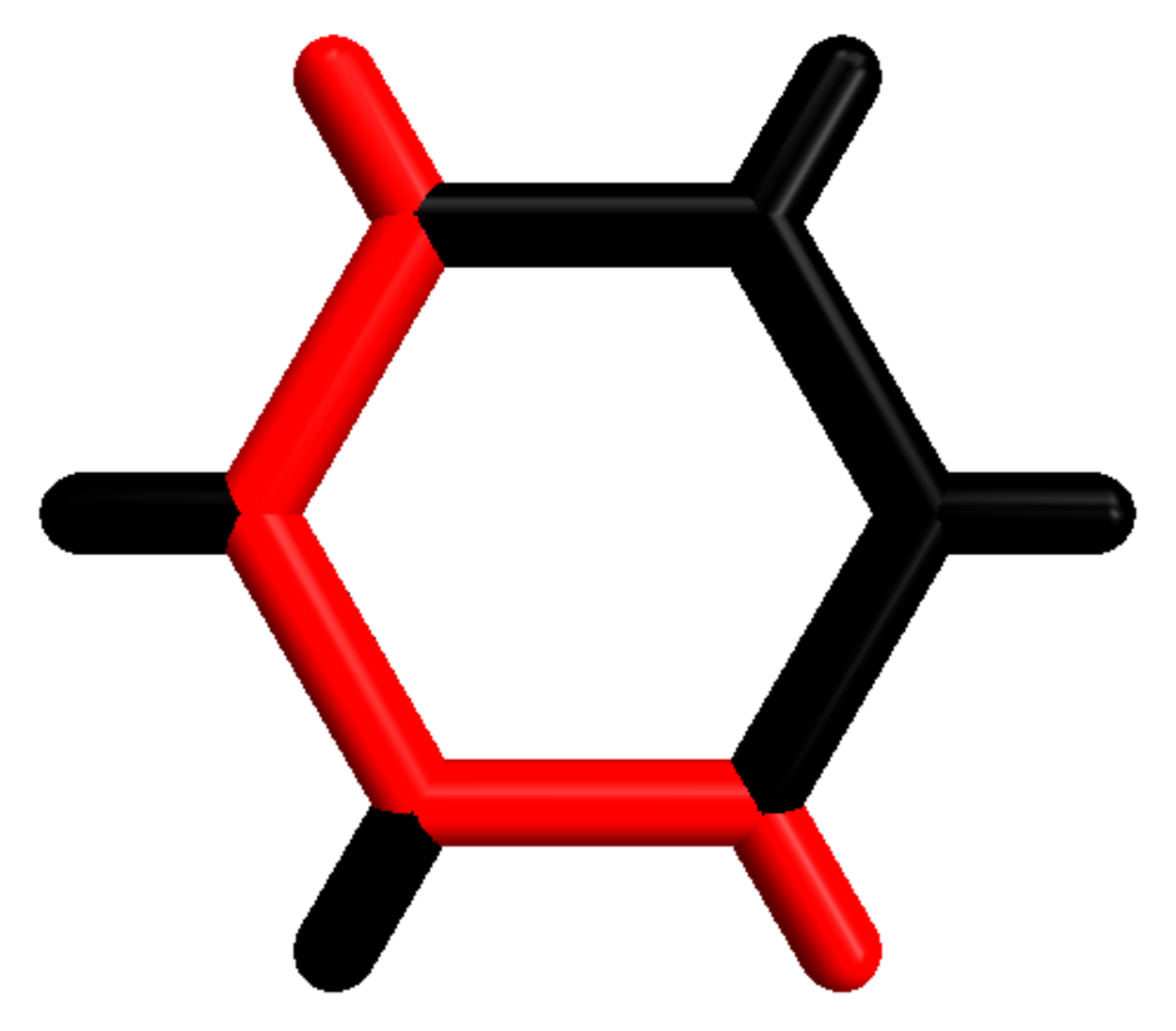}\end{array} 
\right| + \ldots\right)
\right.\nonumber\\
&\phantom{-\Jp\sum\limits_p\ \ \ }\left.
+\left(\frac{1}{\sqrt{2}} \left| 
\begin{array}{c}\includegraphics[width=.75cm]{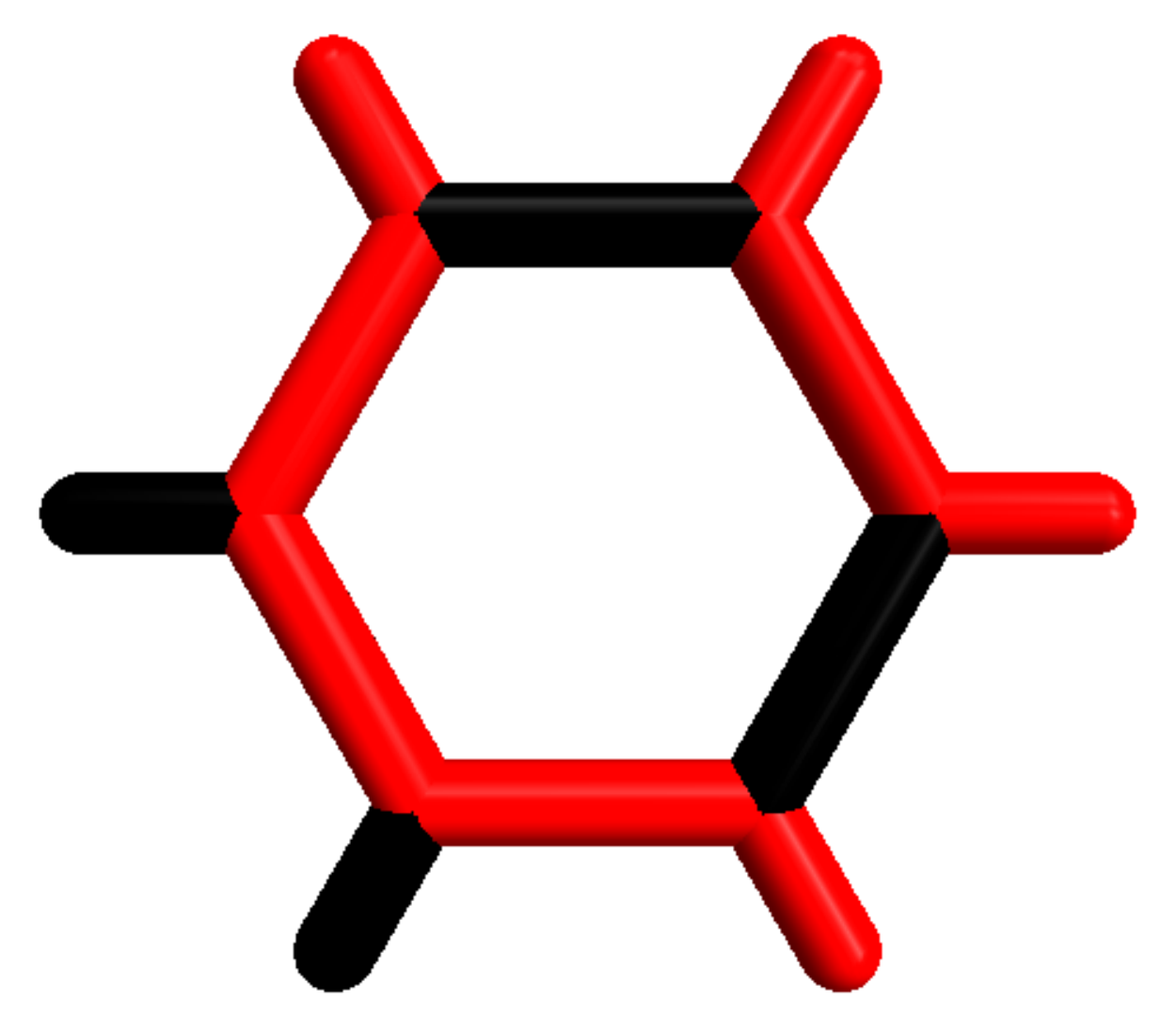}\end{array} 
\right\rangle\left\langle 
\begin{array}{c}\includegraphics[width=.75cm]{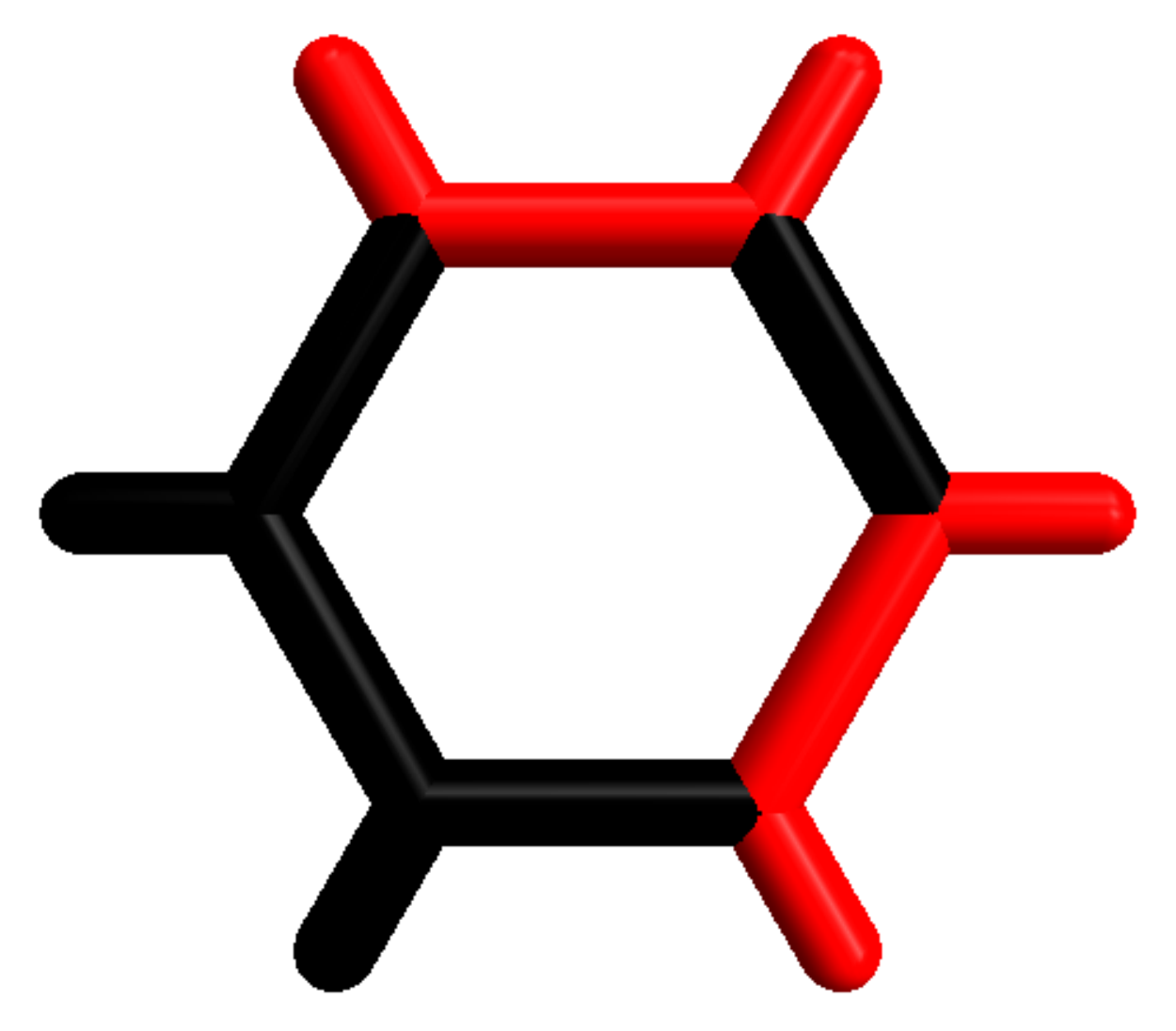}\end{array} 
\right| + \ldots\right)
\right.\nonumber\\
&\phantom{-\Jp\sum\limits_p\ \ \ }\left.
+\left(\frac{1}{2} \left| 
\begin{array}{c}\includegraphics[width=.75cm]{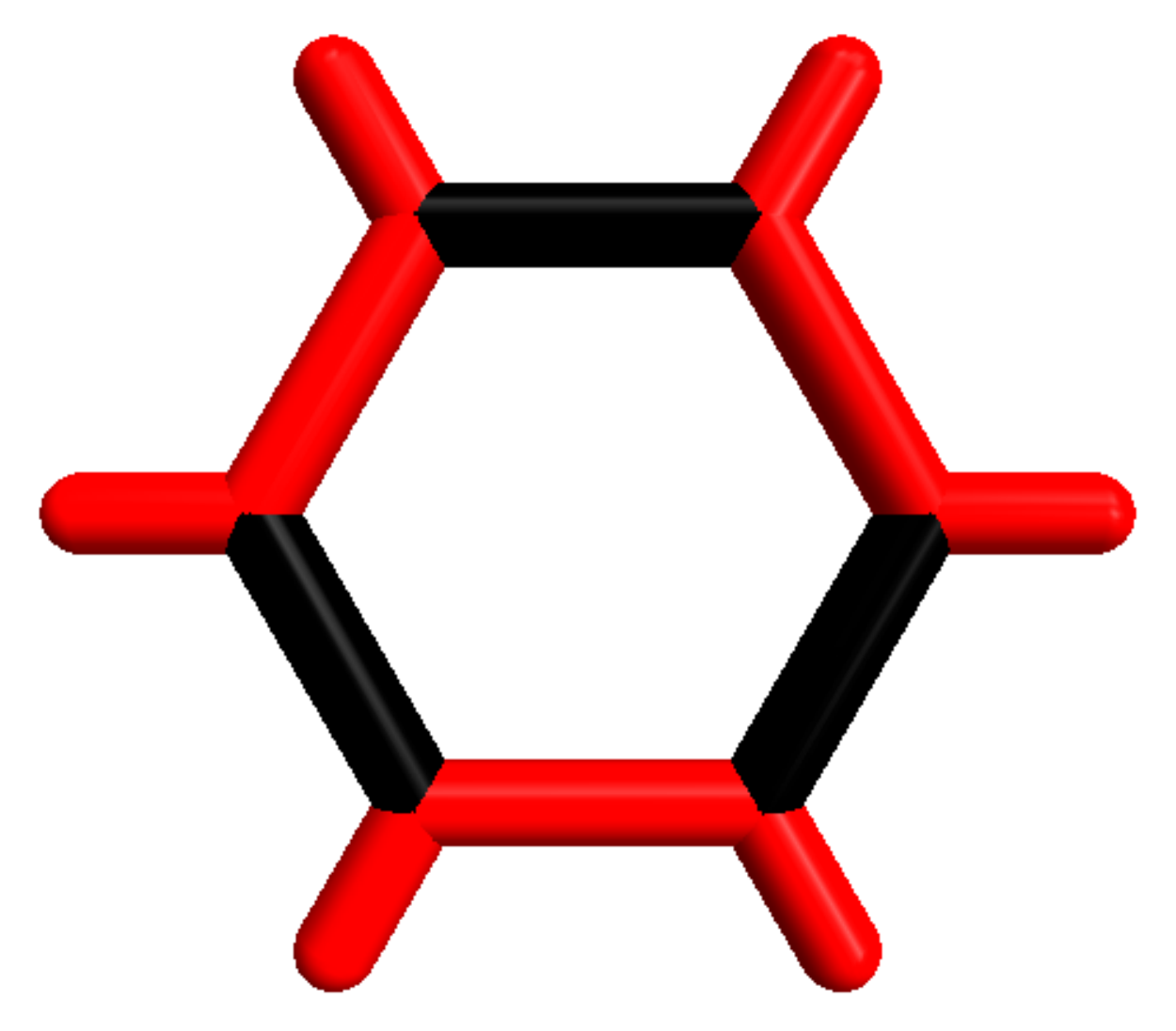}\end{array} 
\right\rangle\left\langle 
\begin{array}{c}\includegraphics[width=.75cm]{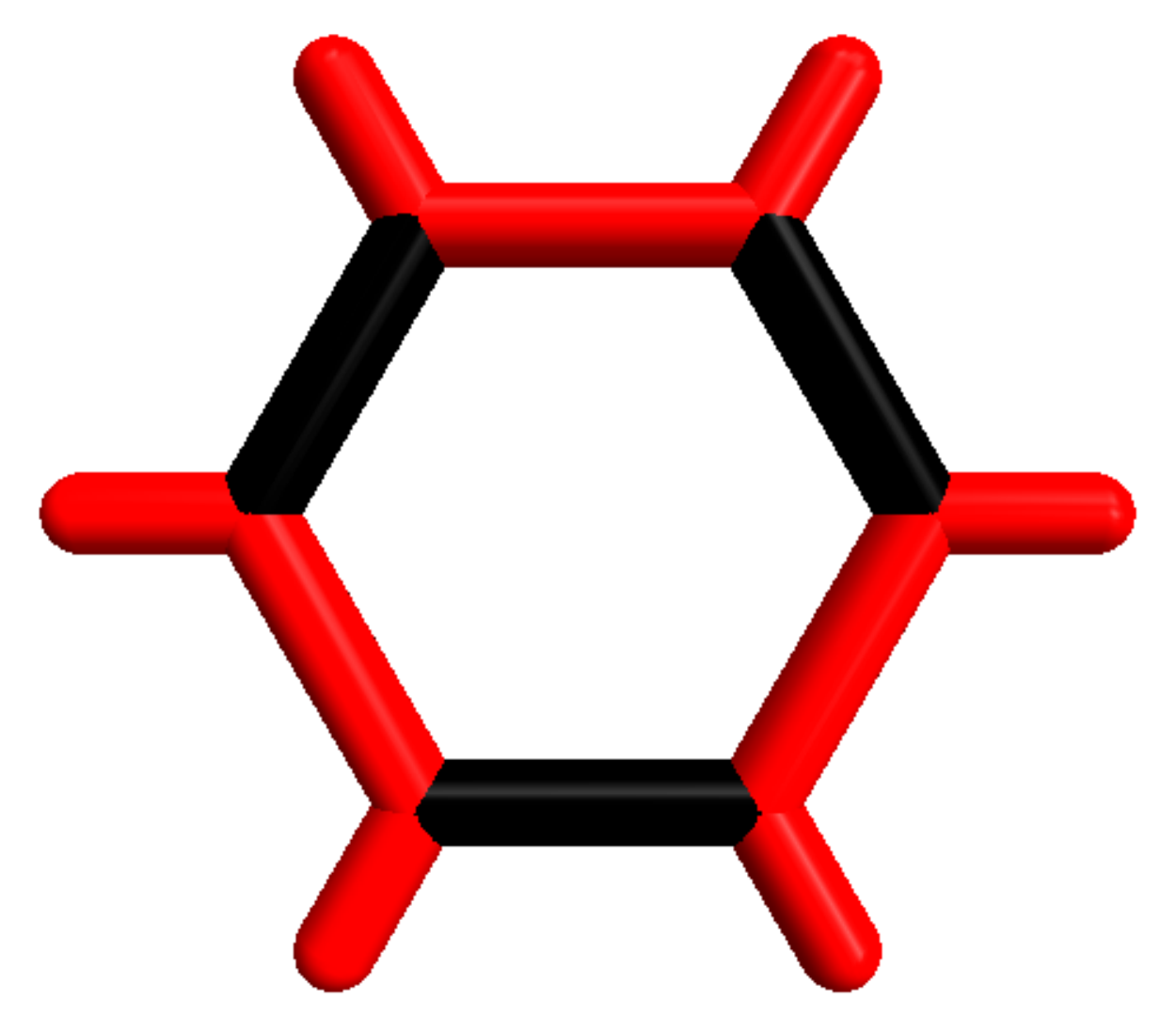}\end{array} 
\right| + \ldots\right) \right)
\nonumber\\
&-\frac{J_{\rm e}^+}{2}\sum_e\left(n_e^{\mone}-n_e^{\msigma}\right)+{\rm const},
\label{eq:hameffc}
\end{align}
where the ``$\ldots$'' include all possible $\msigma$-configurations with the same number of $\msigma$-edges on the outer legs of plaquette $p$. 

The first term in Eq.~(\ref{eq:hameffc}), which results from the action of the operator $B_p^{\mpsi}$, clearly favors the columnar order discussed in the previous section. The last term favors the ``trivial'' state for positive $J_{\rm e}^+$, and maximizes the number of $\msigma$-edges for $-J_{\rm e}^+$. Hence these two terms dictate a trivial ground state for $J_{\rm e}^+ \gg \Jp$, and recover the columnar phase discussed in the previous section for $- J_{\rm e}^+ \gg \Jp$. These correspond to regions labeled by trivial and $\mathrm{col}_{\mone}$ in Fig.~\ref{fig:phase_diag}.

In the regime where $\Jp/J_{\rm e}^+$ is not small, however, the second line of Eq.~(\ref{eq:hameffc}) plays an important role. Indeed, Eq.~(\ref{eq:hameffc}) differs from Eq.~(\ref{eq:hameffpzt}) only in the first term, which selects the columnar ordered state for negative $J_{\rm e}^+$, and in the non-trivial weights of the different terms in the second sum. 
The second term is therefore a deformation of the $\mathbb{Z}_2$ string-net Hamiltonian $H^{\mathbb{Z}_2}_{\rm SN}$ (\ref{eq:hameffpzt}), suggesting that for sufficiently small $J_{\rm e}^+$ a third, $\mathbb{Z}_2$ topologically ordered, phase emerges.

\begin{figure}[htp]%
\includegraphics[width=\columnwidth]{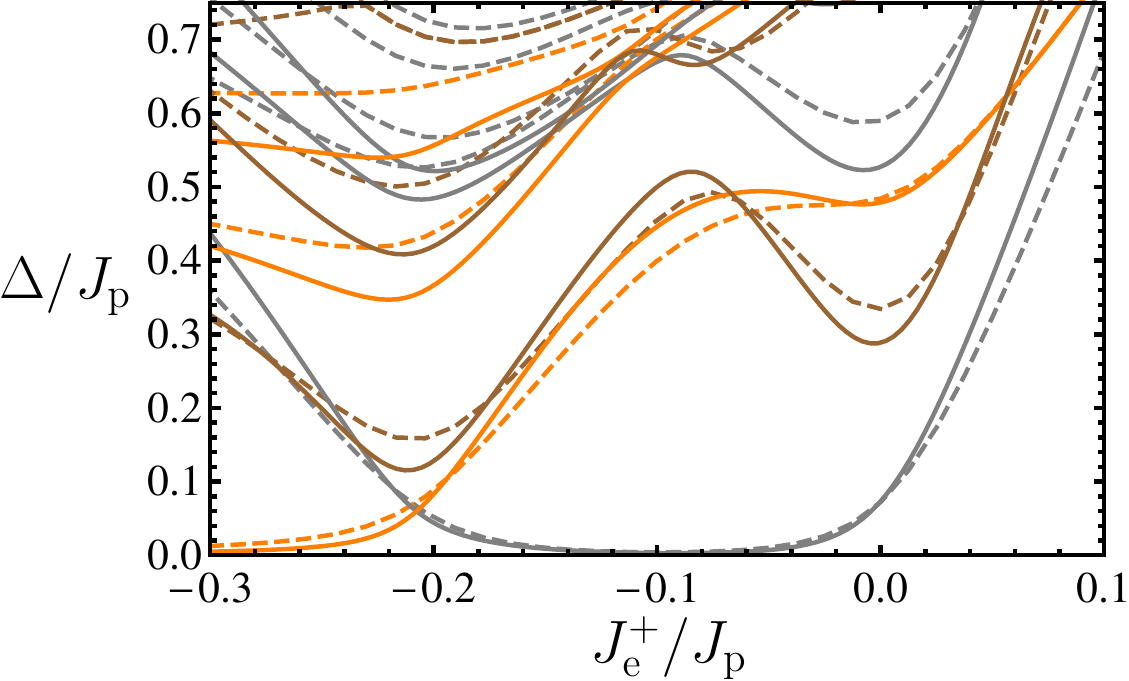}%
\\
\includegraphics[width=\columnwidth]{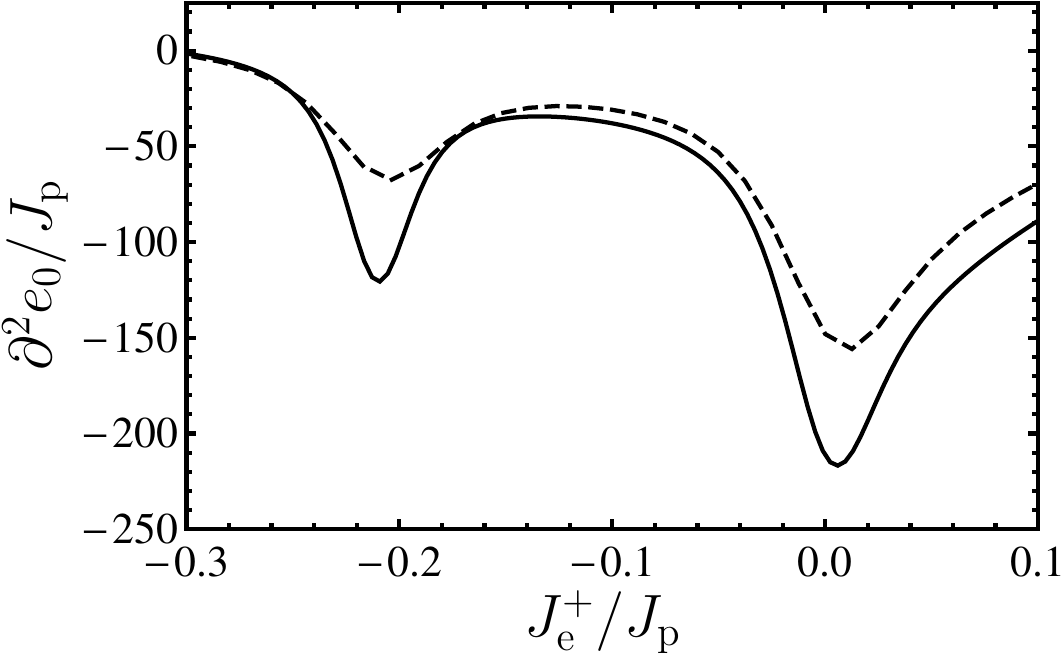}%
\caption{We show in the upper panel the low-energy spectrum of the effective model (\ref{eq:hameffc}) as a function of $J_{\rm e}^+$ (orange $\vec{k}=(\pm\frac{2\pi}{3},\mp\frac{2\pi}{3})$, brown $\vec{k}=(0,0)$ no strings, gray $\vec{k}=(0,0)$ at least one string ) for different system sizes ($N_{\rm p}=21$ plaquettes dashed, $N_{\rm p}=27$ plaquettes solid). The location of the phase transition is obtained from the minima of $\partial^2_{J_{\rm e}^+}e_0$ (lower panel). The corresponding finite size scaling is shown in App.~\ref{app:fss_z2prime}. }%
\label{fig:cornerspectrum}%
\end{figure}

In this section, we will present numerical evidence suggesting that this phase, which we refer to as $\mathbb{Z}_2^{\prime}$ in the following, does indeed exist. 
In Fig.~\ref{fig:cornerspectrum}, we show the low-energy spectrum of $H^{C}_{\rm eff}$ (\ref{eq:hameffc}) derived from exact diagonalization on systems with $N_{\rm p}=21$ (dashed) and $N_{\rm p}=27$ (solid) for periodic boundary conditions, as well as the derivatives of the ground-state energy. Orange lines show modes at the momenta $\vec{k}=(\pm\frac{2\pi}{3},\mp\frac{2\pi}{3})$, in the sector with an even number of non-contractible $\msigma$-loops. The remaining lines have $\vec{k}=(0,0)$, with brown indicating an even number of non-contractible loops, and gray indicating sectors with an odd number of non-contractible $\msigma$-loop in at least one direction. 

For $J_{\rm e}^+>0$, the spectrum is gapped, with a unique ground state. At $J_{\rm e}^+\approx 0$, the lowest-energy mode with an odd number of non-contractible $\msigma$-lines drops in energy to become virtually degenerate with the ground state. 
This degeneracy persists until $J_{\rm e}^+\approx -0.2\Jp$, at which point the states with non-contractible $\msigma$-loops are split from the ground state, while states at non-zero $\vec{k}$ join the ground-state sector. The resulting translational symmetry breaking ground state is adiabatically connected to the columnar ground state discussed in Sec.~\ref{ssec:sac}. 

Thus Fig.~\ref{fig:cornerspectrum} suggests two transitions: one from the trivial phase to a $\mathbb{Z}_2^{\prime}$-topologically ordered phase, and a second from this phase into the three-sublattice ordered columnar phase. 
By extrapolating to the thermodynamic limit as shown in App.~\ref{app:fss_z2prime}, we can estimate the location of these two phase transitions. This yields a transition into the trivial phase at $\left.\frac{{J_{\rm e}^+}}{\Jp}\right|_{c_1}=-0.0276$, and a transition to the columnar phase at $\left.\frac{{J_{\rm e}^+}}{\Jp}\right|_{c_2}=-0.275$.

In the region $- \Jep/J_p \gg 1$, we expect these results for the effective model (\ref{eq:hameffc}) also to be qualitatively correct for the full model, from which it follows that the region labeled by $\mathbb{Z}_2^{\prime}$ in Fig.~\ref{fig:phase_diag} is in the $\mathbb{Z}_2^{\prime}$ topological phase.
Indeed exact diagonalization of the original model indicates the same pattern of ground states (trivial, topologically ordered, and finally translation breaking as $-J_{\rm e}^+$ increases) as seen in Fig.~\ref{fig:cornerspectrum}. However, for the full Hamiltonian we only reach system sizes $N_{\rm p}=13$ ($N_{\rm p}=12$ for systems compatible with three-fold translational symmetry breaking), which is insufficient to obtain a reliable estimate of the positions of the phase boundaries in the thermodynamic limit due to finite size effects, which are already significant for the effective model, where system sizes $N_{\rm p}=27$ can be obtained (cf.~App.~\ref{app:fss_z2prime}). 
This poses a quantitative challenge in the regime where the effective model (\ref{eq:hameffc}) is no longer valid. For example, near the anti-ferromagnetic topological phase (phase ${\rm col}_{\mone}$ in the phase diagram~\ref{fig:phase_diag}), we are not able to reliably extrapolate the location of the phase boundaries between the different types of translational symmetry breaking and topologically ordered phases to the thermodynamic limit. In particular, we are not able to determine how the different phase boundaries connect in this regime, i.e.~ around point $\rm d$ in Fig.~\ref{fig:phase_diag}.

Nevertheless, our results suggest that there is a phase boundary separating the $\mathbb{Z}_2^{\prime}$ topological phase and the non-Abelian $\rm{Ising}\times \overline{\rm{Ising}}$ topological phase (line $\rm d$-$\rm c$ in Fig.~\ref{fig:phase_diag}). Though we cannot resolve the nature of this transition, we conjecture that it is either first order or unconventional for the following reason. 
In contrast to the other $\mathbb{Z}_2$ topological phases, the $\mathbb{Z}_2^{\prime}$-phase cannot be obtained by condensing flux excitations in the \Isi phase. This follows from the fact that the $\mathbb{Z}_2^{\prime}$ string operators cannot be obtained from the operators $W_{\mathcal{C}_i}^{({\alpha},{\beta})}$ of the non-Abelian phase using the prescription of Refs.~\cite{bais09b,burnell12}. For example, in the $\mathbb{Z}_2^{\prime}$-phase the string operator that creates vertex defects is 
\begin{align}
W_{\mathcal{C}_i}^{\me'}= \prod\limits_{e\in\mathcal{C}_i} \left(\begin{array}{c c c} 0 & 1 & 0\\ 1 & 0 & 0\\ 0 & 0 & * \end{array}\right),
\label{eq:Z2primeelectric}
\end{align}
i.e.~it flips between the states $\left|\mone\right\rangle_e$ and $\left|\msigma\right\rangle_e$ and can have any (diagonal) action onto the high-energy states $\left|\mpsi\right\rangle_e$.

In the $\mathbb{Z}_2^{\prime}$ phase, the corresponding excitation should be either a boson or a fermion; however, even if we relax the vertex constraint the \Isi phase contains no bosonic or fermionic excitations associated with string operators that raise the edge labels by $\msigma$. 
Further, squaring the operator $W_{\mathcal{C}_i}^{\msigma}$ that raise edges by $\msigma$ in the \Isi phase gives $1 + W_{\mathcal{C}_i}^{\mpsi}$, which creates extended $\mpsi$-strings and is therefore confined in the $\mathbb{Z}_2^{\prime}$ phase.
This incompatibility of the string operators in the two phases implies that they cannot be related by condensing bosonic excitations (see Ref.~\onlinecite{bais09b}). 

\section{Conclusion}\label{sec:conclusion}
\subsection{Summary}
In this work, we have studied the phase diagram of the perturbed Ising string-net model (\ref{eq:fullham}), and described several new phases not previously discussed in the literature. Notably, we have identified a frustrated phase in which $\mathbb{Z}_2$ topological order coexists with translational symmetry breaking, and outlined the interplay between the topological and symmetry-breaking defects. We have also identified a new $\mathbb{Z}_2^\prime$ phase, separating a columnar ordered phase with three-fold breaking of translational symmetry from the trivial phase. 
In addition, in some cases we have identified the corresponding phase transitions analytically, using effective mappings between our full Hamiltonian and various reduced Hamiltonians whose phase transitions are known. 

For each of the phases identified above, we have presented an effective model, valid in some region of the phase diagram, from which the defining features of each phase can be derived exactly. Needless to say these effective descriptions are not valid over the entire parameter regime that we study, and our phase diagram (Fig.~\ref{fig:phase_diag}) is based partly on numerical analysis (Lanczos exact diagonalization).
In particular, we do not find any numerical evidence for additional (intermediate) phases beyond those described here. Needless to say, with the small system sizes attainable numerically, this does not definitively rule out the possibility of additional structure in the phase diagram.

\begin{figure}[htp]%
\includegraphics[width=\columnwidth]{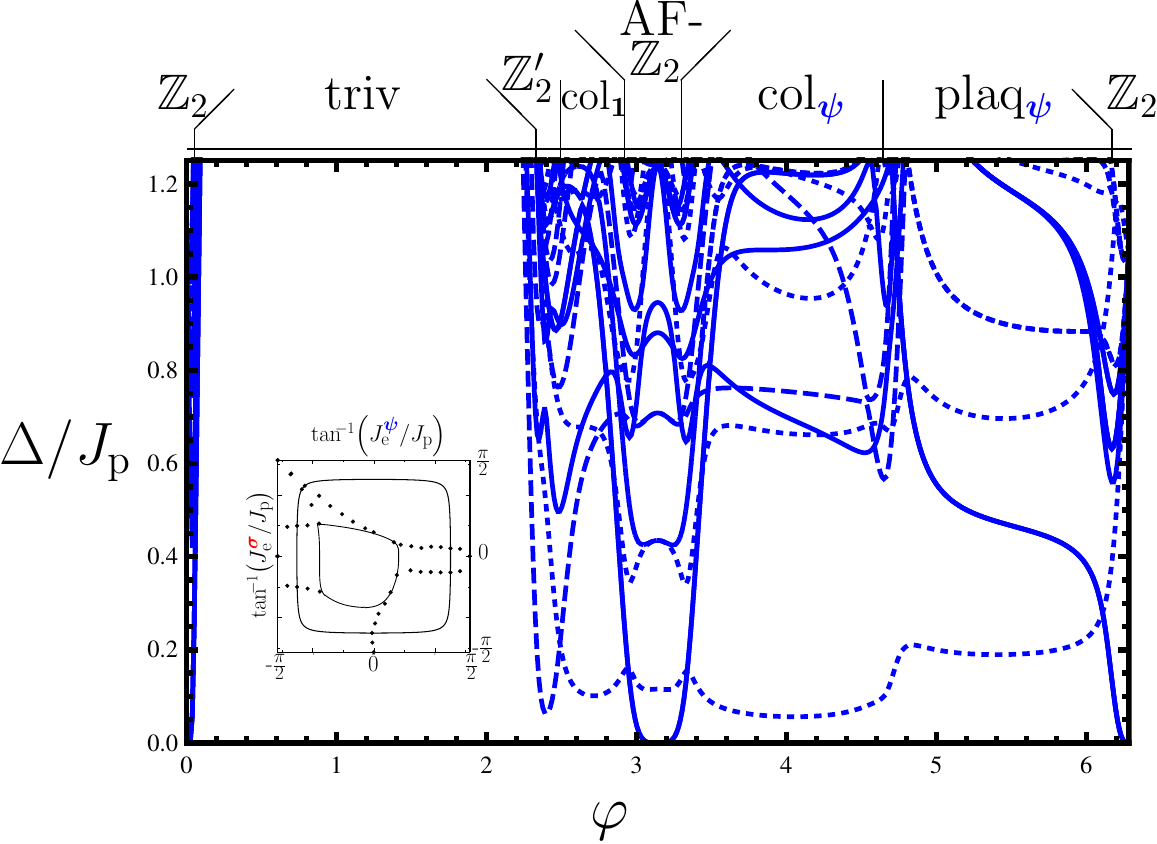}\\
\caption{Low-energy spectrum of the full model (\ref{eq:fullham}) as a function of $\varphi$ for the relevant $\vec{k}$-values, where $\tan\varphi=\frac{\Jes}{\Jep}$ for ${{\Jes}^2+{\Jep}^2}=\Jp\tan\left(\frac{\pi}{5}\right)$. The corresponding cut in the phase diagram is indicated by the solid line in the inset. The low-energy gap $\Delta/\Jp$ is shown here for the momenta $\vec{k}=(0,0)$ and for $\vec{k}=(\pm\frac{2\pi}{3},\mp\frac{2\pi}{3})$ (dotted) for system size $N_{\rm p}=12$. Additionally, we distinguish for $\vec{k}=(0,0)$ the different topological sectors (see text), by solid and dashed lines. The different ground state degeneracies (up to finite size splitting) indicate the different phases labeled above the plot.}%
\label{fig:phi_plot}%
\end{figure}

\begin{figure}[htp]%
\includegraphics[width=\columnwidth]{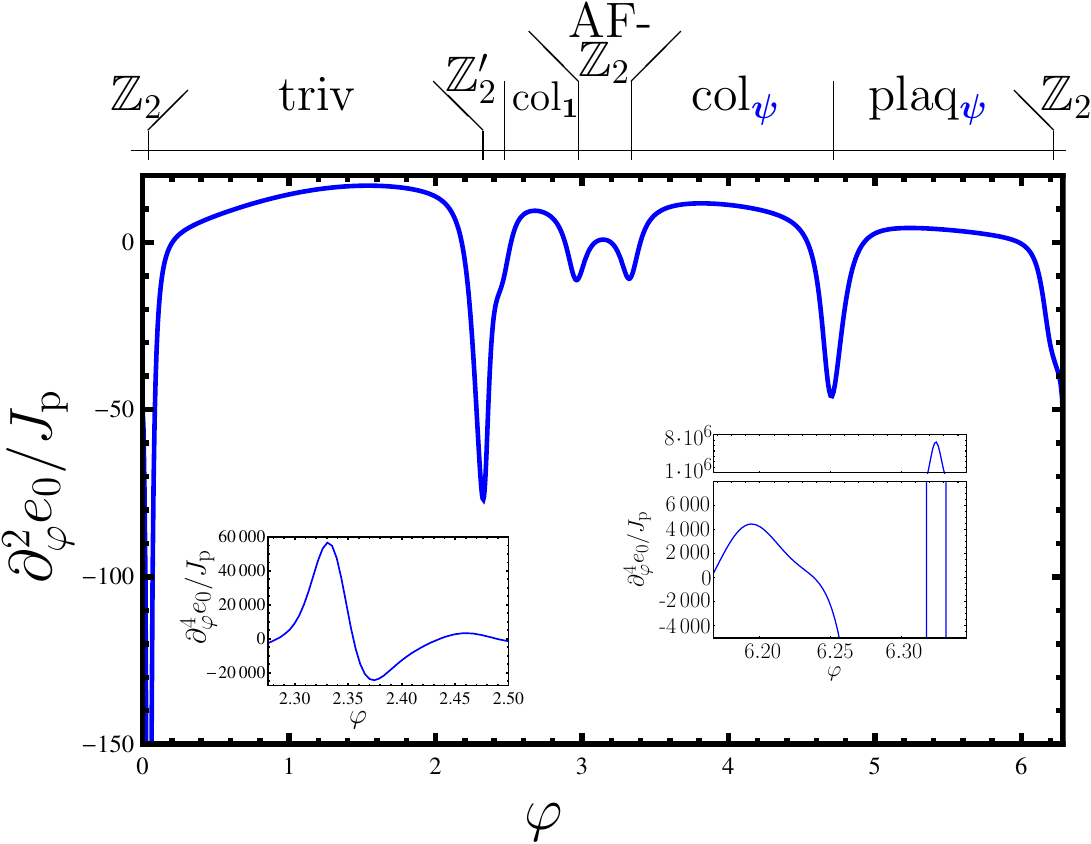} %
\caption{Second derivative of the ground-state energy per plaquette $\partial^2_{\varphi}e_0$ of the full model (\ref{eq:fullham}) (for the $N_{\rm p}=12$-system) as a function of $\varphi$, where $\tan\varphi=\frac{\Jes}{\Jep}$. The divergences (dips for the finite-size system) of the second derivative allow us to estimate the location of the phase transitions between the phases indicated above the plot. The insets show $\partial^4_{\varphi}e_0$, used to separate nearby phase transitions, which allows to locate the phase transitions between the topological to the translational symmetry broken phases, i.e.~the transition from $\mathbb{Z}_2^{\prime}$ to ${\rm col}_{\mone}$ and $\mathbb{Z}_2$ to ${\rm plaq}_{\mpsi}$.}%
\label{fig:phi_e0plot}%
\end{figure}

The numerical results are exemplified in Figs.~\ref{fig:phi_plot} and \ref{fig:phi_e0plot}, where we show the low-energy spectrum (Fig.~\ref{fig:phi_plot}) and ground-state energy derivative (Fig.~\ref{fig:phi_e0plot}) along one cut for
 ${{\Jes}^2+{\Jep}^2}=\Jp\tan\left(\frac{\pi}{5}\right)={\rm const}$ for the phases arising in the limit of large $\Jes,\Jep$. Shown are our results for the largest system ($N_{\rm p}=12$) sizes allowing for the three-sublattice symmetry breaking.
Additionally, we distinguish different topological sectors and different $\vec{k}$-values to show the nature of the different ground state degeneracies (up to finite size splitting), which are in agreement with the conclusions drawn from the effective models in the above sections. 

\subsection{Outlook}
The present work illustrates how topological order can become intertwined with long-ranged order, leading to phases in which the topological excitations are affected by translational symmetry breaking in a non-trivial way. Though we have studied only one example, there are many related lattice models in which condensation transitions that only partially break the topological order are possible\cite{burnell11b}; many of these should admit translation-breaking frustrated phases with topological order similar to the one described here, in which the residual topological order and symmetry-breaking pattern interact in a nontrivial way. 

All of these examples share the common property that the phases simultaneously exhibiting topological and long-ranged order descend from a parent phase with additional anyonic excitations that become confined across the ordering transition. A more experimentally tantalizing question, however, is whether similar phases can emerge in systems where such a parent phase is not natural -- which are much more likely to arise in physically realistic Hamiltonians.  In this context, it is interesting to note that spin liquid states which break lattice rotational symmetries (but not translational ones) are relatively natural in the context of certain frustrated spin models.\cite{ReadSachdev}

Finally, though many of the phase transitions in our model can be deduced from our various effective Hamiltonians, a number of the (possibly second-order) transitions remain inaccessible through the approaches presented here. Notably, the transitions out of the \Isi phase in which the non-Abelian $\msigma$-fluxes condense are a topic worthy of further study. 

\acknowledgments
{\em Acknowledgments --}
We like to thank S.~Dusuel, J.~Romers, F.~Pollmann, K.~Shtengel, S.~Simon, and J.~Vidal for fruitful discussions. FJB is supported by NSF DMR-1352271 and by the Sloan foundation, grant no.~FG-2015-65927. This work was carried out in part using computing resources at the University of Minnesota Supercomputing Institute.
This research was supported in part by Perimeter Institute for Theoretical Physics. REsearch at Perimeter Institute is supported by the Government of Economic Development \& Innovation.

%
\appendix
\section{Technical details of the string-net model}\label{app:stringnetdetails}
In this section we give the definition of the operator $B_p^{\boldsymbol{s}}$ used to define the string-net Hamiltonian $H_{\rm SN}$ (\ref{eq:HSN_def}).

For the sake of generality, we introduce two sets of coefficients, known as $F$- and $R$-symbols, which can be used to define the string-net Hamiltonian $H_{\rm SN}$ (\ref{eq:HSN_def}) and the various string operators $W_{\mathcal{C}_i}$ for a general topological order characterized by a unitary modular tensor category.  For a more comprehensive introduction to unitary modular tensor categories, see e.g. Refs. ~\onlinecite{bonderson_thesis,kitaev06,wang08}.

The $F$-symbols dictate how string operators raise and lower labels on a given edge. They are defined by the pictorial relation
\begin{align}
\begin{array}{c}\includegraphics[width=1.5cm]{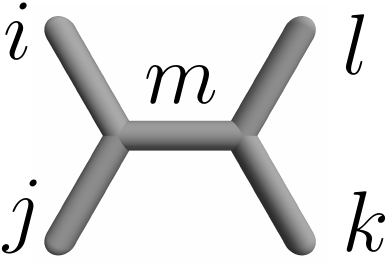}\end{array} = \sum\limits_n F^{i\,j\,m}_{k\,l\,n} \begin{array}{c}\includegraphics[width=1.5cm]{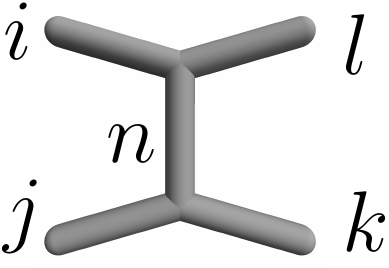}\end{array}.
\label{eq:fmove}
\end{align}
For the Ising theory, they vanish unless all four involved vertices obey the vertex constraints shown in Fig.~\ref{fig:lattice_constraints}. The non-zero $F$-symbols for the Ising theory all equal $1$, except the following six: 
\begin{align}
F^{\msigma\,\mpsi\,\msigma}_{\,\msigma\,\mpsi\,\msigma}&=F^{\mpsi\,\msigma\,\msigma}_{\,\mpsi\,\msigma\,\msigma}=-1,\\
F^{\msigma\,\msigma\,\mone}_{\,\msigma\,\msigma\,\mone}&=F^{\msigma\,\msigma\,\mpsi}_{\,\msigma\,\msigma\,\mone}=F^{\msigma\,\msigma\,\mone}_{\,\msigma\,\msigma\,\mpsi}=-F^{\msigma\,\msigma\,\mpsi}_{\,\msigma\,\msigma\,\mpsi}=\frac{1}{\sqrt{2}}.
\label{eq:fsymbols}
\end{align}
In order to define string operators, we will also use the so-called $R$-symbols. These are needed to define the action of a string operator where it crosses an edge, in such a way that it commutes with $B_p$\cite{levin05,burnell10}. The corresponding pictorial definitions read
\begin{align}
\begin{array}{c}\includegraphics[width=1cm]{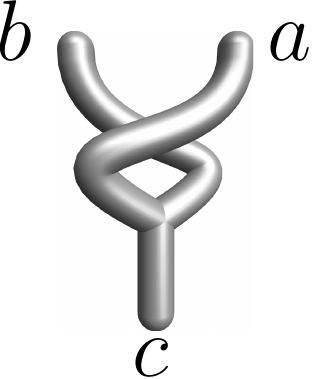}\end{array} = \ & \phantom{\left(R^{ab}_c\right)^{-1}} \hspace{-.75cm}R^{ab}_c \hspace{.25cm} \begin{array}{c}\includegraphics[width=1cm]{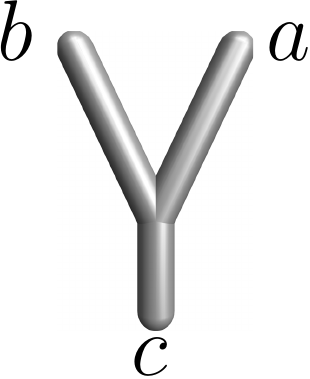}\end{array},\\
\begin{array}{c}\includegraphics[width=1cm]{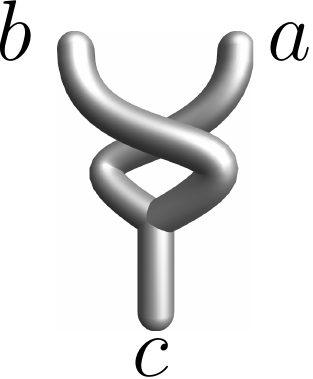}\end{array} = \ & \left(R^{ba}_c\right)^{-1} \begin{array}{c}\includegraphics[width=1cm]{./figures/R_unbraid_gray-crop.pdf}\end{array}.
\label{eq:Rmove}
\end{align}
The coefficients $R^{ab}_c$ also vanish unless the vertex obeys the vertex constraint shown Fig.~\ref{fig:lattice_constraints}. The non-zero $R$-symbols for the Ising theory all equal $1$, except the following:
\begin{align}
R^{\msigma\msigma}_{\mone}=e^{-\frac{\pi\rm{i}}{8}},R^{\mpsi\mpsi}_{\mone}=-1,R^{\mpsi\msigma}_{\msigma}=R^{\msigma\mpsi}_{\msigma}=-\rm{i},R^{\msigma\msigma}_{\mpsi}=e^{\frac{3\pi\rm{i}}{8}}.
\label{eq:rsymbols}
\end{align}

\section{The action of the operators \texorpdfstring{$B_p^{\boldsymbol{s}}$}{Bps}}\label{app:details_bps}
The operators $B_p^s$ in Eq.~(\ref{eq:HSN_def}) are defined by using the $F$-symbols (\ref{eq:fsymbols}) to ``fuse" the string $s$ into each of the edges of the plaquette. 
This can be visualized by the action
\begin{align}
B_p^{\boldsymbol{s}} \left|\begin{array}{c}{\includegraphics[width=1.5cm]{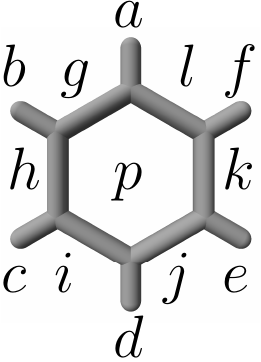}}\end{array}\right\rangle = \left|\begin{array}{c}{\includegraphics[width=1.5cm]{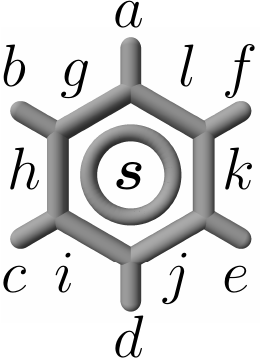}}\end{array}\right\rangle.
\label{eq:action_Bps_orig}
\end{align}

To resolve this to the edge-basis states, one can e.g.~make use of the following relation (for each vertex sequentially):
\begin{align}
\left|\begin{array}{c}{\includegraphics[width=1.5cm]{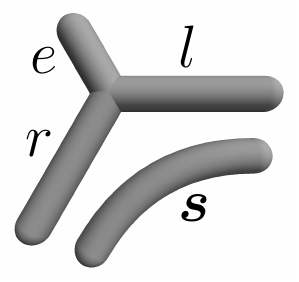}}\end{array}\right\rangle=\underbrace{F^{l\,l\,\mone}_{\,\boldsymbol{s}\boldsymbol{s}l^{\prime}}}_{\Theta^{\boldsymbol{s},l,l^{\prime}}}\left|\begin{array}{c}{\includegraphics[width=1.5cm]{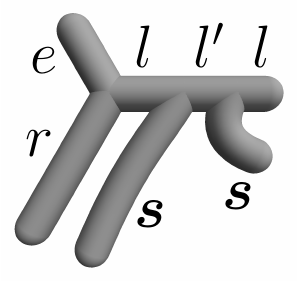}}\end{array}\right\rangle.
\label{eq:resolve_Bps_1}
\end{align}
In a second step, one uses 
\begin{align}
\left|\begin{array}{c}{\includegraphics[width=1.5cm]{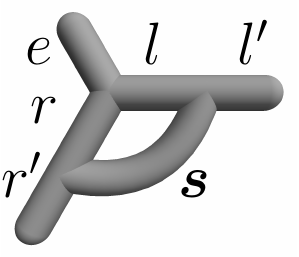}}\end{array}\right\rangle=\frac{F^{e\,l\,r}_{\,\boldsymbol{s}r^{\prime}l^{\prime}}}{\Theta^{\boldsymbol{s},l,l^{\prime}}}\left|\begin{array}{c}{\includegraphics[width=1.5cm]{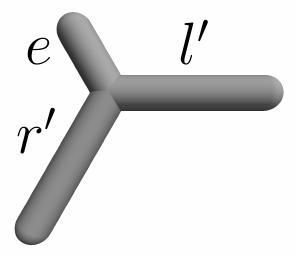}}\end{array}\right\rangle.
\label{eq:resolve_Bps_2}
\end{align}
Combining these two steps, the $\Theta$-factors cancel. Thus the coefficients $\phi(v)$ in Eq.~(\ref{eq:bps_def}) are given by the F-symbols (\ref{eq:fmove}). The non-trivial relations are:
\begin{align}
\begin{array}{c}\includegraphics[width=0.75cm]{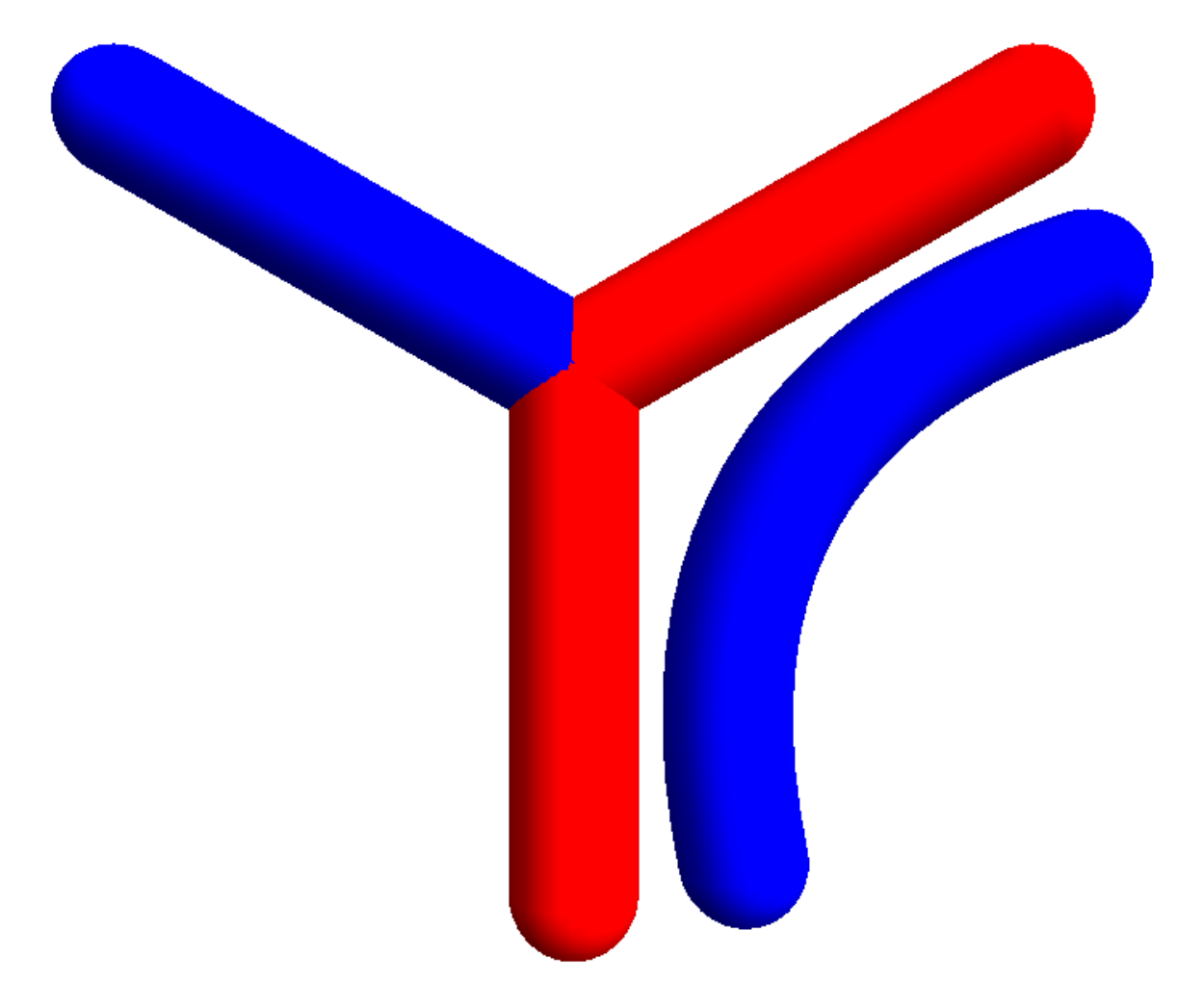}\end{array} &= - \begin{array}{c}\includegraphics[width=0.75cm]{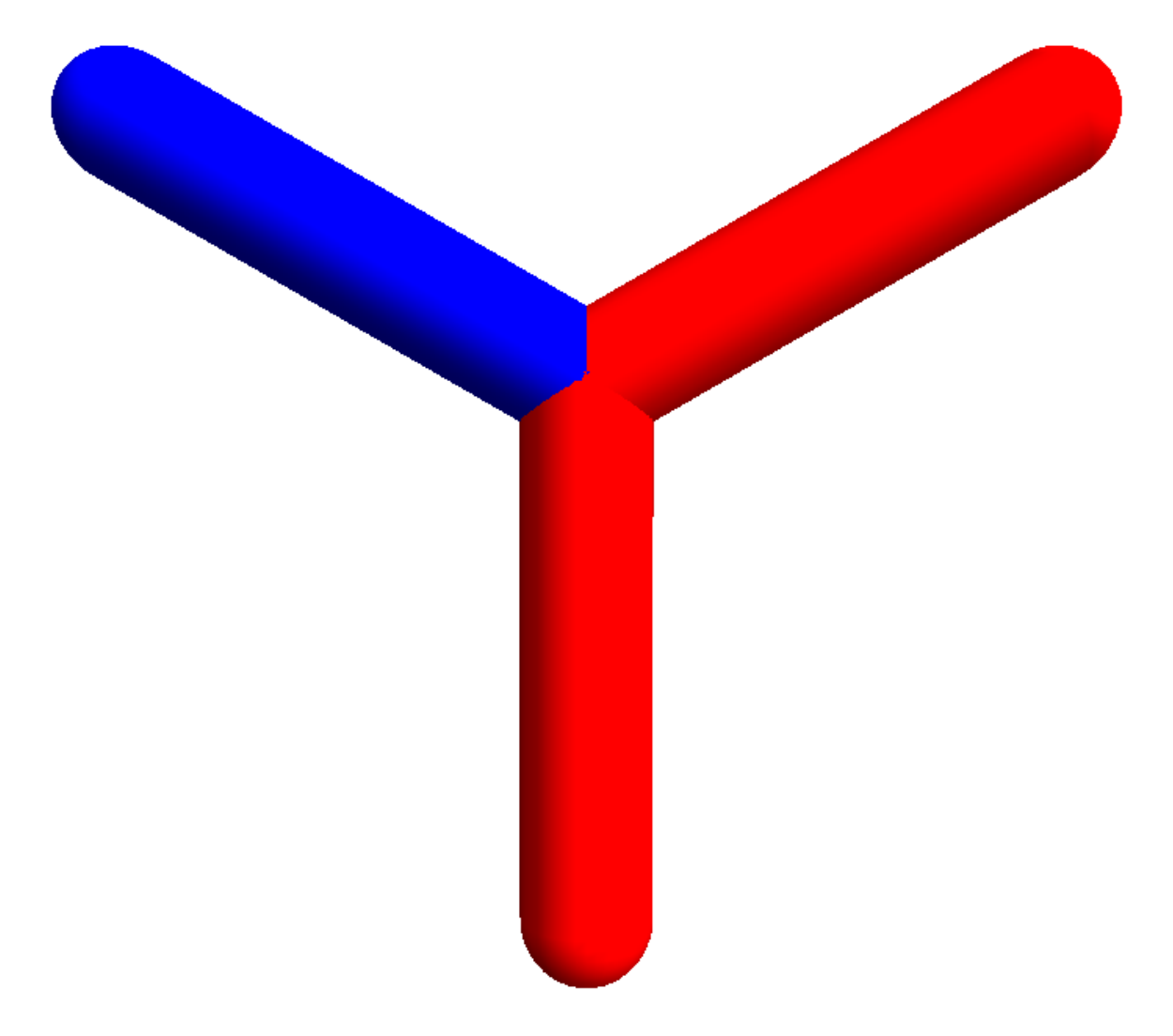}\end{array},\label{eq:fsymbpic1}\\
 \ \ \begin{array}{c}\includegraphics[width=0.75cm]{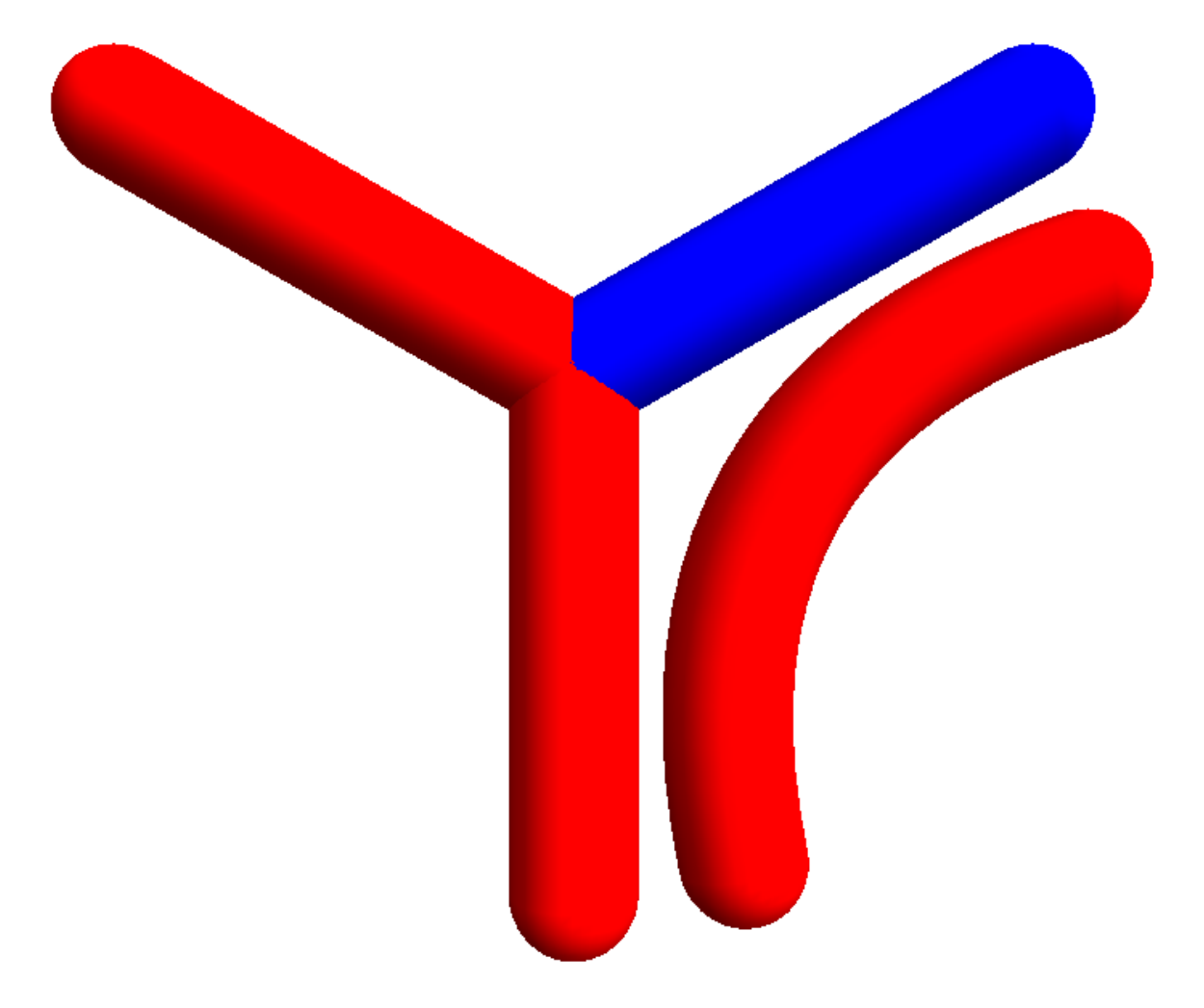}\end{array} &= \begin{array}{c}\includegraphics[width=0.75cm]{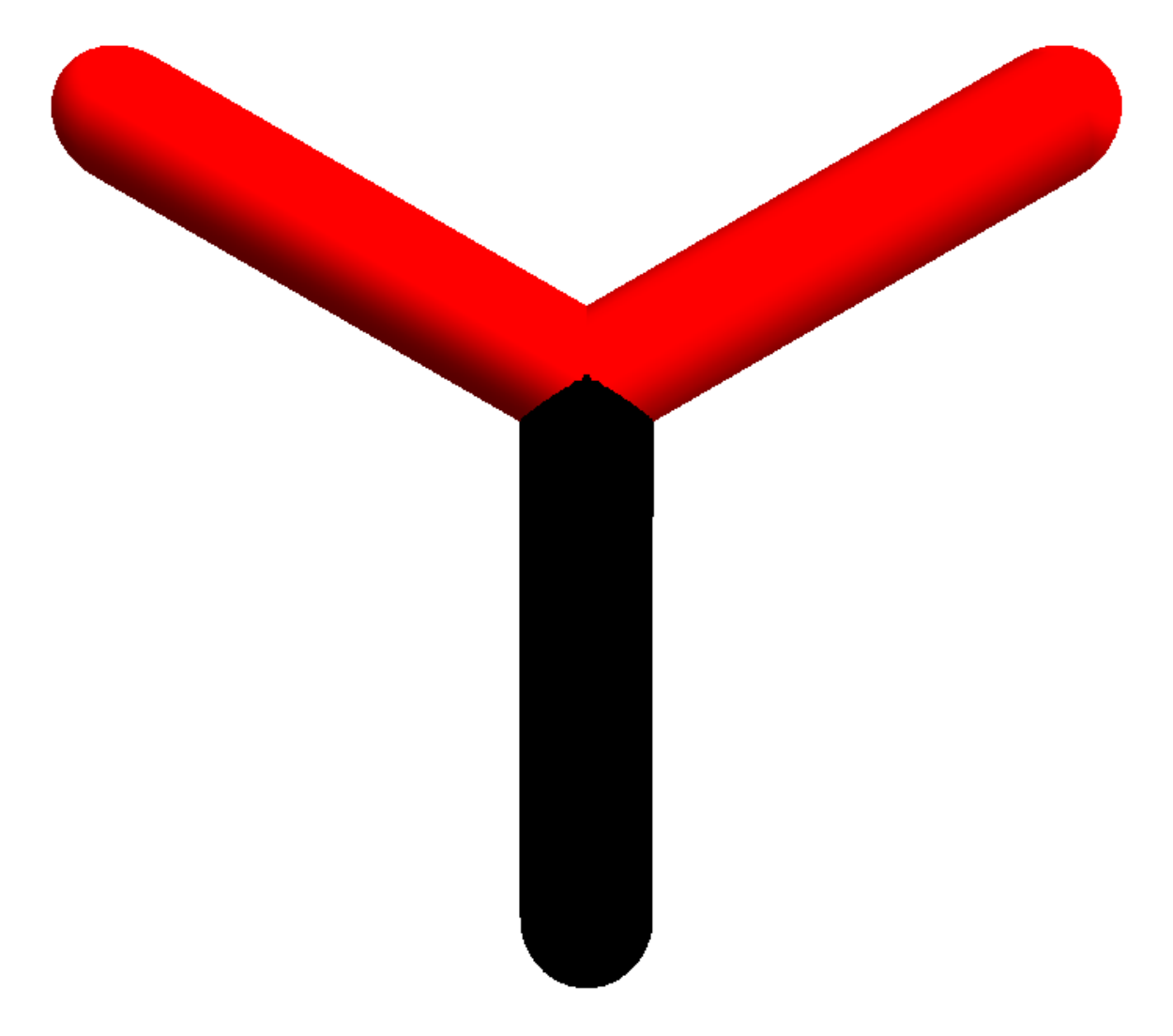}\end{array} - \begin{array}{c}\includegraphics[width=0.75cm]{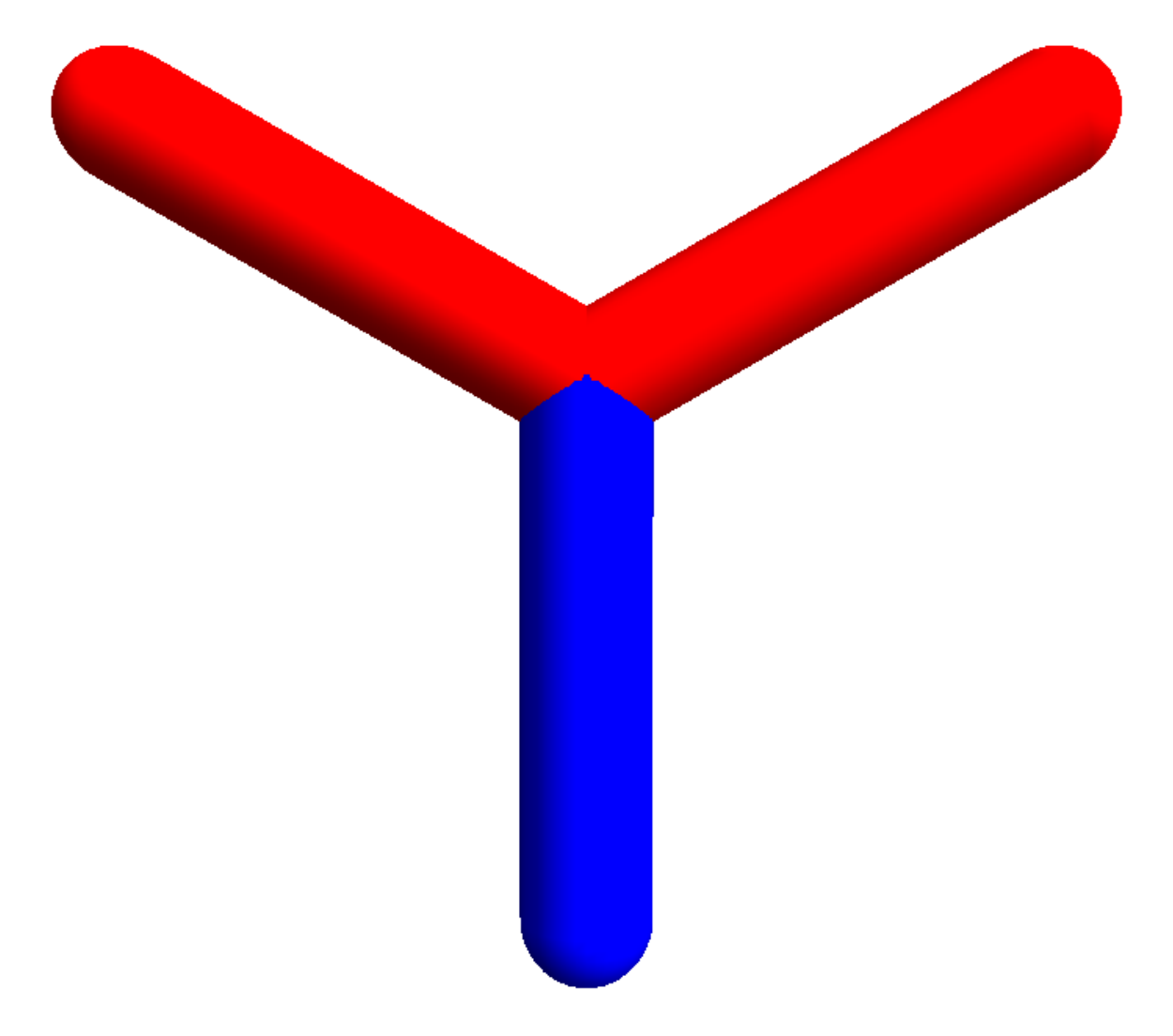}\end{array}, \label{eq:fsymbpic1a}\\
\begin{array}{c}\includegraphics[width=0.75cm]{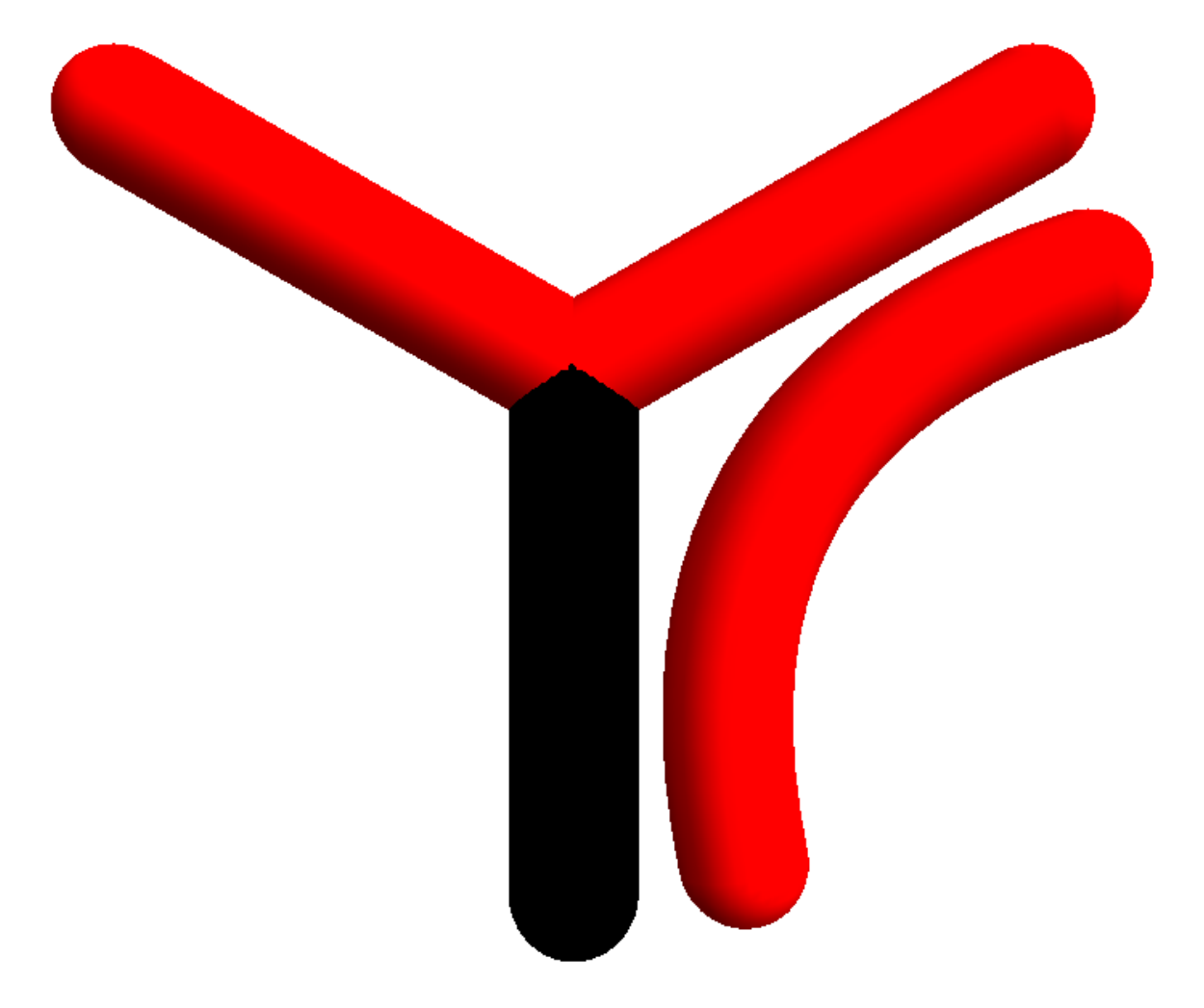}\end{array} &= \frac{1}{\sqrt{2}} \begin{array}{c}\includegraphics[width=0.75cm]{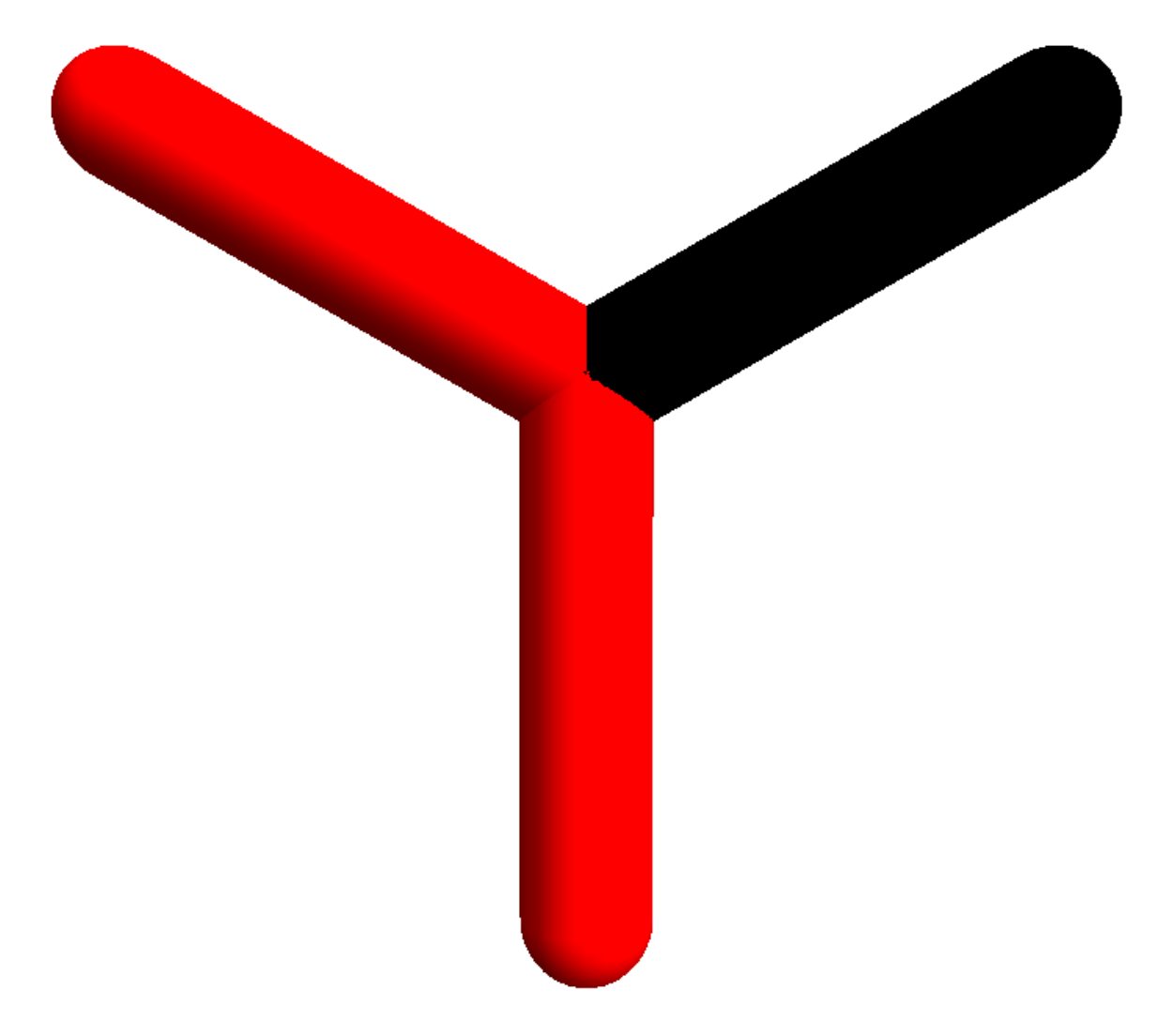}\end{array}+ \frac{1}{\sqrt{2}} \begin{array}{c}\includegraphics[width=0.75cm]{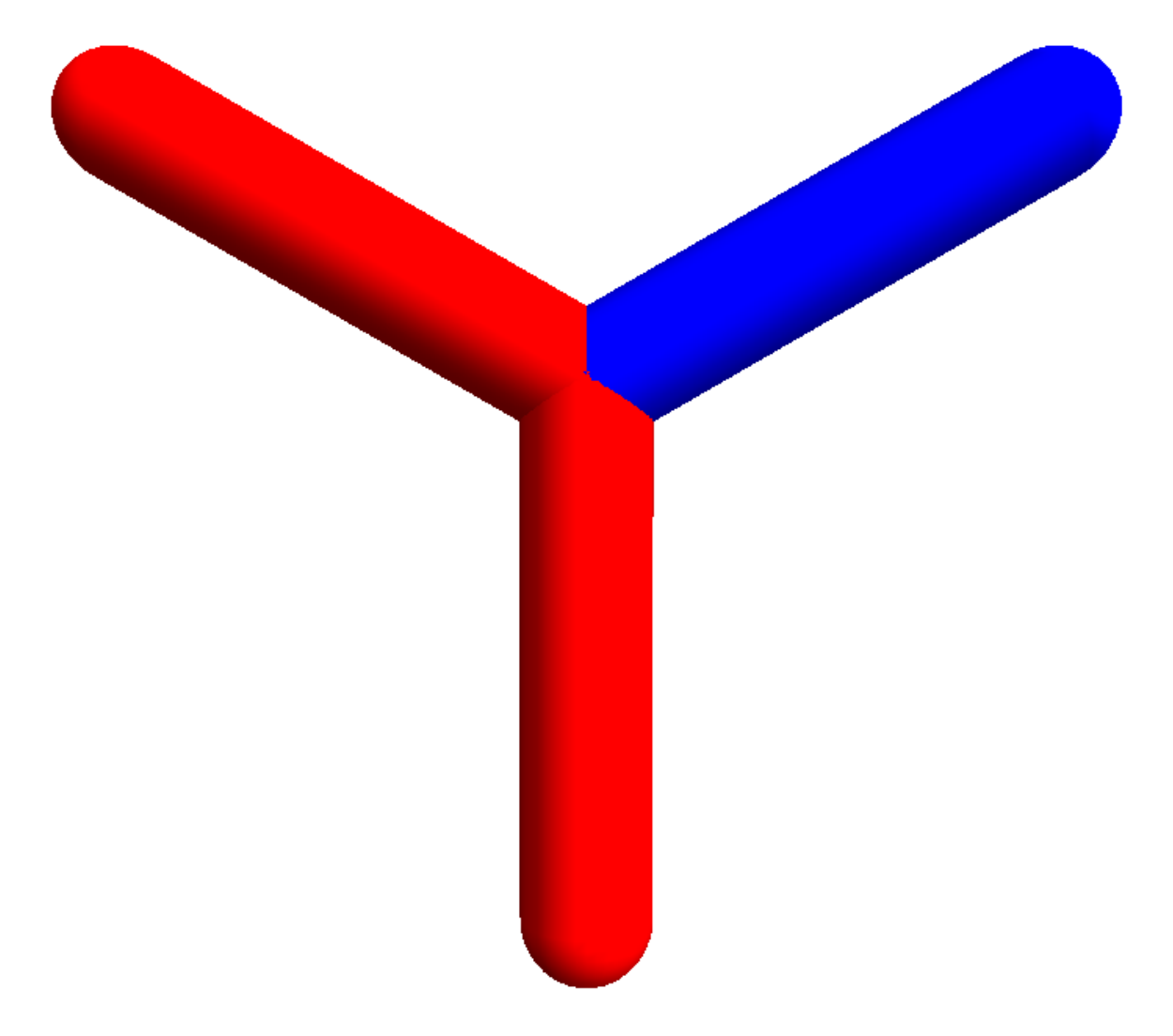}\end{array}, \label{eq:fsymbpic2}\\
\begin{array}{c}\includegraphics[width=0.75cm]{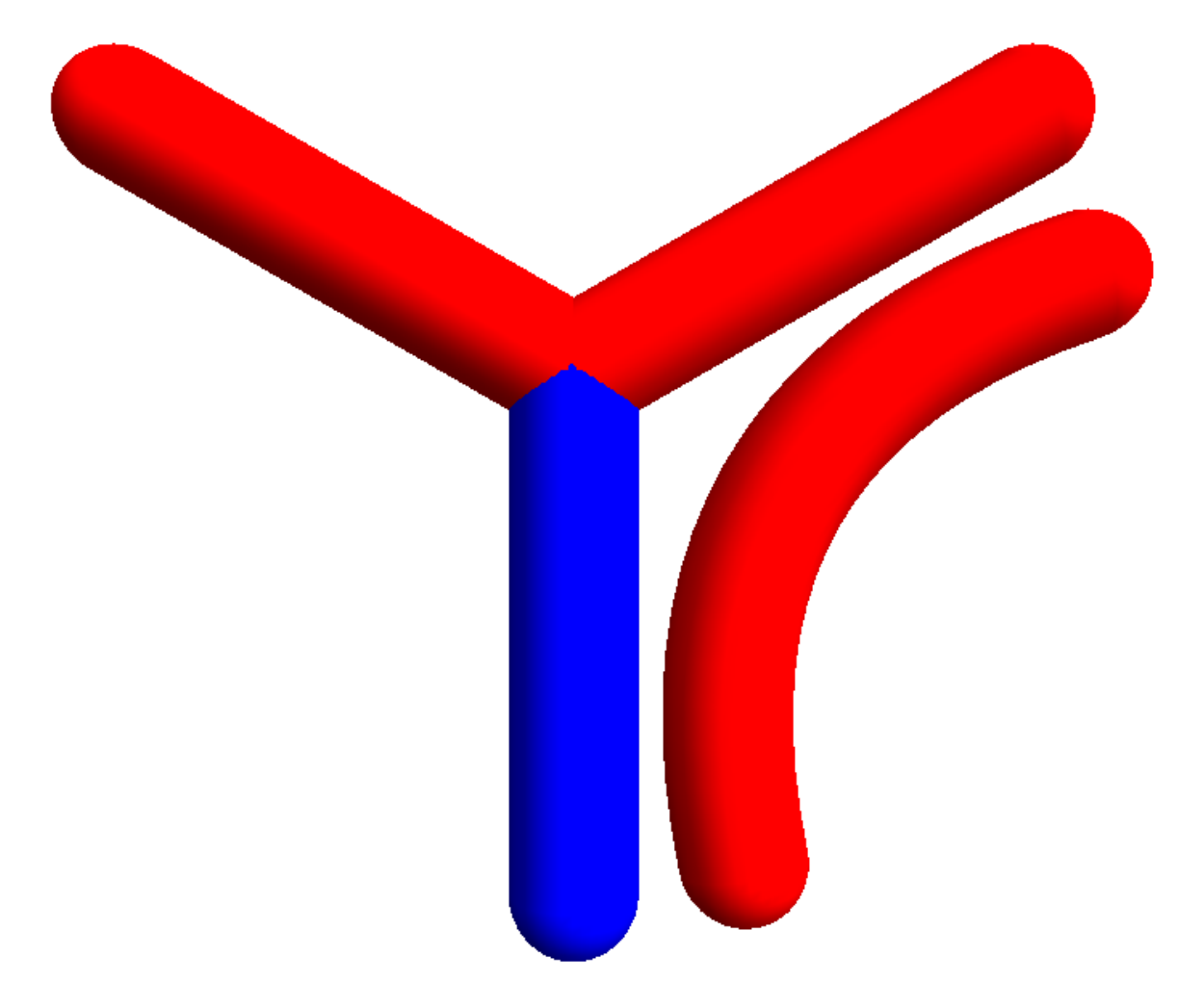}\end{array} &= \frac{1}{\sqrt{2}} \begin{array}{c}\includegraphics[width=0.75cm]{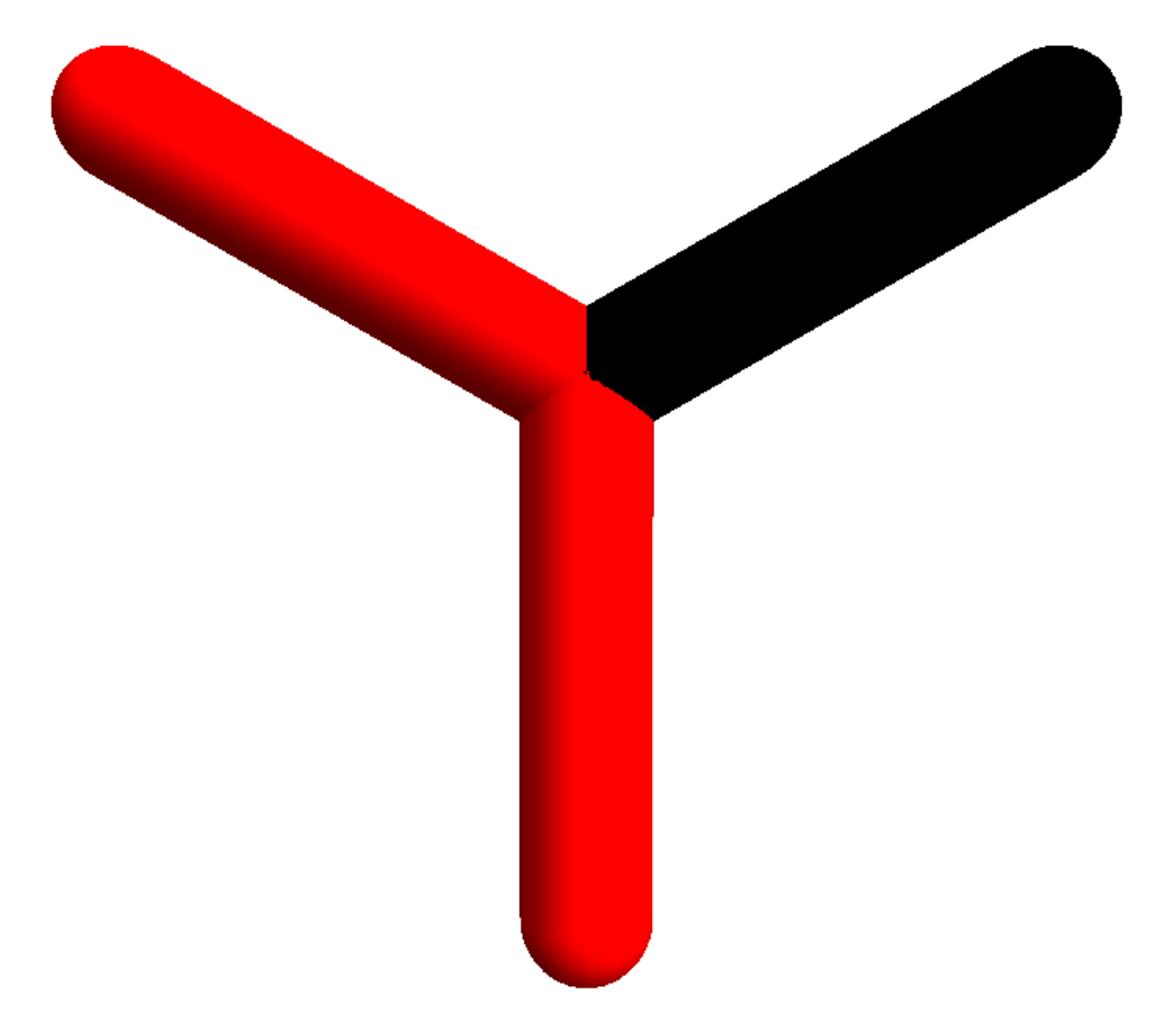}\end{array} -\frac{1}{\sqrt{2}} \begin{array}{c}\includegraphics[width=0.75cm]{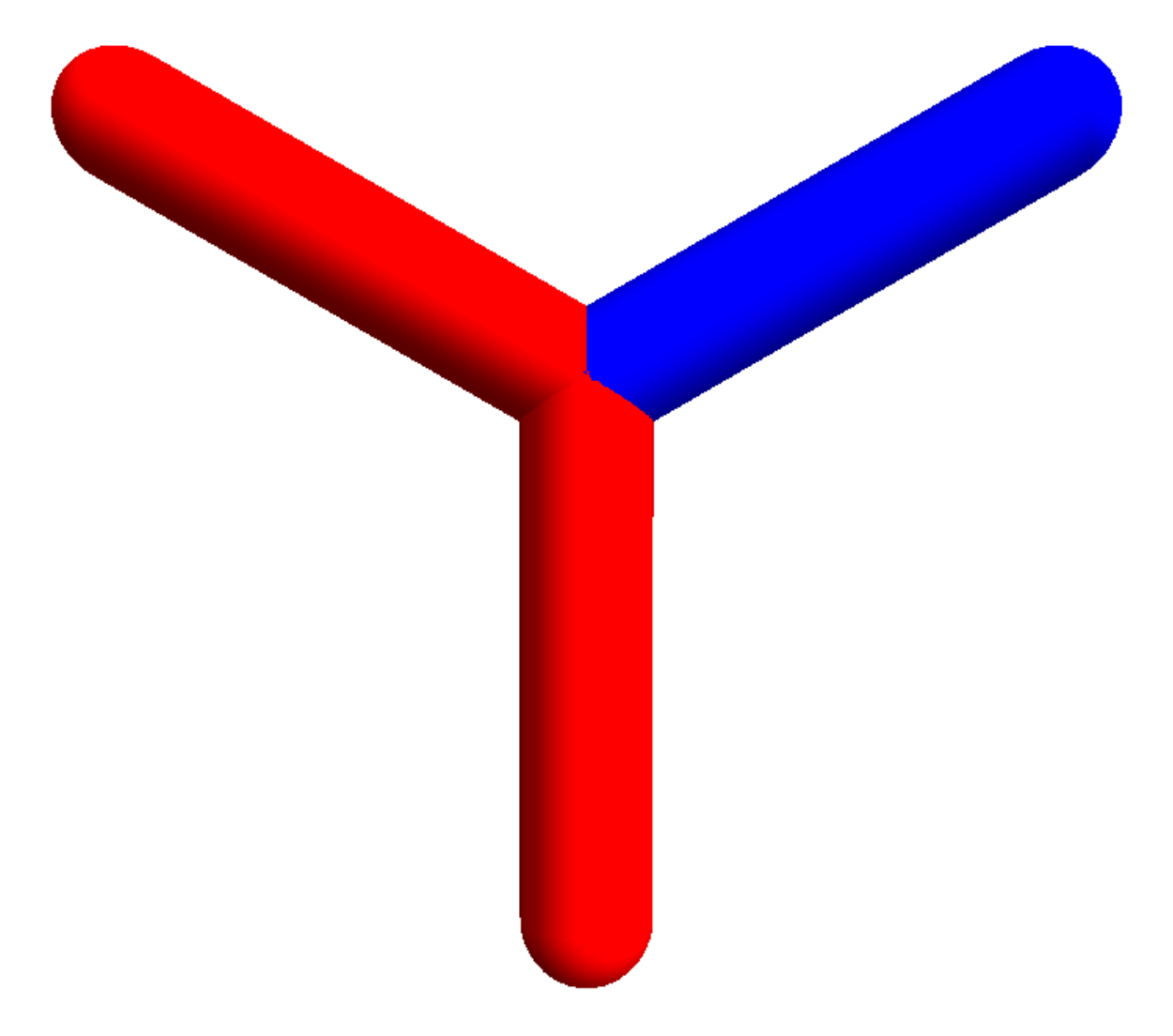}\end{array}. \label{eq:fsymbpic3}
\end{align}
Here we have shown the coefficients for one vertex type, where $s$ acts on the right. The coefficients of the remaining vertices can be obtained by appropriate rotations of the terms shown here. It is convenient to know that the factors of $\sqrt{2}$, which depend on which internal edge is labeled $\msigma$, always give a net amplitude of $1$ if the plaquette move creates a $\msigma$-loop, $1/2$ if it breaks a $\msigma$-loop, and $1/\sqrt{2}$ if it neither breaks nor creates $\msigma$-loops. 

This gives the following matrix elements for $B_p^{\boldsymbol{s}}$:
\begin{align}
\left\langle\begin{array}{c}{\includegraphics[width=1.5cm]{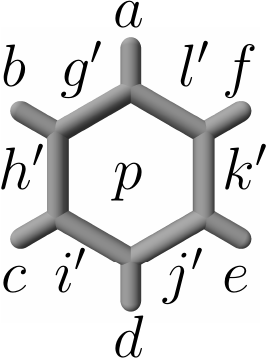}}\end{array}\right|B_p^{\boldsymbol{s}} \left|\begin{array}{c}{\includegraphics[width=1.5cm]{./figures/hexa3d_labels-crop.pdf}}\end{array}\right\rangle \hspace{2cm}\nonumber\\
= 
F^{a\,l\,g}_{\,\boldsymbol{s}g^{\prime}l^{\prime}}
F^{b\,g\,h}_{\,\boldsymbol{s}h^{\prime}g^{\prime}}
F^{c\,h\,i}_{\,\boldsymbol{s}i^{\prime}h^{\prime}}
F^{d\,i\,j}_{\,\boldsymbol{s}j^{\prime}i^{\prime}}
F^{e\,j\,k}_{\,\boldsymbol{s}k^{\prime}j^{\prime}}
F^{f\,k\,l}_{\,\boldsymbol{s}l^{\prime}k^{\prime}}.
\label{eq:action_Bps}
\end{align}
For non-zero matrix elements, the final (primed) link labels differ from the initial (unprimed) ones by a product of the raising operators $S_e^{\boldsymbol{s}}$ (\ref{eq:S_operators}), i.e. $B_p^{\boldsymbol{s}} = \prod\limits_{v \in p} \phi(v) \prod_eS_e^{\boldsymbol{s}}$, where the prefactor $\phi(v)$ is given by the F-symbols stemming from Eq.~(\ref{eq:resolve_Bps_2}).

\section{General form of the string and loop operators}\label{app:stringoperatorsdetails}
In this section, we will review the general mathematical formulation for the string-(or loop-) operators characterizing the topological order realized by the string-net Hamiltonians. We discuss the details for the $\mathbb{Z}_2$ operators in more detail in App.~\ref{app:stringoperatorsz2details}.
\subsection{The action of the loop operators \texorpdfstring{$W_{\mathcal{C}}^{(\alpha,\beta)}$}{WCab}}\label{ssec:strings}
\begin{figure}[pt]%
\centering
\includegraphics[width=.85\columnwidth]{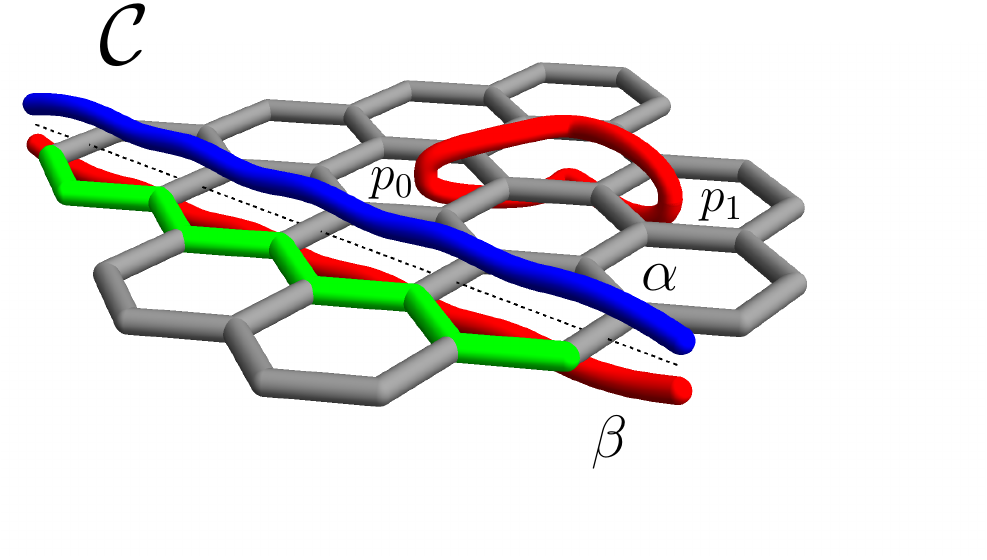}%
\caption{
A pictorial representation of loop operators: The loop operators $W_{\mathcal{C}}^{(\alpha,\beta)}$ for a given cycle $\mathcal{C}$ (depicted in green) is given by adding one loop above (blue, labeled by $\alpha$) and one below (red, labeled by $\beta$) the lattice and subsequently fusing these into the lattice. Additionally one can also depict the open string operators creating particles on plaquettes $p_0$ and $p_1$ (red) in a similar fashion\cite{gils09,dusuel_inprep}.}%
\label{fig:lattice_operators}%
\end{figure}

Just as the operators $B_p^{\boldsymbol{s}}$ can be defined as ``fusing" closed loops into plaquettes of the lattice, the loop operators $W_{\mathcal{C}_i}^{(\alpha,\beta)}$ can be visualized as fusing a pair of closed loops -- an $\alpha$ loop above the lattice and a $\beta$ loop below the lattice-- along the non-contractible cycle $\mathcal{C}_i$, as depicted in Fig.~\ref{fig:lattice_operators}. The action of this fusion process on the edges along the curve $\mathcal{C}_i$ (shown in green in Fig.~\ref{fig:lattice_operators}) is dictated by the $F$- and $R$-symbols. The non-trivial $R$-symbols result in the fact that operators defined on intersecting loops do not commute in general. Loops above (below) the lattice correspond to right-chiral (left-chiral) operators in the \Isi topological theory.\cite{burnell11} 

One way to evaluate the resulting coefficients in practice is given e.g.~in Ref.~\onlinecite{burnell12}: First, the two loops are contracted to one in between the crossed links via
\begin{align}
\begin{array}{c}\includegraphics[height=1.45cm]{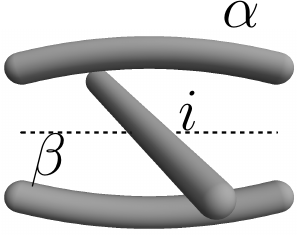}\end{array}
=&\sum_{X,Y}F^{\alpha\,\alpha\,\mone}_{\,\beta\,\beta\,X}F^{\alpha\,\alpha\,\mone}_{\,\beta\,\beta\,Y}
\begin{array}{c}\includegraphics[height=1.5cm]{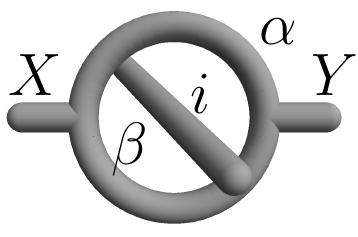}\end{array}.
\label{eq:loopstring}
\end{align}
Second, the resulting rings are resolved to a planar arrangement by
\begin{align}
&\sum_{X,Y}F^{\alpha\,\alpha\,\mone}_{\,\beta\,\beta\,X}F^{\alpha\,\alpha\,\mone}_{\,\beta\,\beta\,Y}
\begin{array}{c}\includegraphics[height=1.5cm]{./figures/resolve_cylces2_labels-crop.pdf}\end{array}\nonumber\\
=&
\sum\limits_{i^{\prime},j,l,X,Y}
F^{\beta\,\beta\,\mone}_{\,i\,i\,j}
F^{\alpha\,\alpha\,\mone}_{\,j\,j\,i^{\prime}}
F^{i\,\beta\,j}_{\,i^{\prime}\,\alpha\,l}
R^{i \, \beta}_{j} \times \nonumber\\
&\left(R^{i\,\alpha}_l\right)^{-1}
F^{i\,\alpha\,l}_{\,\beta\, i^{\prime} \, X}
F^{\alpha\, i^{\prime} \, j}_{\, i \,\beta Y}
\begin{array}{c}\includegraphics[width=2.cm]{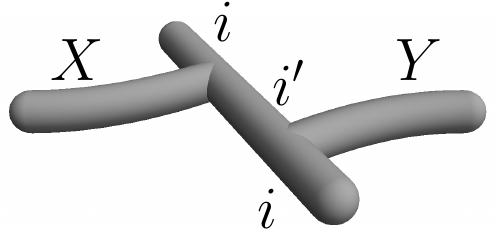}\end{array}\label{eq:fusingstringsphase}\\
\equiv & \sum\limits_{i^{\prime},X,Y} w^{\alpha,\beta,i}_{X,Y,i'}\begin{array}{c}\includegraphics[width=2.cm]{./figures/resolve_cylces3_labels-crop.pdf}\end{array},
\label{eq:loopstring2}
\end{align}
which can then, in a third step not explicitly shown here, be reduced to a new configuration of edges by fusing the remaining strings $X,Y$ into the lattice. This gives after canceling out all remaining $\Theta$ factors a resulting matrix element $\Phi(v)$ for crossed edges
\begin{align}
\sum\limits_{X,Y}\!\!\begin{array}{c}\includegraphics[width=1.5cm]{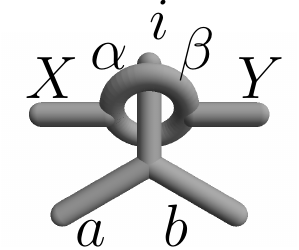}\end{array}\!\!\!=\!\!\!\!\!\!\sum\limits_{X,Y,i^{\prime},a^{\prime},b^{\prime}}\!\!\!\!\!\!w^{\alpha,\beta,i}_{X,Y,i'} \frac{F^{a\, b \, i}_{\, Y \,i^{\prime} b^{\prime}}F^{b^{\prime}\, i \, a}_{\, X \,a^{\prime} i}}{\Theta^{X,i,i^{\prime}}}\!\begin{array}{c}\includegraphics[width=1.cm]{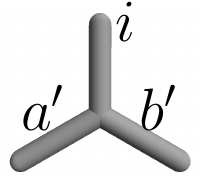}\end{array}\!.\label{eq:fusingloops}
\end{align}

From Eq.~(\ref{eq:fusingloops}) it is clear that 
\begin{align}
W_{\mathcal{C}_i}^{(\alpha,\beta)}=\prod\limits_{\begin{array}{c}\rm\scriptstyle outside\vspace{-.2cm}\\ \vspace{-.05cm} \rm\scriptstyle corners\end{array}} \bar{\phi}(v)\prod\limits_{\rm\begin{array}{c}\rm\scriptstyle  inside \vspace{-.2cm}\\ \vspace{-.05cm} \rm\scriptstyle corners\end{array}}\phi(v)\prod\limits_{e\in\mathcal{C}_i}S_e^{\alpha}S_e^{\beta},
\label{eq:eff_wab}
\end{align}
where ``inside corners" refers to vertices without crossed edges, for which the factor $\phi(v)$ is given as for the operators $B_p^s$ and ``outside corners" are corners with crossed edges, for which the matrix element $\bar{\phi}(v)$ is given in Eq.~(\ref{eq:fusingloops}).
One can then show that 
\begin{align}
W_{\mathcal{C}_i}^{(\alpha,\beta)}W_{\mathcal{C}_i}^{(\gamma,\delta)}=\sum\limits_{\epsilon,\zeta} \delta_{\epsilon,\alpha,\gamma}\delta_{\zeta,\beta,\delta}W_{\mathcal{C}_i}^{(\epsilon,\zeta)},
\label{eq:prodsamecycle}
\end{align}
where $\delta_{\sigma, \rho, \epsilon} =1$ if $(\sigma, \rho, \epsilon)$ is one of the configurations shown in Fig.~(\ref{fig:lattice_constraints}) and $0$ otherwise.

\subsection{Loop operators and the ground state degeneracy}
We can find the ground state degeneracy on the torus (or, through similar means, on any closed manifold) by studying the algebra of loop operators on the two non-contractible cycles ($\mathcal{C}_1$ and $\mathcal{C}_2$). 
$\{ W_{\mathcal{C}_1} \}$ is a set of mutually commuting operators that may be simultaneously diagonalized, as is $\{ W_{\mathcal{C}_2} \}$. 
However, since any path around $\mathcal{C}_1$ must intersect a path around $\mathcal{C}_2$ an odd number of times, any (non-identity) operator $W_{\mathcal{C}_1}^{(\alpha, \beta)}$ will fail to commute with at least one operator in $\{ W_{\mathcal{C}_2} \}$. Here we choose to label states by eigenvalues of appropriate combinations of $W_{\mathcal{C}_1}$ strings. We will show that all nine possible sets of eigenvalues can be obtained by acting with $W_{\mathcal{C}_1}$ operators on a given eigenstate.

It is convenient to label our states by defining projectors $\{ \mathcal{P}^\alpha \}$ onto a fixed flux $\alpha$ through the cycle $\mathcal{C}_1$. For left-chiral strings (acting below the lattice), the appropriate projectors are\cite{levin05,schulz15}
\begin{align}
\mathcal{P}^{\mone}=&\frac{1}{4}W_{\mathcal{C}_1}^{(\mone,\mone)}+\frac{\sqrt{2}}{4}W_{\mathcal{C}_1}^{(\mone,\msigma)}+\frac{1}{4}W_{\mathcal{C}_1}^{(\mone,\mpsi)},\\
\mathcal{P}^{\msigma}=&\frac{1}{2}W_{\mathcal{C}_1}^{(\mone,\mone)}\phantom{+\frac{\sqrt{2}}{4}W_{\mathcal{C}_1}^{(\mone,\msigma)}}-\frac{1}{2}W_{\mathcal{C}_1}^{(\mone,\mpsi)},\\
\mathcal{P}^{\mpsi}=&\frac{1}{4}W_{\mathcal{C}_1}^{(\mone,\mone)}-\frac{\sqrt{2}}{4}W_{\mathcal{C}_1}^{(\mone,\msigma)}+\frac{1}{4}W_{\mathcal{C}_1}^{(\mone,\mpsi)}.
\label{eq:projs}
\end{align}
(Projectors for right-chiral strings, which act above the lattice, are obtained from analogous expressions with the order of the two indices in the superscripts exchanged. The remaining projectors are constructed from products of left- and right- chiral projectors). 

One can show that, for any reference configuration $|\Psi_{\text{Ref}}\rangle$, 
\begin{align}
\mathcal{P}^{\alpha} \mathcal{P}^{\beta}|\Psi_{\text{Ref}}\rangle = \delta_{\alpha, \beta} \mathcal{P}^{\beta}|\Psi_{\text{Ref}}\rangle \nonumber \\
\mathcal{P}^{\alpha}W_{\mathcal{C}_2}^{(\mone,\alpha)}\mathcal{P}^{\mone}|\Psi_{\text{Ref}}\rangle =W_{\mathcal{C}_2}^{(\mone,\alpha)}\mathcal{P}^{\mone}|\Psi_{\text{Ref}}\rangle,
\label{eq:raising}
\end{align}
i.e.~the three left-chiral projectors project onto orthogonal Hilbert spaces, and the operator $W_{\mathcal{C}_2}^{(\mone,\beta)}$ acts as a raising/lowering operator for the corresponding conserved flux quantum numbers. In particular, three distinct left-chiral flux eigenstates can be constructed in this way. 

Similarly, one can show that $W_{\mathcal{C}_2}^{(\alpha,\mone)}$ is a raising/lower operators for the right-chiral flux. Since all operators in the right-chiral sector commute with all operators in the left-chiral sector, each of the nine possible ground states can be obtained from a state in the trivial sector by acting with 
an operator $W_{\mathcal{C}_2}^{(\alpha,\beta)}=W_{\mathcal{C}_2}^{(\alpha,\mone)}W_{\mathcal{C}_2}^{(\mone,\beta)}$.

This construction can in principle also be applied to determine the ground-state degeneracy in the $\mathbb{Z}_2$ topologically ordered phases. In that case, however, we can no longer separate the loop operators into products of left- and right-chiral components. Instead, we construct our projectors from a maximally commuting set of loop operators (for example, $W^{\mone}_{\mathcal{C}_i}$ and $W^{\mm}_{\mathcal{C}_i}$), and use the remaining non-commuting loop operators (i.e. $W^{\me}_{\mathcal{C}_{3-i}}$, or equivalently $W^{\epsilon}_{\mathcal{C}_{3-i}})$ as raising/ lowering operators. 

\section{Technical details of the string-operators of the \texorpdfstring{$\mathbb{Z}_2$}{Z(2)}-phase}\label{app:stringoperatorsz2details}
Here we will give matrix elements associated with the loop operators $W_{\mathcal{C}_i}^{(\alpha,\beta)}$ in the {(anti-)} ferromagnetic $\mathbb{Z}_2$-topological phases (point Z (A) in Fig.~\ref{fig:phase_diag}). Our starting point will be the nine loop operators in the \Isi phase, whose matrix elements as detailed in Sec.~\ref{ssec:strings}. Since long $\msigma$-strings create confined defects in the three-sublattice order, the possible loop operators in this phase are constructed from $W_{\mathcal{C}_i}^{(\alpha,\beta)}$ for $(\alpha, \beta) = (1, \mpsi), (\mpsi, 1), (\mpsi, \mpsi)$, and $(\msigma, \msigma)$. 

As discussed in Sec.~\ref{sec:AIZ}, open $(\mpsi, \mpsi)$ strings are associated with vison-like defects in the dimer model. The corresponding loop operator $W^{(\mpsi, \mpsi)}_{\mathcal{C}_1}$ counts the parity of the number of $\msigma$-edges around the closed curve $\mathcal{C}_1$. This is necessarily even if $\mathcal{C}_1$ is contractible. If $\mathcal{C}_1$ is not contractible, it depends on the parity of the number of non-contractible $\msigma$-loops parallel to ${\mathcal{C}}_2$. However in the $\mathbb{Z}_2$-phases discussed here, the non-contractible $\msigma$-loops are absent and therefore this parity is fixed to be even. Thus the operator $W^{(\mpsi, \mpsi)}_{\mathcal{C}_1}$ coincides in this phase with the identity as stated in Eqs.~(\ref{eq:stringoperatorsz2_1a},\ref{eq:afmops1}).

Since $W_{\mathcal{C}}^{ (1, \mpsi)} = W_{\mathcal{C}}^{ (\mpsi,1)} W_{\mathcal{C}}^{ (\mpsi, \mpsi)}$, this leaves two string operators that are purely topological: $W_{\mathcal{C}_i}^ {(\mpsi, 1)}$, and $W_{\mathcal{C}_i}^ {(\msigma, \msigma)}$. Here we will show the following: 
(1) open $W_{\mathcal{C}_i}^ {(\mpsi, 1)} \equiv W_{\mathcal{C}_i}^ {(\epsilon)}$ strings create fermions; 
(2) $W_{\mathcal{C}_i}^ {(\msigma, \msigma)}$ strings come in two types, which we will call $W_{\mathcal{C}_i}^ {(b_1)}$ and $W_{\mathcal{C}_i}^ {(b_2)}$; 
(3) $W_{\mathcal{C}_i}^ {(b_1)}$ and $W_{\mathcal{C}_i}^ {(b_2)}$ are both bosons, and are mutual semions; and
(4) $W_{\mathcal{C}_i}^ {(\meps)} =W_{\mathcal{C}_i}^ {(b_1)}W_{\mathcal{C}_i}^ {(b_2)} $. In other words, we will identify a set of purely topological string operators creating exactly the quasi-particle spectrum (for open strings) and ground-state degeneracy (for closed non-contractible strings) of the Toric code. 

\subsection{Matrix elements of string operators away from their endpoints}
We begin by giving the details of the relevant string operators, valid everywhere except near the string endpoints, which we will discuss separately in the next subsection. In general, the curve $\mathcal{C}$ will contain both inside corners (where the string does not cross over any edges, as depicted in Eqs.~(\ref{eq:fsymbpic1}-\ref{eq:fsymbpic3})) and outside corners (where it does, as depicted in Eqs.~(\ref{eq:PsiString1}-\ref{eq:PsiString3})). At inside corners, there are only two choices: either the operator raises the edges along the string's path by $\mpsi$, in which case the coefficients are given in Eq.~(\ref{eq:fsymbols}), or it acts as the identity (with coefficient $1$). 

At outside corners the action is more involved, as it requires both $F$- and $R$-symbols, as described in Sec.~\ref{ssec:strings}. For $\mpsi$-strings, the matrix elements relevant to the frustrated phase are:
\begin{align} 
\begin{array}{c}\includegraphics[width=0.75cm]{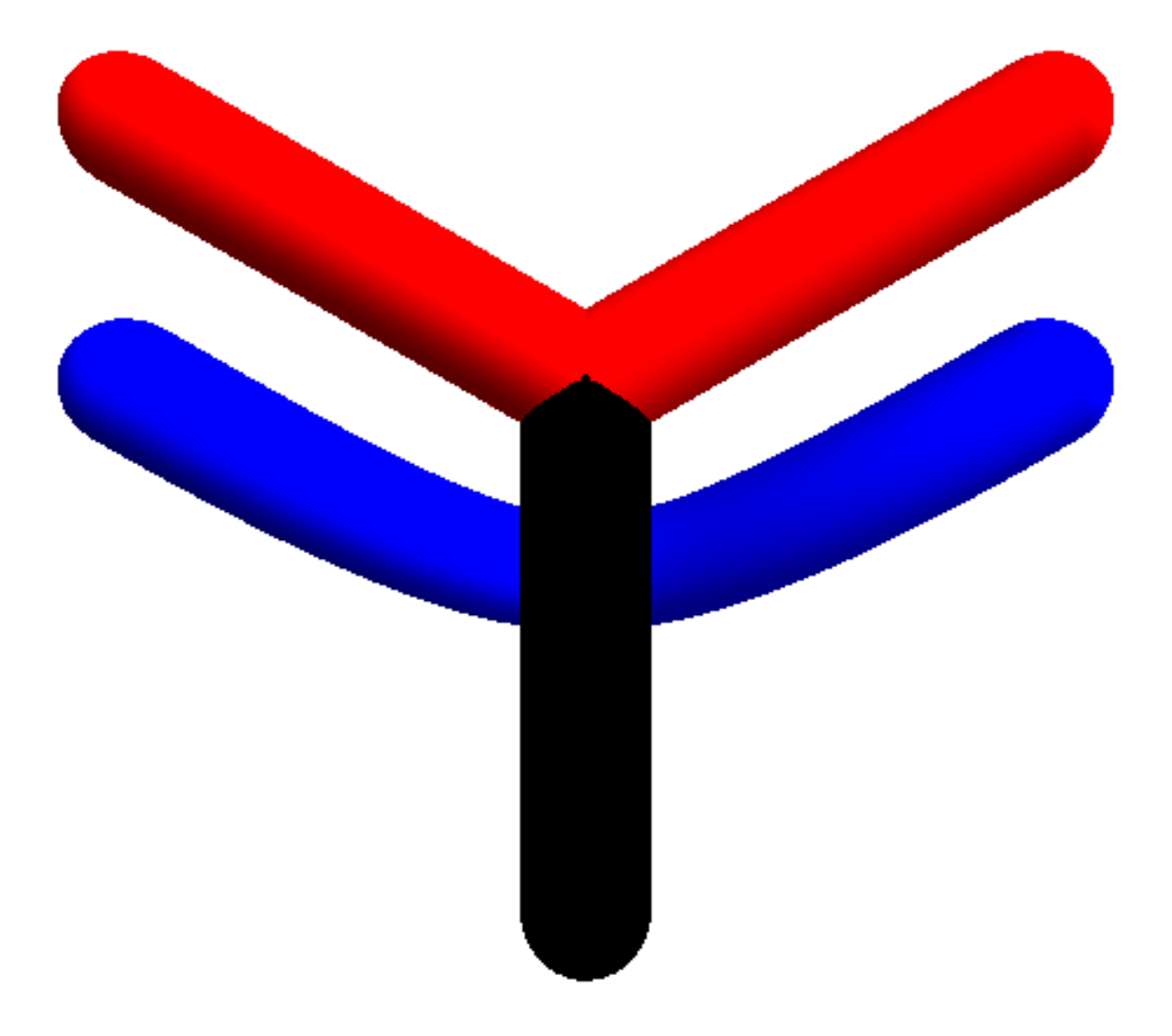}\end{array} = \begin{array}{c}\includegraphics[width=0.75cm]{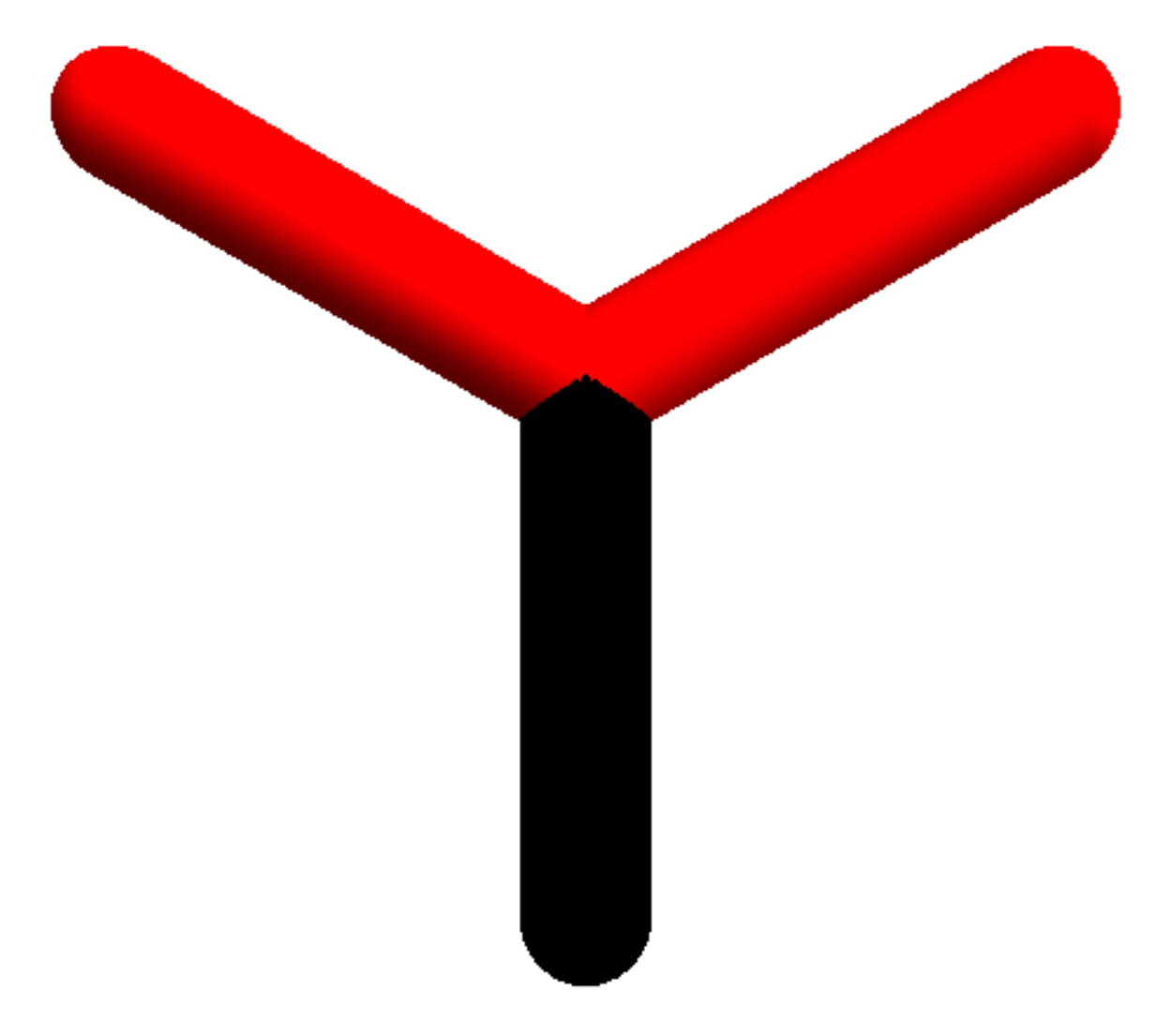}\end{array}, &\ \ \begin{array}{c}\includegraphics[width=0.75cm]{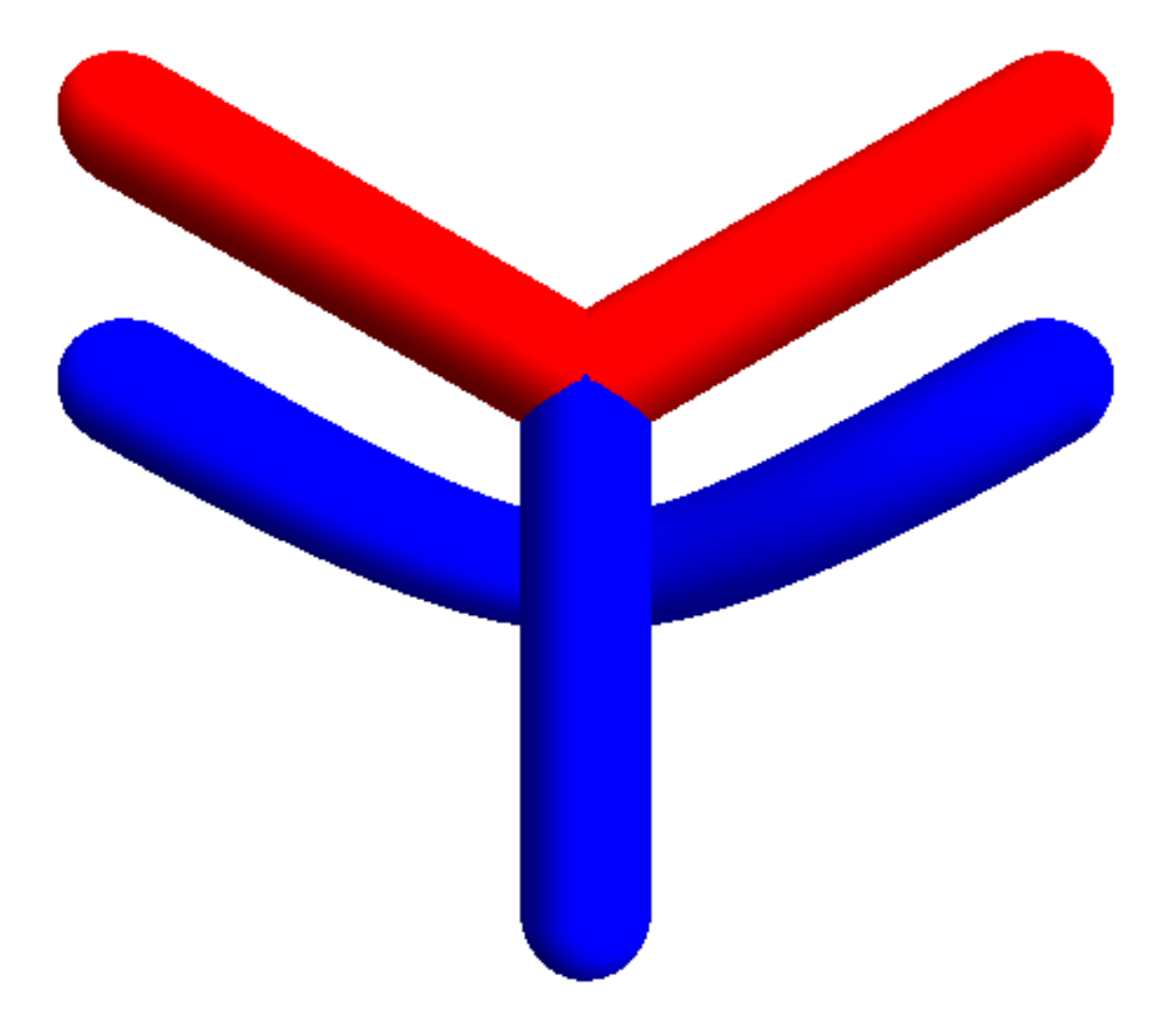}\end{array} = - \begin{array}{c}\includegraphics[width=0.75cm]{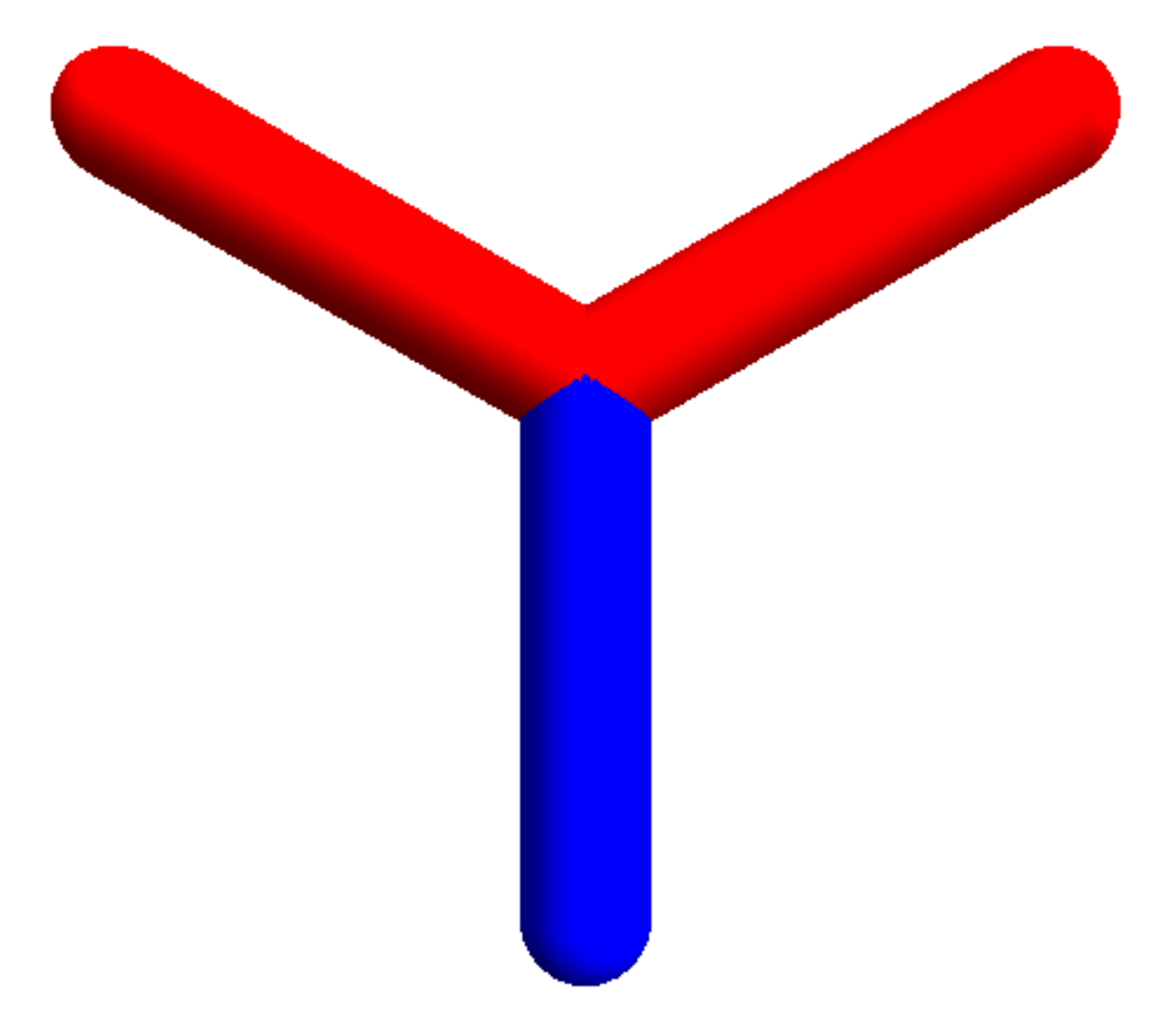}\end{array}, \label{eq:PsiString1}\\
\begin{array}{c}\includegraphics[width=0.75cm]{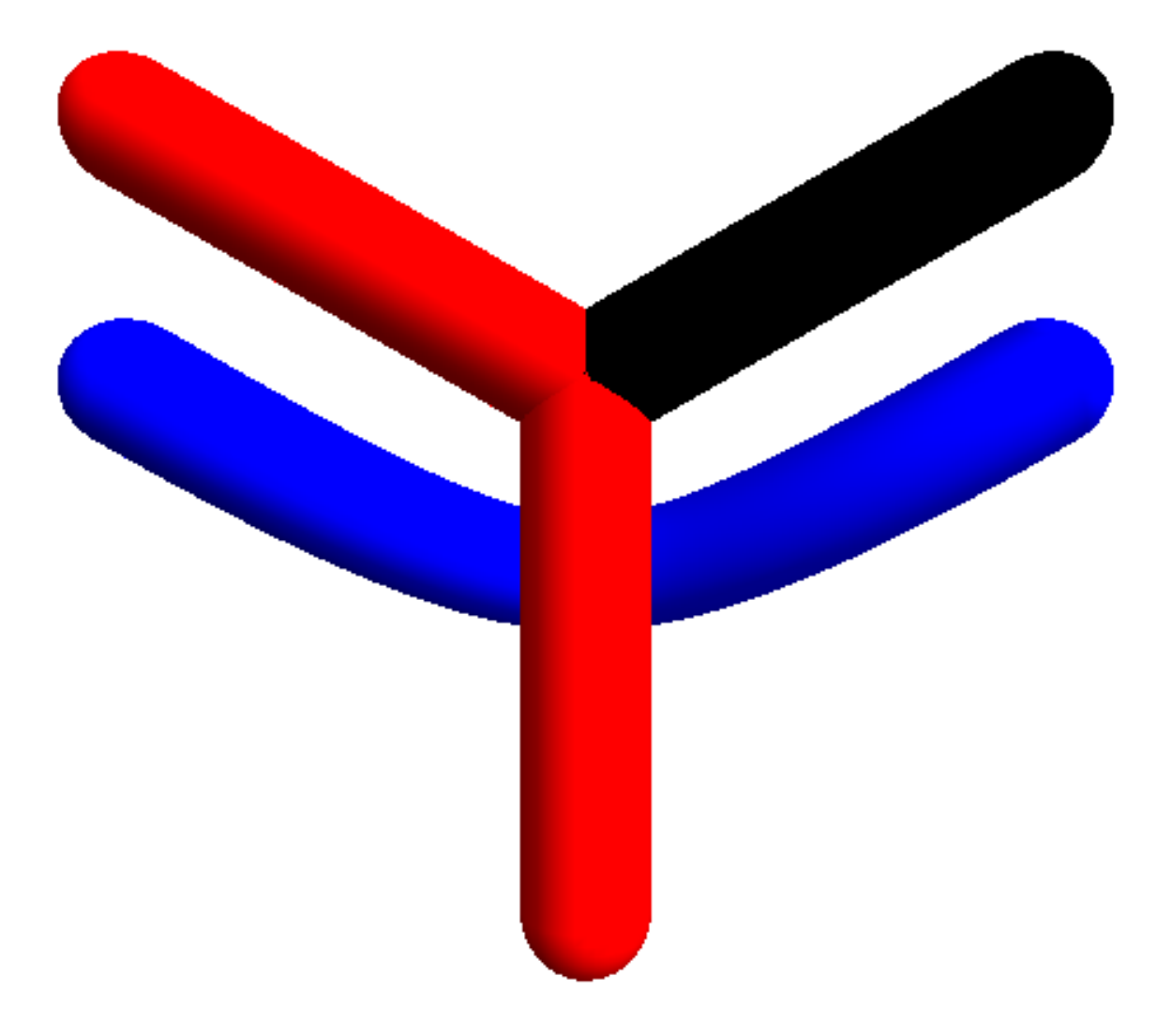}\end{array} = R^{\msigma,\mpsi}_{\msigma} \begin{array}{c}\includegraphics[width=0.75cm]{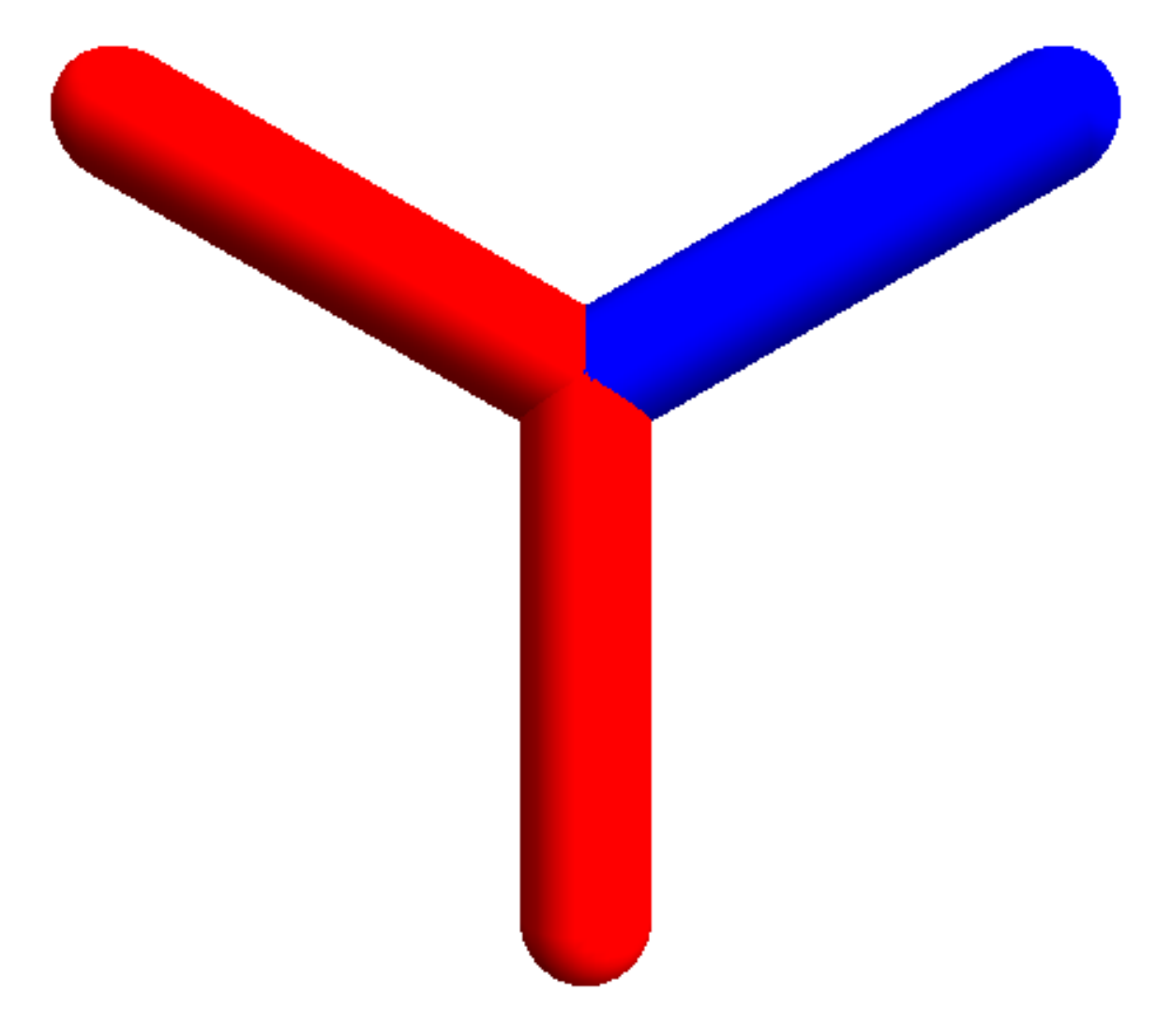}\end{array}, & \ \ \begin{array}{c}\includegraphics[width=0.75cm]{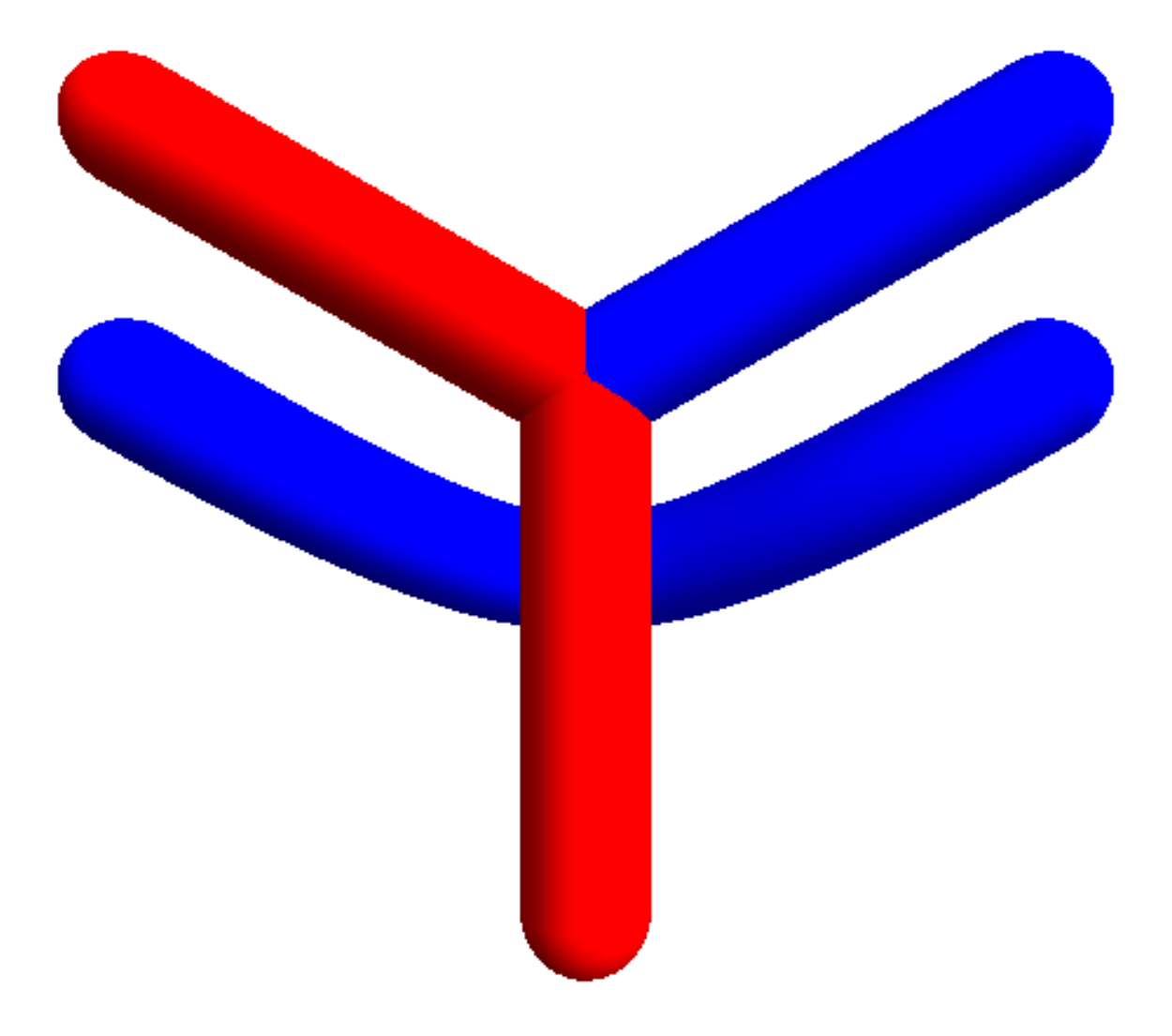}\end{array} =- R^{\msigma,\mpsi}_{\msigma} \begin{array}{c}\includegraphics[width=0.75cm]{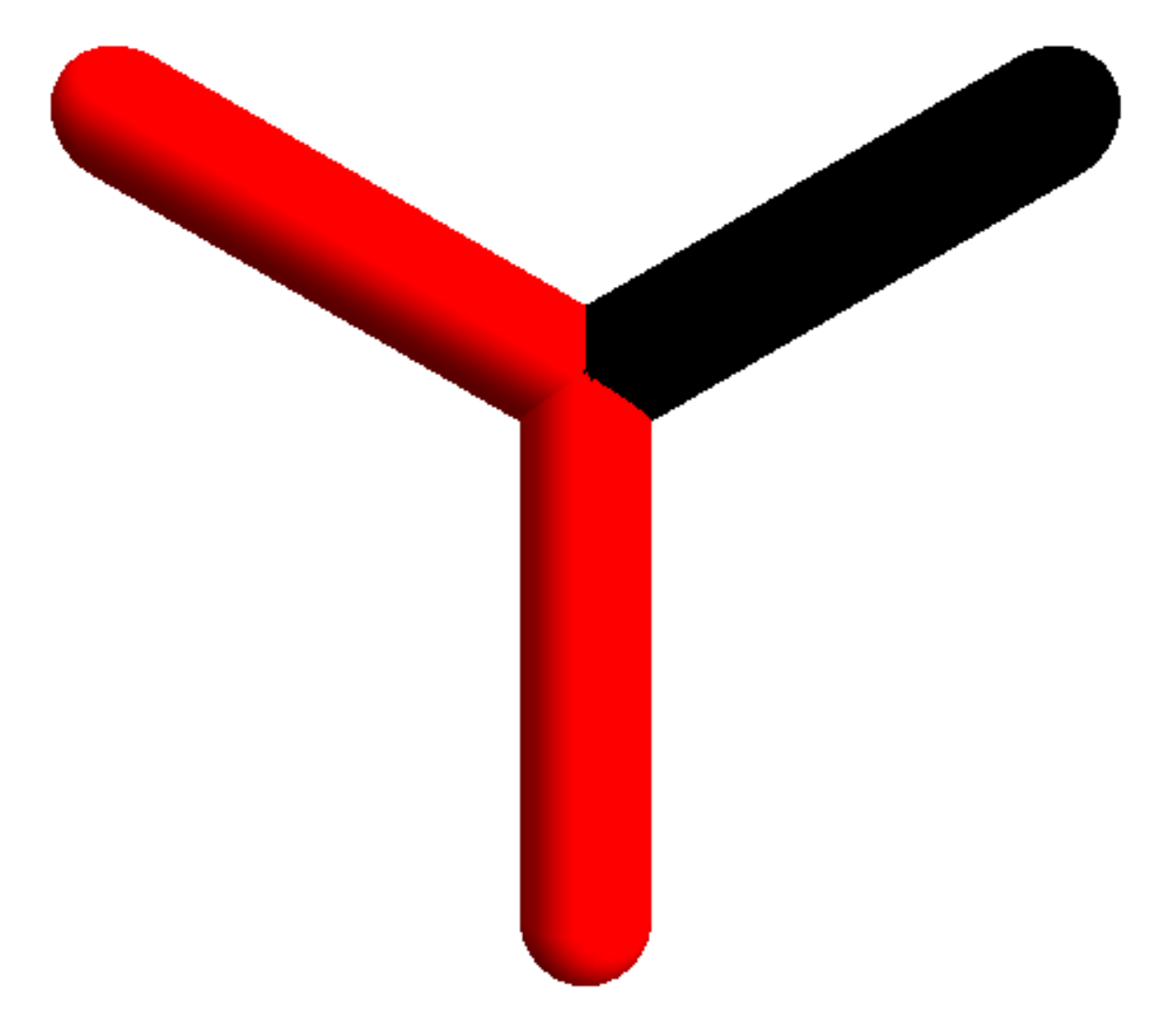}\end{array}, \label{eq:PsiString2}\\
 \begin{array}{c}\includegraphics[width=0.75cm]{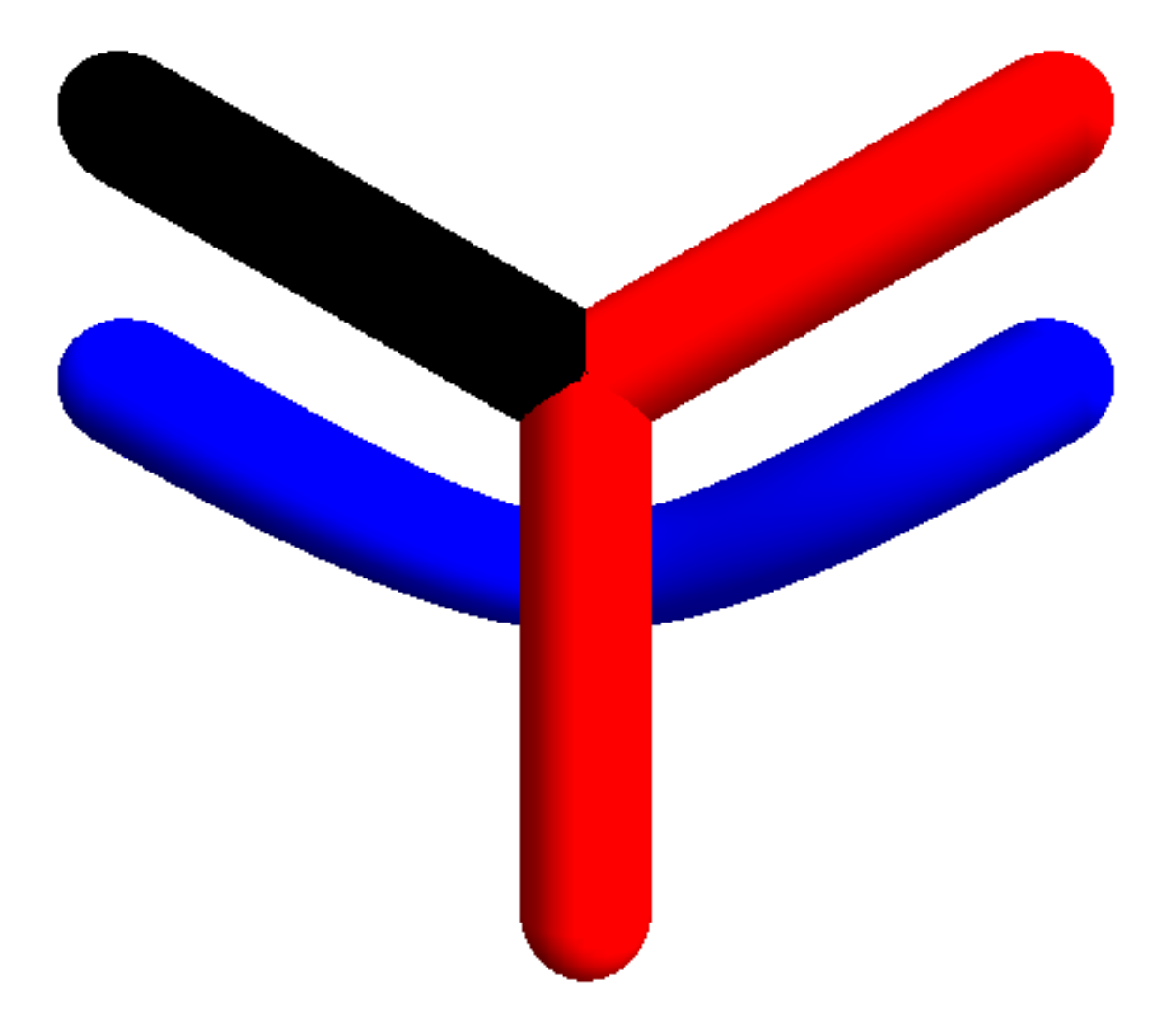}\end{array} = -R^{\msigma,\mpsi}_{\msigma} \begin{array}{c}\includegraphics[width=0.75cm]{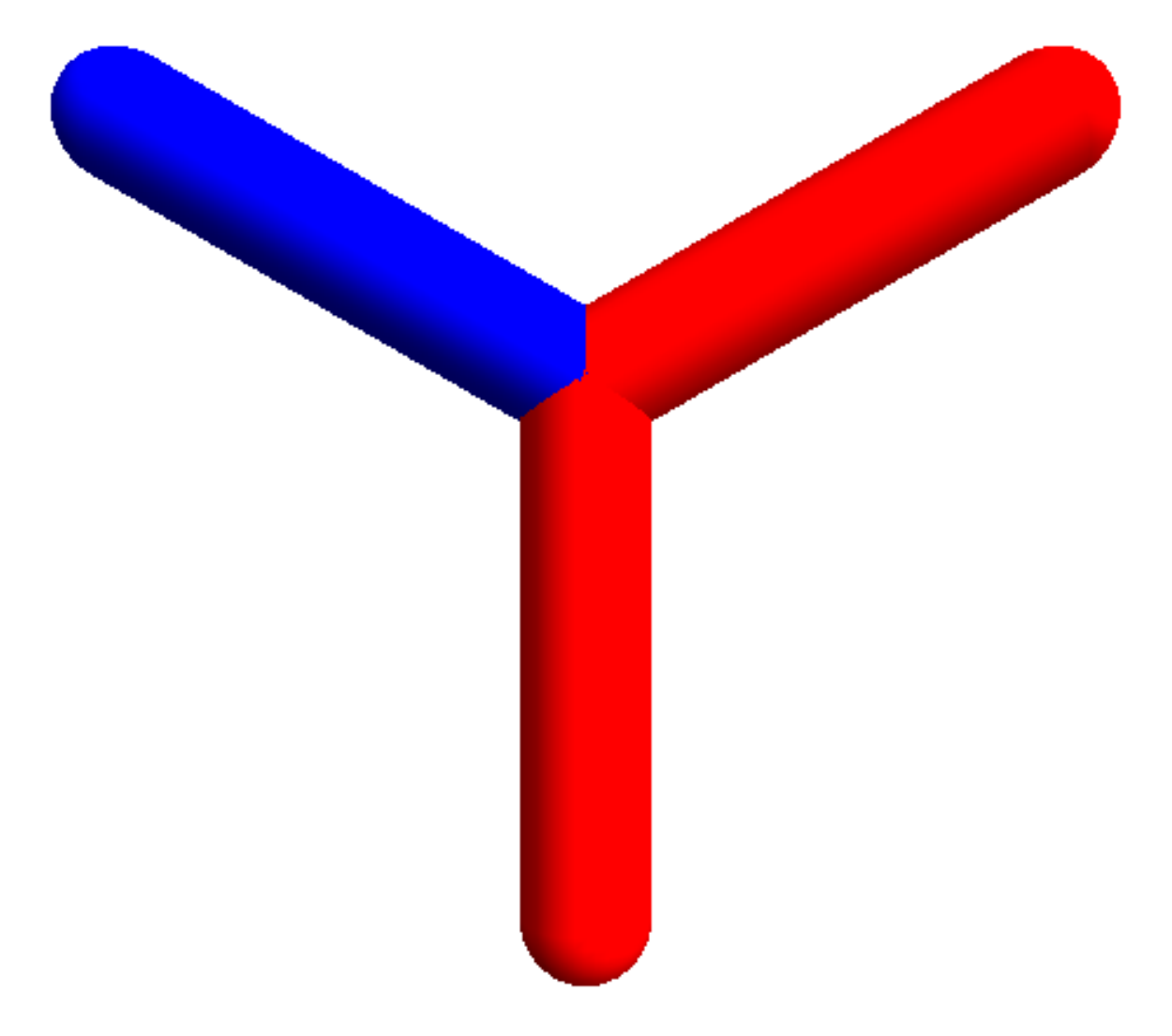}\end{array}, 
& \ \ \begin{array}{c}\includegraphics[width=0.75cm]{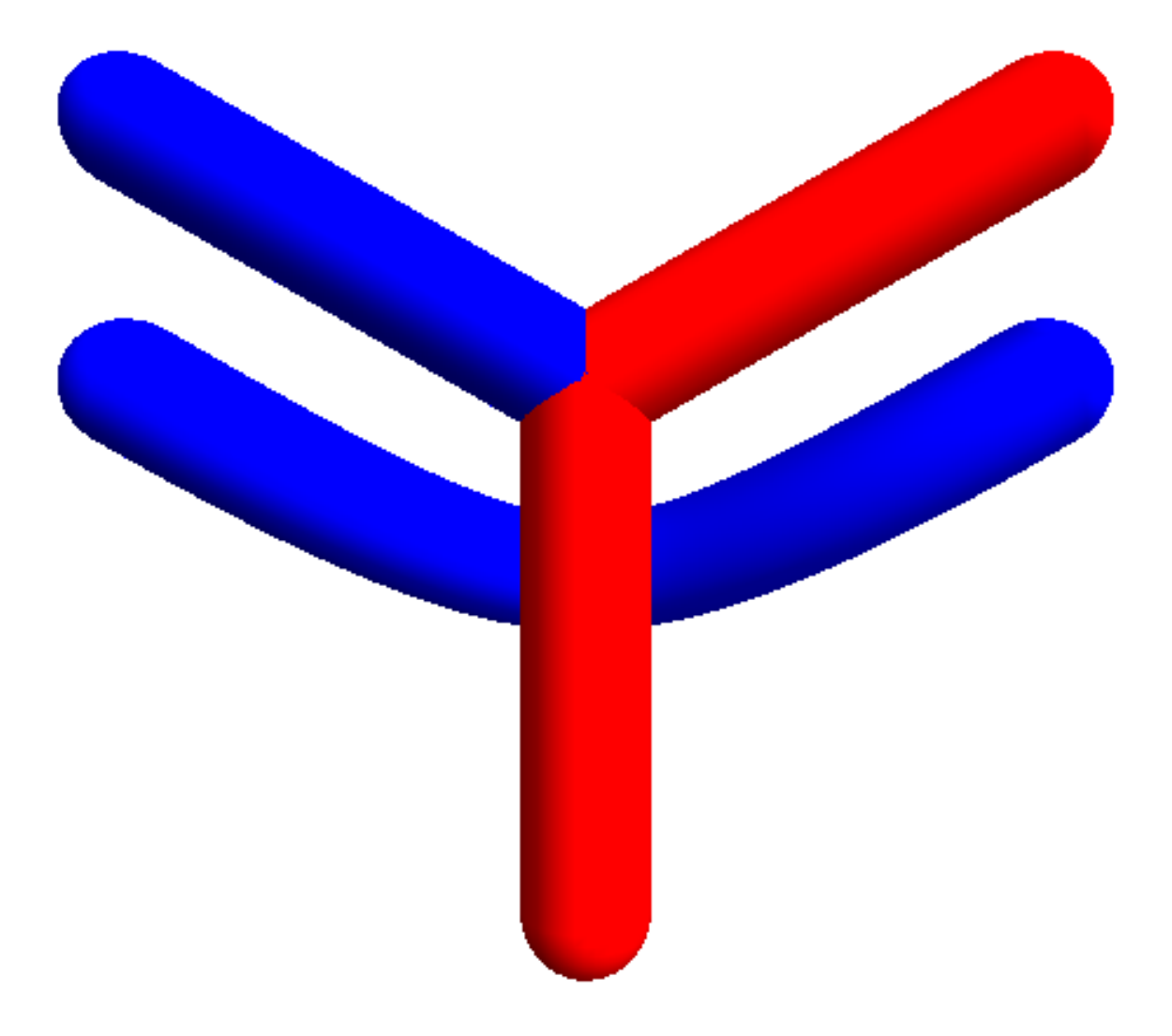}\end{array} =R^{\msigma,\mpsi}_{\msigma} \begin{array}{c}\includegraphics[width=0.75cm]{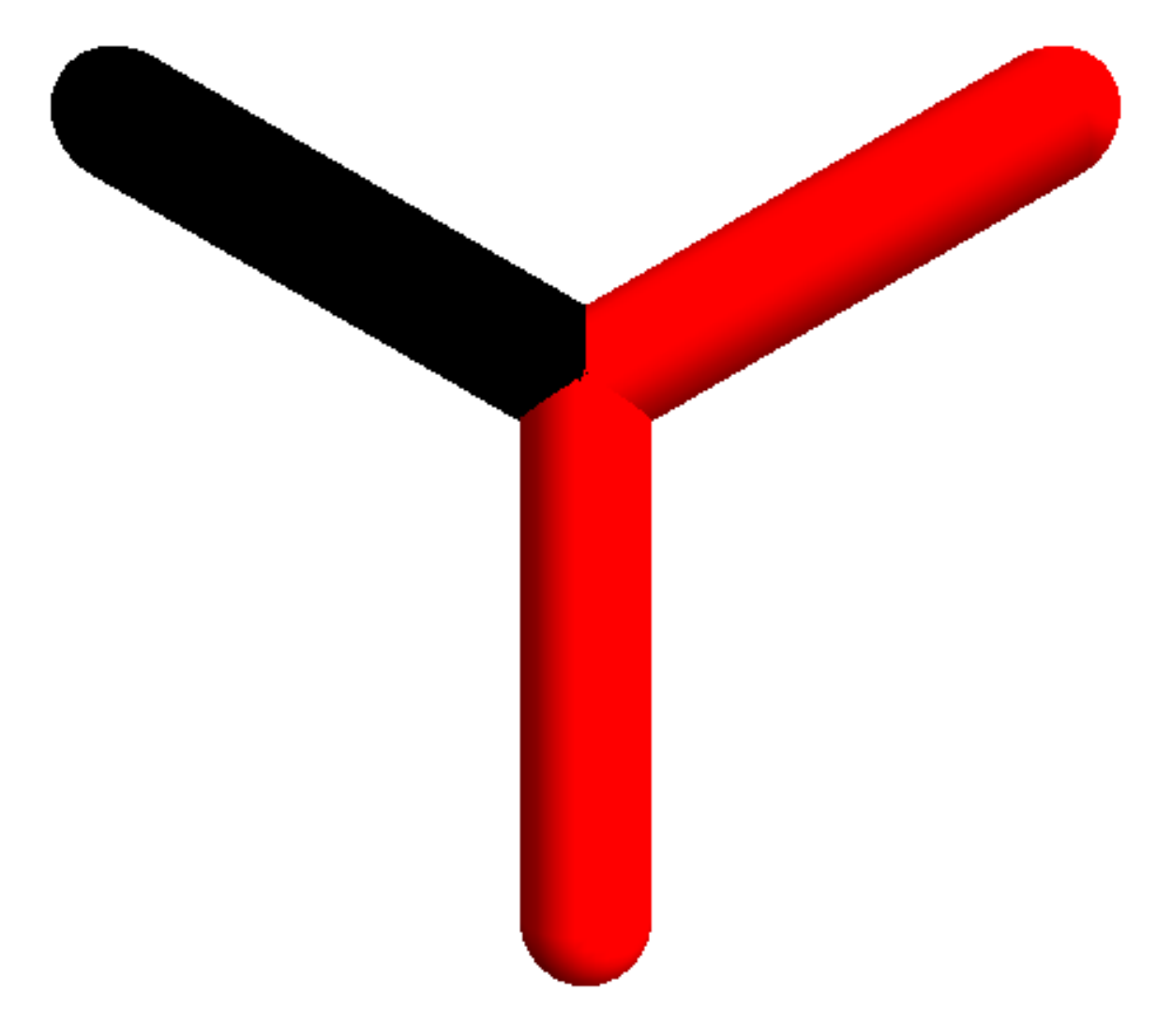}\end{array}, \label{eq:PsiString3}
\end{align}
where $R^{\msigma,\mpsi}_{\msigma}= -\mathrm{i}$ (\ref{eq:rsymbols}), and we have used the fact that the relevant $\Theta$-symbols are all $1$. 

For closed $\mpsi$-strings, however, an equal number of $R^{\msigma,\mpsi}_{\msigma}$ and $-R^{\msigma,\mpsi}_{\msigma}$ phases occur, such that they cancel and can be dropped entirely without altering the action of the string operator (one can check that these phases do not alter the commutation relations with the other operators). 

In the dimer Hilbert space for $-\Jep \gg J_p$, the net effect of the $\mpsi$-string can be compactly represented as follows. In the basis $(1, \msigma, \mpsi)$, we define the three matrices:
\begin{align}
	\Sigma_X = \begin{pmatrix} 0 & 0 & 1 \\ 0 & 1 & 0 \\ 1 & 0 & 0 \\ \end{pmatrix}\!\!,\ 
	\Sigma_Y = \begin{pmatrix} 0 & 0 & -\mathrm{i} \\ 0 & 1 & 0 \\ \mathrm{i} & 0 & 0 \\ \end{pmatrix}\!\!,\ 
	\Sigma_Z = \begin{pmatrix} 1 & 0 & 0 \\ 0 & 1 & 0 \\ 0 & 0 & -1 \\ \end{pmatrix}. 
	\label{eq:raisingops}
\end{align}
Then, in our projected Hilbert space, the string operator takes the form
\begin{align}
	W_{\mathcal{C}_i}^ {(\meps)} =\prod_{i \in \text{crossed edges} } \Sigma_{Z,i} \prod_{j \in \text{raised edges}} M_j
\end{align}
where $M_j$ depends on how the string turns (relative to its starting point) at the two vertices adjacent to the edge $j$, via
\begin{align}
 M_j =\begin{cases} \Sigma_{X,j} & \mathcal{C} \text{ turns left, left or right, right} \\
 \Sigma_{Y,j} & \mathcal{C} \text{ turns right, left} \\
 \Sigma_{Y,j}^* & \mathcal{C} \text{ turns left, right}\\
\end{cases},
\end{align}
i.e.~$\Sigma_X$ acts on edges neighboring inside corners and $\Sigma_Y$ acts on edges neighboring one outside corner.

To understand how $\bar{P}_pW_{\mathcal{C}}^{(\msigma,\msigma)}\bar{P}_p$ acts on outside corners, we observe that for $\alpha = \beta = \msigma$ in Eq.~(\ref{eq:loopstring}), the labels $X$ and $Y$ can each be either $\mone$ or $\mpsi$. If the crossed edge (labeled $i$ in Eq.~(\ref{eq:loopstring})) is labeled $\mone$ or $\mpsi$, the coefficients in Eq.~(\ref{eq:loopstring2}) vanish unless $X = Y$, giving the following possibilities:\footnote{Readers may notice that we have omitted certain factors of $1/\sqrt{2}$ in our definition of the string operators, relative to what is natural from the Ising CFT.  This is because these exactly cancel with a factor of $\sqrt{2}$ which arises when evaluating the matrix element, which we have also omitted here.}
\begin{align}
\begin{array}{c}\includegraphics[width=1cm]{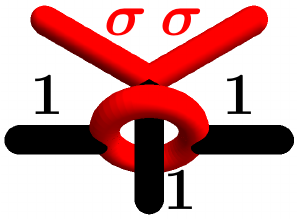}\end{array} = \begin{array}{c}\includegraphics[width=0.75cm]{./figures/openstringvertex32-crop.pdf}\end{array}\ , \ \ &\begin{array}{c}\includegraphics[width=1cm]{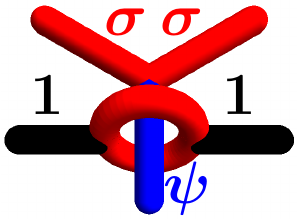}\end{array} = - \begin{array}{c}\includegraphics[width=0.75cm]{./figures/openstringvertex42-crop.pdf}\end{array},\\
\begin{array}{c}\includegraphics[width=1cm]{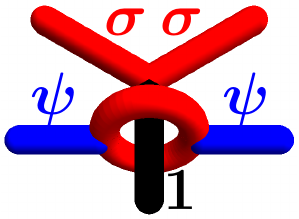}\end{array} = \begin{array}{c}\includegraphics[width=0.75cm]{./figures/openstringvertex32-crop.pdf}\end{array}\ , \ \ &\begin{array}{c}\includegraphics[width=1cm]{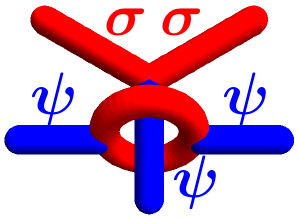}\end{array} = \begin{array}{c}\includegraphics[width=0.75cm]{./figures/openstringvertex42-crop.pdf}\end{array}.
\end{align}
On inside corners, the first operator acts as the identity, while the second operator acts by raising by a $\mpsi$-string. 

If the crossed edge is labeled $\msigma$, however, we have either $X=\mone$, $Y=\mpsi$ or $X=\mpsi$, $Y=\mone$. For the second choice, the relevant matrix elements are:
\begin{align}
\begin{array}{c}\includegraphics[width=1cm]{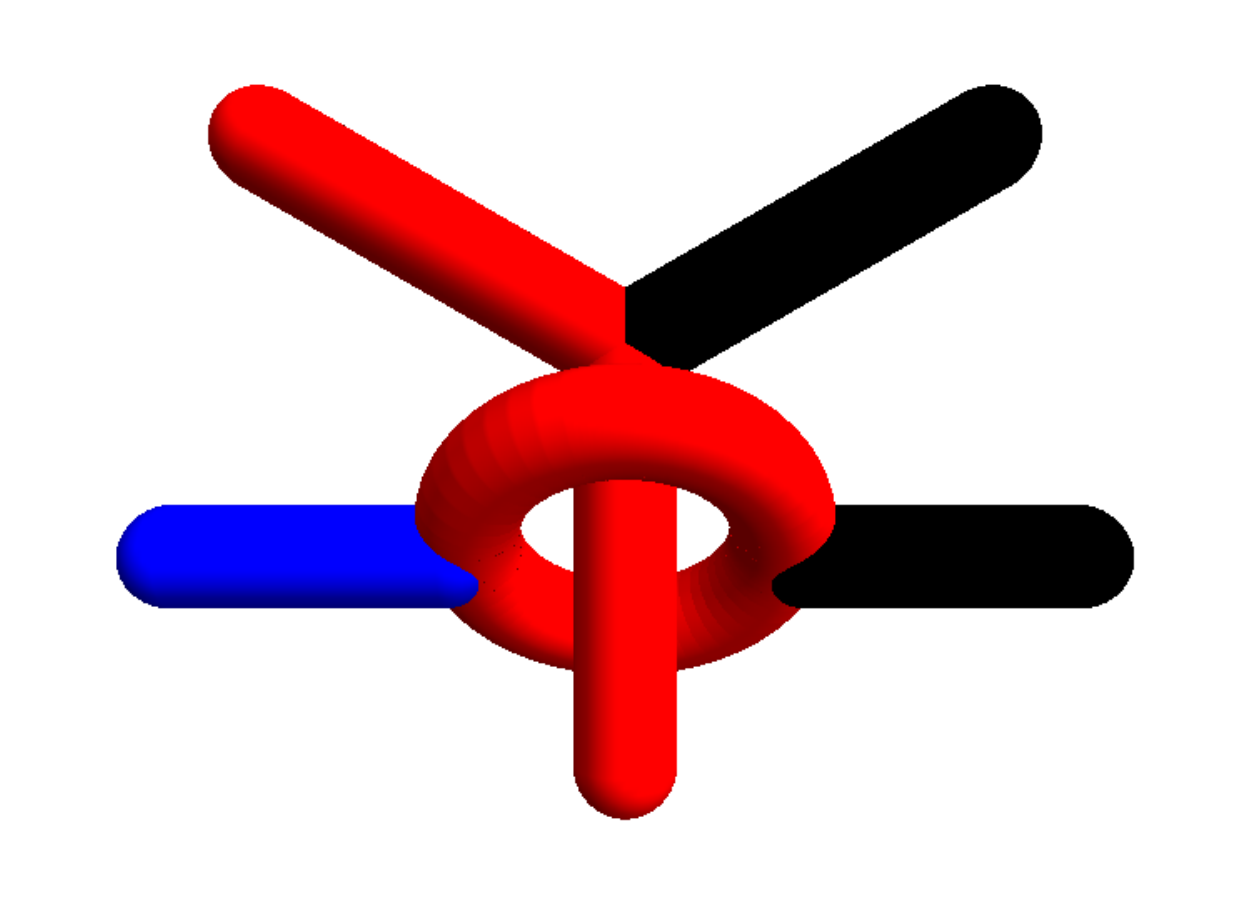}\end{array}\! = (R^{\msigma,\msigma}_{\mone})^{-2}\begin{array}{c}\includegraphics[width=0.75cm]{./figures/openstringvertex22-crop.pdf}\end{array}\!\!\!,&\ 
\begin{array}{c}\includegraphics[width=1cm]{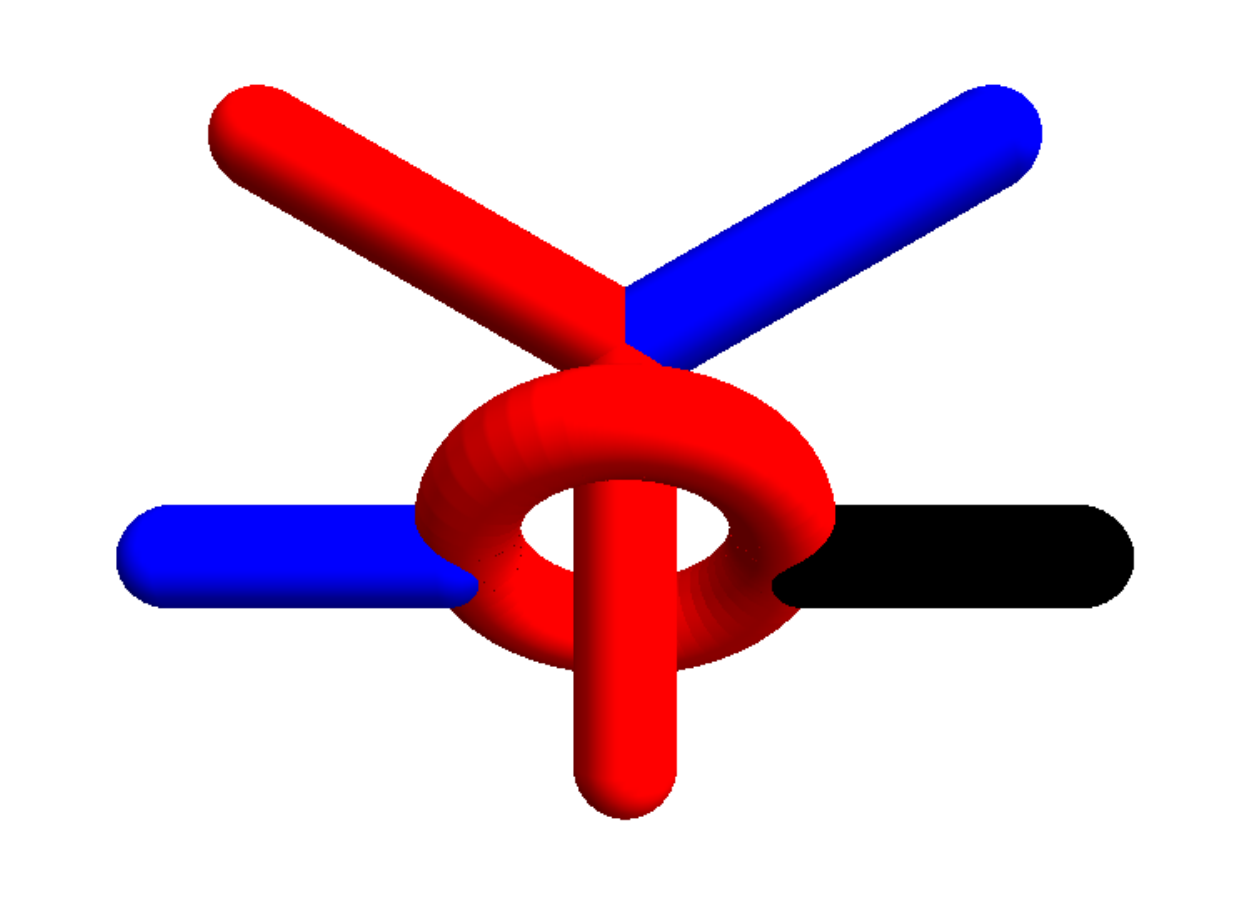}\end{array}\! =- (R^{\msigma,\msigma}_{\mone} )^{-2}\begin{array}{c}\includegraphics[width=0.75cm]{./figures/openstringvertex12-crop.pdf}\end{array}\!\!\!,\\
\begin{array}{c}\includegraphics[width=1cm]{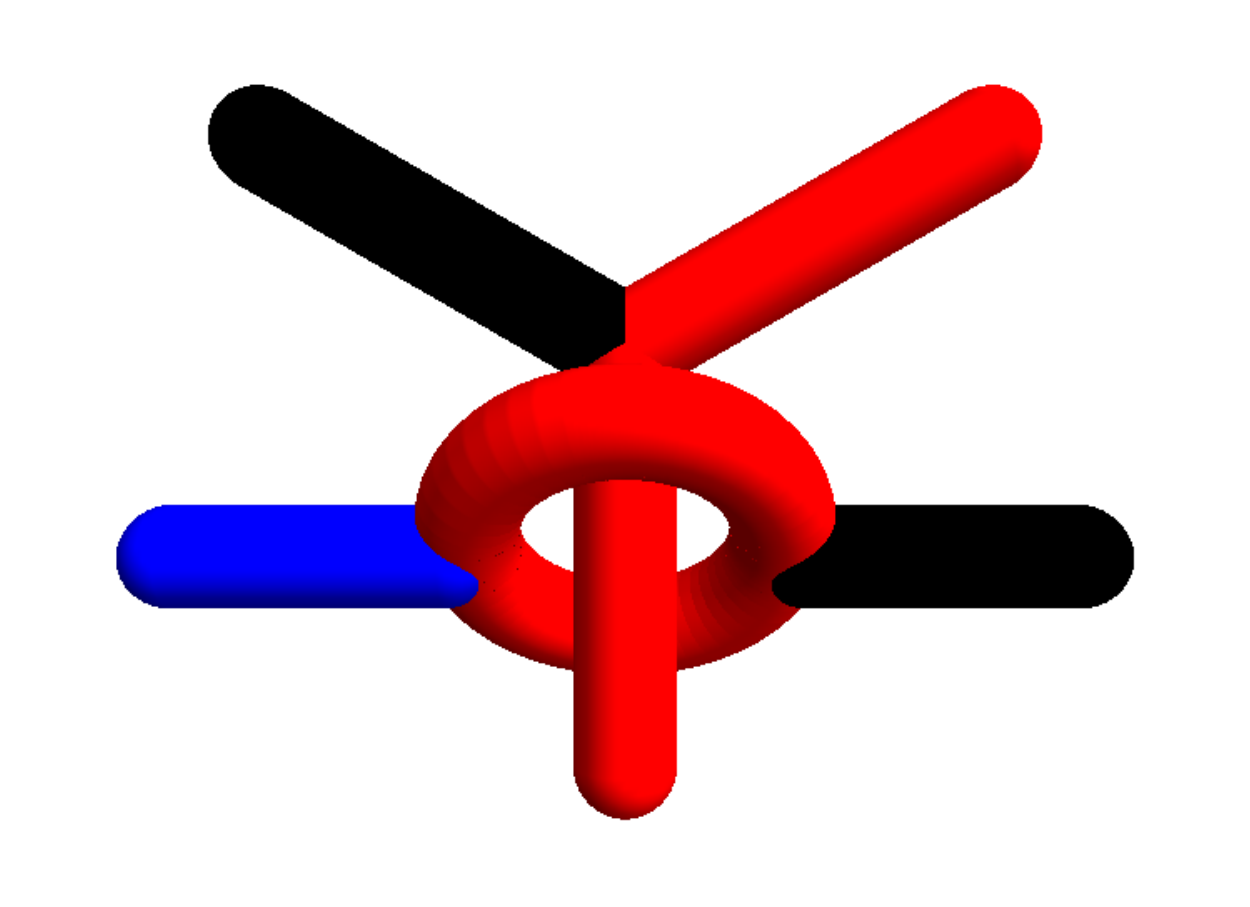}\end{array}\! = (R^{\msigma,\msigma}_{\mone})^{-2} \begin{array}{c}\includegraphics[width=0.75cm]{./figures/openstringvertex62-crop.pdf}\end{array}\!\!\!,&\ 
\begin{array}{c}\includegraphics[width=1cm]{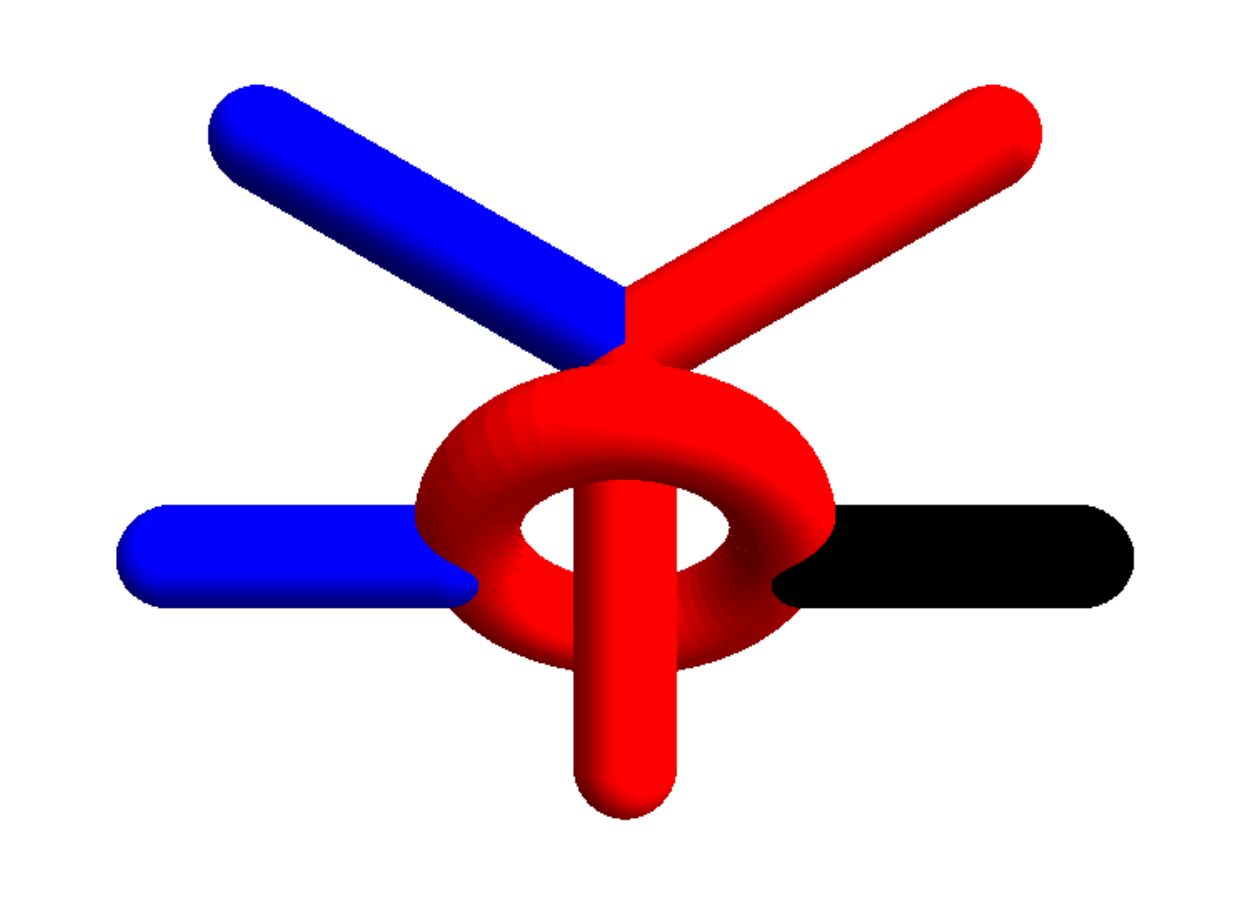}\end{array}\! = \phantom{-}(R^{\msigma,\msigma}_{\mone})^{-2} \begin{array}{c}\includegraphics[width=0.75cm]{./figures/openstringvertex52-crop.pdf}\end{array}\!\!\!.
\end{align}
where $R^{\msigma,\msigma}_{\mone} = e^{ - i \pi/8}$.  The matrix elements for the other choice are similar, with $R^{\msigma,\msigma}_{\mone} \leftrightarrow (R^{\msigma,\msigma}_{\mone})^*$. 

We may now define two distinct closed string operator types from $W_{\mathcal{C}}^{(\msigma, \msigma)}$ as follows. First, for any low- energy edge configuration we may partition the lattice into ``white" and ``black" regions by picking an initial plaquette colored white, with $\msigma$-loops forming domain walls between black and white regions (Fig.~3 of Ref.~\onlinecite{burnell11b}). We then let $W_{\mathcal{C}}^{(b_1)}$ be the component of $\bar{P}_pW_{\mathcal{C}}^{(\msigma, \msigma)}\bar{P}_p$ for which $X = Y = \mpsi$ in the white region, and $X= Y = \mone$ in the black region. $W_{\mathcal{C}}^{(b_2)}$ is defined analogously, with black and white reversed. 

Thus (when acting on a fixed edge configuration) the operator $\bar{P}_pW^{(\msigma,\msigma)}_{\mathcal{C}_i}\bar{P}_p$ splits into two distinct string operators, which we labeled $W^{\me},W^{\mm}$ and $W^{b_1},W^{b_2}$, respectively, in Eqs.~(\ref{eq:stringoperatorsz2_2},\ref{eq:afmops3}). 
Further, the product of these two string types gives the $\mpsi$-string, as can be checked by comparing the relevant matrix elements. (For closed strings, recall that we can drop the phases $R^{\msigma,\mpsi}_{\msigma}$ from $W_{\mathcal{C}_i}^{\meps}$).

One might worry that resonance moves, which change the configuration of $\msigma$-edges, will mix these two candidate string types as they reconfigure the black and white regions.
However, since the closed string operators commute with $\bar{B}_p^{\msigma}$, these resonance moves cannot alter their commutation relations, which fix the topological order; hence the topological properties of the string operators are independent of the configuration acted upon. 
One can easily check that in any reference configuration, for $\mathcal{C}_1, \mathcal{C}_2$ being the two non-contractible curves on the torus, 
\begin{align}
W_{\mathcal{C}_1,\rm{(A)FM}}^{\alpha}W_{\mathcal{C}_2,\rm{(A)FM}}^{\beta}=e^{{\rm i}\phi_{\alpha, \beta}}W_{\mathcal{C}_2,\rm{(A)FM}}^{\beta}W_{\mathcal{C}_1,\rm{(A)FM}}^{\alpha},
\label{eq:z2comm}
\end{align}
where the phase $\phi_{\alpha, \beta}$ is given by
\begin{align}
\begin{array}{c | c c c c}
\phi{\ \rm for \ } \beta\backslash\alpha & b_1 & b_2 & \meps \\ \hline
b_1 & 0 & \pi & \pi \\
b_2 & \pi & 0 & \pi \\
{\meps}& \pi & \pi & 0
\end{array}.
\label{eq:commuphase}
\end{align}
In other words, the resulting closed string operators correspond exactly to those of the toric code: $W_{\mathcal{C}_i}^{(b_1)}$, $W_{\mathcal{C}_i}^{(b_2)}$ to the two bosons (usually called $\me$ and $\mm$) and $W_{\mathcal{C}}^{\meps}$ to the fermion\cite{kitaev03}. \footnote{As noted above, to verify that $W^{b_1}\times W^{b_2} =W^{\meps}$ when acting on states with no non-contractible $\msigma$ loops, we utilize the fact that in this case the factors of $\pm R^{\msigma,\mpsi}_{\msigma}$ cancel, and can be dropped.}

\subsection{Open string operators}
Unlike in most of the phase diagram, in the frustrated $\mathbb{Z}_2$-phase the vertex constraint does not limit the potential quasi-particle types, and all three of the closed string operators discussed above can also exist as open strings with anyonic excitations at their end-points. These open strings differ from the closed strings only at their end-points, which we describe here.

\subsubsection{The fermionic open string}
We begin with the $\mpsi$-string. The action of the bulk string operator dictates everything except the coefficient induced at the first $(v_i)$ and last ($v_f)$ vertices along the string. We will choose these coefficients to be $1$ for all vertex configurations for the purpose of this presentation.

\begin{figure}[htp]%
\includegraphics[width=.2\columnwidth]{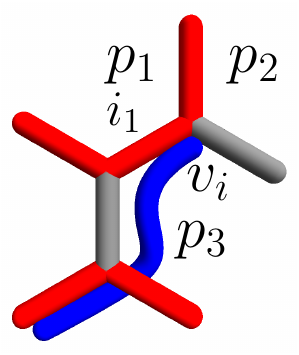}%
\caption{Open $\mpsi$-(or $\meps$-) string ending at vertex $v_i$ creates two fluxes on adjacent plaquettes (in the given configuration at plaquette $p_1$ and $p_3$). }%
\label{fig:openpsi}%
\end{figure}

With this choice, the $\mpsi$-string operator creates a $\msigma$-flux defect (i.e. anti-commutes with $B_p^{\mpsi}$) in two of the three plaquettes adjoining $v_{i,f}$. However, which two plaquettes are violated depends on the configuration that the string acts on. This is illustrated in Fig.~\ref{fig:openpsi}: since the $\mpsi$-string crosses only one edge of $p_3$, the string always anti-commutes with $B_{p_3}^{\mpsi}$, as can be verified from the relations (\ref{eq:PsiString1}-\ref{eq:PsiString3}). Further at $v_i$, we see that if edge $i_1$ (the first edge that is raised by the $\mpsi$-string) is labeled $\mone$ or $\mpsi$, then the string anti-commutes with $B_{p_2}^{\mpsi}$, while if $i_1$ is labeled $\msigma$ then the string anti-commutes with $B_{p_1}^{\mpsi}$. This follows from Eq.~(\ref{eq:fsymbpic1}). 

In Sec.~\ref{ssec:afmz2}, we showed that plaquettes where $\bar{B}_p^{\mpsi}$ has an eigenvalue of $-1$ cannot resonate. For the plaquettes at the end-points of $W^{(\meps)}$, this implies that resonance terms cannot mix configurations in which $\msigma$-fluxes are on different plaquettes. On the remaining plaquette at the string's endpoint, the net phase induced during resonance is clearly unaffected by the string operator, meaning that there are also no vison-like defects bound to $\meps$. 

\subsubsection{The bosonic open strings}
Next, we turn to the bosonic strings. To define these, we first pick a plaquette $p_1$ on which the string starts running to the right, and designate the interior of this plaquette as ``white". We then have:
\begin{align}
W^{(b_1)}\begin{array}{c}\includegraphics[width=.75cm]{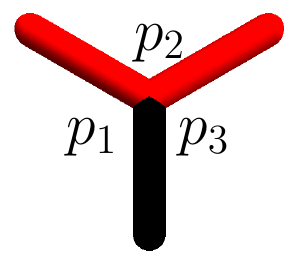}\end{array}\! &= \phantom{-}\begin{array}{c}\includegraphics[width=0.75cm]{./figures/openstringvertex32_plables-crop.pdf}\end{array}\!\!\!,\ \\
W^{(b_1)}\begin{array}{c}\includegraphics[width=.75cm]{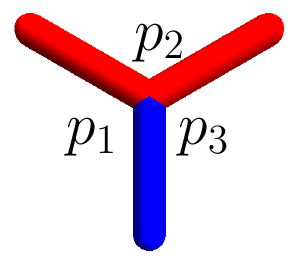}\end{array}\! &= - \begin{array}{c}\includegraphics[width=0.75cm]{./figures/openstringvertex42_plables-crop.pdf}\end{array}\!\!\!,\ \\
W^{(b_1)}\begin{array}{c}\includegraphics[width=.75cm]{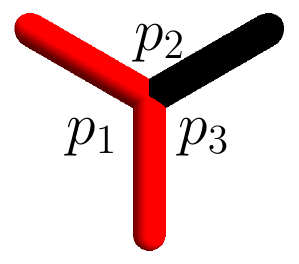}\end{array}\! &= \phantom{-}(R^{\msigma,\msigma}_{\mone})^{-2}\begin{array}{c}\includegraphics[width=0.75cm]{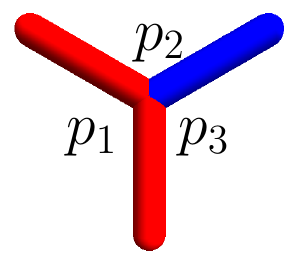}\end{array}\!\!\!,\ \\
W^{(b_1)}\begin{array}{c}\includegraphics[width=.75cm]{./figures/openstringvertex12_plables-crop.pdf}\end{array}\! &= \phantom{-}(R^{\msigma,\msigma}_{\mone})^{-2}\begin{array}{c}\includegraphics[width=0.75cm]{./figures/openstringvertex22_plables-crop.pdf}\end{array}\!\!\!,\ \\
W^{(b_1)}\begin{array}{c}\includegraphics[width=.75cm]{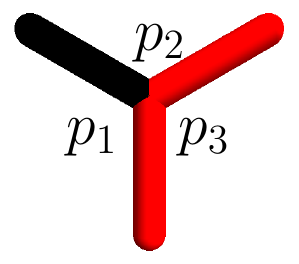}\end{array}\! &= \phantom{-}(R^{\msigma, \msigma}_{\mone})^{-2}\begin{array}{c}\includegraphics[width=0.75cm]{./figures/openstringvertex52_plables-crop.pdf}\end{array}\!\!\!,\\
 W^{(b_1)}\begin{array}{c}\includegraphics[width=.75cm]{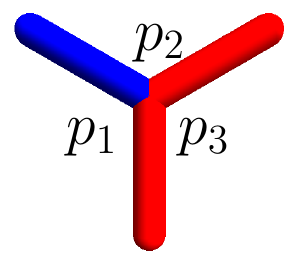}\end{array}\! &= -(R^{\msigma,\msigma}_{\mone})^{-2}\begin{array}{c}\includegraphics[width=0.75cm]{./figures/openstringvertex62_plables-crop.pdf}\end{array}\!\!\!.
\end{align}
It is easy to check that this operator anti-commutes with $B_{p_1}^{\mpsi}$, and commutes with $B_{p_2}^{\mpsi}, B_{p_3}^{\mpsi}$. Hence this operator generates a $\msigma$-flux on $p_1$. 

We may define the second boson via $W^{(b_2)} = W^{(b_1)} W^{\meps}$, where if $W^{(b_1)}$ anti-commutes with $B_{p_1}^{\mpsi}$, then the terminal vertex of the $\meps$ string is chosen such that $W^{\meps}$ also always anti-commutes with $B_{p_1}^{\mpsi}$. This gives: 
\begin{align}
W^{(b_2)}\begin{array}{c}\includegraphics[width=.75cm]{./figures/openstringvertex32_plables-crop.pdf}\end{array}\! &= \phantom{-}\begin{array}{c}\includegraphics[width=0.75cm]{./figures/openstringvertex32_plables-crop.pdf}\end{array}\!\!\!,\ \\
W^{(b_2)}\begin{array}{c}\includegraphics[width=.75cm]{./figures/openstringvertex42_plables-crop.pdf}\end{array}\! &= \phantom{-}\begin{array}{c}\includegraphics[width=0.75cm]{./figures/openstringvertex42_plables-crop.pdf}\end{array}\!\!\!,\ \\
W^{(b_2)}\begin{array}{c}\includegraphics[width=.75cm]{./figures/openstringvertex22_plables-crop.pdf}\end{array}\! &= \phantom{-}R^{\msigma,\mpsi}_{\msigma}(R^{\msigma,\msigma}_{\mone})^{-2}\begin{array}{c}\includegraphics[width=0.75cm]{./figures/openstringvertex22_plables-crop.pdf}\end{array}\!\!\!,\ \\
W^{(b_2)}\begin{array}{c}\includegraphics[width=.75cm]{./figures/openstringvertex12_plables-crop.pdf}\end{array}\! &= -R^{\msigma,\mpsi}_{\msigma}(R^{\msigma,\msigma}_{\mone})^{-2}\begin{array}{c}\includegraphics[width=0.75cm]{./figures/openstringvertex12_plables-crop.pdf}\end{array}\!\!\!,\ \\
W^{(b_2)}\begin{array}{c}\includegraphics[width=.75cm]{./figures/openstringvertex52_plables-crop.pdf}\end{array}\! &= \phantom{-}R^{\msigma,\mpsi}_{\msigma}(R^{\msigma, \msigma}_{\mone})^{-2}\begin{array}{c}\includegraphics[width=0.75cm]{./figures/openstringvertex62_plables-crop.pdf}\end{array}\!\!\!,\\
W^{(b_2)}\begin{array}{c}\includegraphics[width=.75cm]{./figures/openstringvertex62_plables-crop.pdf}\end{array}\! &= \phantom{-}R^{\msigma,\mpsi}_{\msigma}(R^{\msigma,\msigma}_{\mone})^{-2}\begin{array}{c}\includegraphics[width=0.75cm]{./figures/openstringvertex52_plables-crop.pdf}\end{array}\!\!\!.
\end{align}
This operator creates one $\msigma$-flux defect, on either $p_2$ or $p_3$ (whichever is excited by $W^{\meps}$). 
Again, since the violated plaquettes cannot resonate, resonance terms only mix configurations with $\msigma$-fluxes on the same plaquettes. 

Though these matrix elements are different from those relevant to $ W^{(b_1)}$, the defect that is being created is also on a different plaquette.  In fact, one can show that the full string operator $W^{(b_2)}$ is proportional to a string operator $W^{(b_1)}$ that terminates on either $p_2$ or $p_3$, as appropriate.  The form of $W^{\meps}$ ensures that if $ W^{(b_1)}$ excites a plaquette in the ``white" region,  $W^{(b_1)} W^{\meps}$ excites a plaquette in the ``black" region (and vice versa).  In other words, for a given configuration of $\msigma$-labels, we may identify two distinct types of bosons, distinguished by which of the two regions they inhabit.  

This raises two puzzles: first, can the action of the resonance term turn $b_1$ into $b_2$?  The answer is no, because plaquettes bearing these excitations cannot resonate.  Further, the dynamics induced on these excitations for $\Jes\neq 0$ only allow hopping within the same region (the matrix element for hopping across an $\msigma$-edge is $0$).  Second, can a given open string create one $b_1$-type excitation and one $b_2$-type excitation?  Here again, the answer turns out to be no: there is no consistent termination of the string operators which will create defects on plaquettes separated by an odd number of $\msigma$-edges.  Hence $b_1$ and $b_2$ are truly distinct excitations -- as we expect from the fact that $W^{(b_2)} = W^{(b_1)} W^{\meps}$.  

Indeed, it is easy to check that not only are $b_1$ and $b_2$ distinct excitations, but they are mutual semions.  The statistics of the excitations are fixed by the commutation relations of the various string operators, as computed away from the string end-points (see e.g. Ref.~\onlinecite{kitaev03,levin05}). From Eqs.~(\ref{eq:z2comm}, \ref{eq:commuphase}), we see that the end-points of $W^{b_1}$, $W^{b_2}$ are mutually semionic bosons, while $W^{\meps}$ creates fermions, as expected for the toric code.

This division of particle type by region, though cumbersome, is in fact inevitable. The dimension of the Hilbert space associated with $\msigma$-fluxes on $N$ plaquettes can grow at most as $2^N$, whereas we wish to accommodate three distinct excitations, all of which are characterized by their eigenvalue of $-1$ under $B_p^{\mpsi}$.  Since for a given plaquette there is only one linearly independent state with this eigenvalue, to accommodate all three particle types requires excitations on multiple plaquettes, as we have found here.

\section{Correspondence of the \texorpdfstring{$\mathbb{Z}_2$}{Z(2)}- and the original dimer model}\label{app:dimer2dimer}
Here, we show that in the absence of $\msigma$-fluxes (provided $\Jes=0$) the signs appearing in certain resonance terms in Eq.~(\ref{eq:hameffa}) do not invalidate a many-to-one mapping in which the internal dimer labels, as well as the phases $\beta(i,f), \gamma(i,f)$, are all set to $1$. We will see that this holds not only in the dimer limit, but more generally. (Similar arguments were made in Ref.~\onlinecite{chandran13}). Hence the extra signs play no role in determining the ordering, or the critical properties, of the frustrated $\mathbb{Z}_2$ topological phase.

The phases $\beta(i,f), \gamma(i,f)$ appearing in the coefficients of the effective Hamiltonian (\ref{eq:hameffa}) are determined by the matrix elements for the $B_p^{\msigma}$ and $B_p^{\mpsi}$, respectively, which are given in Eq.~(\ref{eq:action_Bps}). For the phases $\beta(i,f)$, one can verify from Eq.~(\ref{eq:action_Bps}) that $\beta(i,f)=\prod_{e\in\circ p}\left(1-2n_{e+1}^{\mpsi}(i)n_e^{\mpsi}(f)\right)$, where $\circ p$ denotes the inner edges of plaquette $p$, $e+1$ is the edge following edge $e$ in a counterclockwise sense and $n_e^{\mpsi}(i)$ ($n_e^{\mpsi}(f)$) denotes the eigenvalue of $n_e^{\mpsi}$ when acting on the initial (final) state $i$ ($f$). 
The phases $\gamma(i,f)$ are given by $\gamma(i,f)=\prod_{e\in *p}\left(1-2n_e^{\mpsi}\right)$, where $*p$ denotes the outgoing edges of plaquette $p$.
These phases thus fix the relative coefficients of the different edge configurations in the ground state(s).

To argue that these phases have no impact on the resulting order, we proceed as follows:
First, we observe that any operator diagonal in the edge labels (and specifically the operator $n_e^{\msigma}$ which we use to tune the model to the dimer limit) cannot detect these relative signs. 

Second, at least within each topological sector, the relative phases are uniquely fixed in the ground state. 
By definition, a ground state $\left|\Psi\right\rangle$ obeys $\left|\Psi\right\rangle\propto\prod\limits_p\frac{1}{2}\left(1+B_p^{\mpsi}\right)\left|\Psi\right\rangle$. 
For any given configuration of $\msigma$-edges, this fixes the relative phases of all configurations in the internal dimer space. Hence the dimer configuration is unique, up to the $\mathbb{Z}_2$ topological degeneracy.

Third, we argue that phases induced by changing the pattern of $\msigma$-loops are compatible with those fixed by $B_p^{\mpsi}$, and therefore within a given topological sector the amplitude for resonance moves connecting any two configurations of $\msigma$-loops is simply a sum over all internal dimer configurations (since the relative phases of the internal dimer configurations are fixed). 

Specifically, if $B_p^{\mpsi} \left|\Psi\right\rangle = \left|\Psi\right\rangle$, then (as $B_p^{\mpsi}$ and $B_p^{\msigma}$ commute), 
\begin{align}
B_p^{\mpsi} ( B_p^{\msigma} \left|\Psi\right\rangle ) = B_p^{\msigma} \left|\Psi\right\rangle.
\end{align}
This remains true whether or not we project our Hilbert space onto that of the dimer model. 

Further, $\left( B_p^{\msigma}\right)^2=\left(\mathds{1}+B_p^{\mpsi}\right)$ (an identity which also holds for the operator $\bar{P}B_p^{\msigma}$, where $\bar{P}$ is the projector into the dimer Hilbert space). Hence by acting an even number of times on any plaquette (or group of plaquettes) by $\bar{P}B_p^{\msigma}\bar{P}$, one obtains an eigenstate of $B_p^{\mpsi}$ with eigenvalue $1$. In particular, the relative signs potentially introduced by the different matrix elements of $\bar{P}B_p^{\msigma}\bar{P}$ can never cause any interference at any finite order; hence we must simply add all transition amplitudes that bring a configuration of $\msigma$-loops back to itself, as we would in the absence of internal dimer states. 

The only other way to act non-trivially onto a state $\left|\Psi\right\rangle$ while maintaining the same location of the dimers is to change the topological sector by the action of the operator $\prod\limits_p\bar{P}B_p^{\msigma}\bar{P}$ (note that the single terms in the product have to be ordered in a certain way to yield a non-vanishing matrix element when acting on a given state). As this operator does involve an action on the whole system, its effect will be suppressed in the thermodynamic limit and can thus be neglected. 

In conclusion, the critical properties and ordering pattern of our dimer model are identical to those of Ref.~\onlinecite{moessner01}.

\section{Series expansions for the anyons in the perturbed Ising string-net (Leading orders)}\label{app:seriesising}
In this section we give the leading orders of series expansion for the Ising string-net phase in terms of $ {J^{\msigma}}=\frac{\Jes}{\Jp}$ and $ {J^{\mpsi}}=\frac{\Jep}{\Jp}$. Higher orders and individual hopping elements are available upon request.

For the ground-state energy per plaquette $e_0$, we get
\begin{align}
\frac{e_0}{\Jp}=&-1+\frac{3\,{{J^{\mpsi}}}}{4}-\frac{3\,{{{J^{\msigma}}}}^{2}}{16}-{\frac {3\,{{{J^{\mpsi}}}}^{2}}{32}}-{\frac {9\,{{{J^{\msigma}}}}^{2}{{J^{\mpsi}}}}{128}}
-{\frac {3\,{{{J^{\msigma}}}}^{3}}{32}}\nonumber\\&
-{\frac {3\,{{{J^{\mpsi}}}}^{3}}{64}}-{\frac {41\,{{{J^{\msigma}}}}^{2}{{{J^{\mpsi}}}}^{2}}{512}}-{\frac {7\,{{{J^{\msigma}}}}^{3}{{J^{\mpsi}}}}{64}}-{\frac {25\,{{{J^{\msigma}}}}^{4}}{256}}\nonumber\\&
-{\frac {87\,{{{J^{\mpsi}}}}^{4}}{2048}}-{\frac {2429\,{{{J^{\msigma}}}}^{2}{{{J^{\mpsi}}}}^{3}}{24576}}-{\frac {65\,{{{J^{\msigma}}}}^{3}{{{J^{\mpsi}}}}^{2}}{512}}\nonumber\\&
-{\frac {1187\,{{{J^{\msigma}}}}^{4}{{J^{\mpsi}}}}{6144}}-{\frac {761\,{{{J^{\msigma}}}}^{5}}{6144}}-{\frac {99\,{{{J^{\mpsi}}}}^{5}}{2048}}.
\label{eq:isinggse}
\end{align}

For positive $J^{\msigma}$, the dispersion of a single non-Abelian anyon $\omega^{\msigma}(\vec{k})$ is minimized for $\vec{k}=\vec{0}$ at leading order and we obtain for the corresponding gap $\Delta_{\msigma}^+$:
\begin{align}
\frac{\Delta_{\msigma}^+}{\Jp}=&1-\frac{3\,{{J^{\msigma}}}}{2}-\frac{3\,{{{J^{\mpsi}}}}^{2}}{16}-\frac{3\,{{J^{\msigma}}}\,{{J^{\mpsi}}}}{4}-{\frac {15\,{{{J^{\msigma}}}}^{2}}{16}}-{\frac {15\,{{{J^{\mpsi}}}}^{3}}{64}}\nonumber\\&
-{\frac {87\,{{J^{\msigma}}}\,{{{J^{\mpsi}}}}^{2}}{128}}-{\frac {9\,{{{J^{\msigma}}}}^{2}{{J^{\mpsi}}}}{8}}-{\frac {99\,{{{J^{\msigma}}}}^{3}}{128}}-{\frac {189\,{{{J^{\mpsi}}}}^{4}}{512}}\nonumber\\&
-{\frac {243\,{{J^{\msigma}}}\,{{{J^{\mpsi}}}}^{3}}{256}}-{\frac {991\,{{{J^{\msigma}}}}^{2}{{{J^{\mpsi}}}}^{2}}{512}}-{\frac {77\,{{{J^{\msigma}}}}^{3}{{J^{\mpsi}}}}{32}}\nonumber\\&
-{\frac {367\,{{{J^{\msigma}}}}^{4}}{256}}-{\frac {159\,{{{J^{\mpsi}}}}^{5}}{256}}-{\frac {5739\,{{J^{\msigma}}}\,{{{J^{\mpsi}}}}^{4}}{4096}}\nonumber\\&
-{\frac {765\,{{{J^{\msigma}}}}^{2}{{{J^{\mpsi}}}}^{3}}{256}}-{\frac {113567\,{{{J^{\msigma}}}}^{3}{{{J^{\mpsi}}}}^{2}}{24576}}\nonumber\\&
-{\frac {18997\,{{{J^{\msigma}}}}^{4}{{J^{\mpsi}}}}{4096}}
-{\frac {55391\,{{{J^{\msigma}}}}^{5}}{24576}}.
\label{eq:deltasigmaplus}
\end{align}

For negative $J^{\msigma}$, the dispersion of a single non-Abelian anyon $\omega^{\msigma}(\vec{k})$ is minimized for $\vec{k}=(\pm \frac{2\pi}{3},\mp \frac{2\pi}{3})$ at leading order and we obtain for the corresponding gap $\Delta_{\msigma}^-$:
\begin{align}
\frac{\Delta_{\msigma}^-}{\Jp}=&1+\frac{3\,{{J^{\msigma}}}}{4}-\frac{3\,{{{J^{\mpsi}}}}^{2}}{16}+\frac{3\,{{J^{\msigma}}}\,{{J^{\mpsi}}}}{8}-{\frac {3\,{{{J^{\msigma}}}}^{2}}{32}}-{\frac {15\,{{{J^{\mpsi}}}}^{3}}{64}}\nonumber\\&
+{\frac {87\,{{J^{\msigma}}}\,{{{J^{\mpsi}}}}^{2}}{256}}+{\frac {63\,{{{J^{\msigma}}}}^{2}{{J^{\mpsi}}}}{128}}-{\frac {45\,{{{J^{\msigma}}}}^{3}}{256}}-{\frac {189\,{{{J^{\mpsi}}}}^{4}}{512}}\nonumber\\&
+{\frac {243\,{{J^{\msigma}}}\,{{{J^{\mpsi}}}}^{3}}{512}}-{\frac {269\,{{{J^{\msigma}}}}^{2}{{{J^{\mpsi}}}}^{2}}{1024}}+{\frac {121\,{{{J^{\msigma}}}}^{3}{{J^{\mpsi}}}}{128}}\nonumber\\&
-{\frac {857\,{{{J^{\msigma}}}}^{4}}{2048}}-{\frac {159\,{{{J^{\mpsi}}}}^{5}}{256}}+{\frac {5739\,{{J^{\msigma}}}\,{{{J^{\mpsi}}}}^{4}}{8192}}\nonumber\\&
-{\frac {2661\,{{{J^{\msigma}}}}^{2}{{{J^{\mpsi}}}}^{3}}{8192}}+{\frac {12845\,{{{J^{\msigma}}}}^{3}{{{J^{\mpsi}}}}^{2}}{49152}}+{\frac {3061\,{{{J^{\msigma}}}}^{4}{{J^{\mpsi}}}}{2048}}\nonumber\\&
-{\frac {40795\,{{{J^{\msigma}}}}^{5}}{49152}}.
\label{eq:deltasigmaminus}
\end{align}

For positive $J^{\mpsi}$, the dispersion of a single hardcore boson $\omega^{\mpsi}(\vec{k})$ is minimized for $\vec{k}=\vec{0}$ at leading order and we obtain for the corresponding gap $\Delta_{\mpsi}^+$:
\begin{align}
\frac{\Delta_{\mpsi}^+}{\Jp}=&1-\frac{3\,{{J^{\mpsi}}}}{2}-\frac{3\,{{{J^{\mpsi}}}}^{2}}{4}-{\frac {9\,{{{J^{\msigma}}}}^{2}}{8}}-{\frac {21\,{{{J^{\mpsi}}}}^{3}}{32}}-{\frac {3\,{{{J^{\msigma}}}}^{2}{{J^{\mpsi}}}}{32}}\nonumber\\&
-{\frac {33\,{{{J^{\msigma}}}}^{3}}{16}}-{\frac {63\,{{{J^{\mpsi}}}}^{4}}{64}}-{\frac {215\,{{{J^{\msigma}}}}^{2}{{{J^{\mpsi}}}}^{2}}{128}}+\frac{\,{{{J^{\msigma}}}}^{3}{{J^{\mpsi}}}}{8}\nonumber\\&
-{\frac {583\,{{{J^{\msigma}}}}^{4}}{128}}-{\frac {3153\,{{{J^{\mpsi}}}}^{5}}{2048}}-{\frac {6127\,{{{J^{\msigma}}}}^{2}{{{J^{\mpsi}}}}^{3}}{4096}}\nonumber\\&
-{\frac {23375\,{{{J^{\msigma}}}}^{3}{{{J^{\mpsi}}}}^{2}}{6144}}+{\frac {585\,{{{J^{\msigma}}}}^{4}{{J^{\mpsi}}}}{2048}}-{\frac {15313\,{{{J^{\msigma}}}}^{5}}{1536}}.
\label{eq:deltapsiplus}
\end{align}

For negative $J^{\mpsi}$, the dispersion of a single hardcore boson $\omega^{\mpsi}(\vec{k})$ is minimized for $\vec{k}=(\pm \frac{2\pi}{3},\mp \frac{2\pi}{3})$ at leading order and we obtain for the corresponding gap $\Delta_{\mpsi}^-$:
\begin{align}
\frac{\Delta_{\mpsi}^-}{\Jp}=&1+\frac{3\,{{J^{\mpsi}}}}{4}+{\frac {3\,{{{J^{\mpsi}}}}^{2}}{32}}+{\frac {15\,{{{J^{\mpsi}}}}^{3}}{128}}-{\frac {3\,{{{J^{\msigma}}}}^{2}{{J^{\mpsi}}}}{128}}+{\frac {21\,{{{J^{\msigma}}}}^{3}}{64}}\nonumber\\&
+{\frac {243\,{{{J^{\mpsi}}}}^{4}}{2048}}+{\frac {7\,{{{J^{\msigma}}}}^{2}{{{J^{\mpsi}}}}^{2}}{512}}+{\frac {31\,{{{J^{\msigma}}}}^{3}{{J^{\mpsi}}}}{128}}-{\frac {5\,{{{J^{\msigma}}}}^{4}}{1024}}\nonumber\\&
+{\frac {1671\,{{{J^{\mpsi}}}}^{5}}{8192}}+{\frac {221\,{{{J^{\msigma}}}}^{2}{{{J^{\mpsi}}}}^{3}}{8192}}+{\frac {6317\,{{{J^{\msigma}}}}^{3}{{{J^{\mpsi}}}}^{2}}{12288}}\nonumber\\&
-{\frac {1673\,{{{J^{\msigma}}}}^{4}{{J^{\mpsi}}}}{8192}}+{\frac {215\,{{{J^{\msigma}}}}^{5}}{1536}}.
\label{eq:deltapsiminus}
\end{align}

\section{Details for the standard (\texorpdfstring{$\mathbb{Z}_2$}{Z(2)}) case}\label{app:z2}
\subsection{Benchmarking the numerical results using the \texorpdfstring{$\mathbb{Z}_2$}{Z(2)} case}
\begin{figure}[htp]%
\includegraphics[width=\columnwidth]{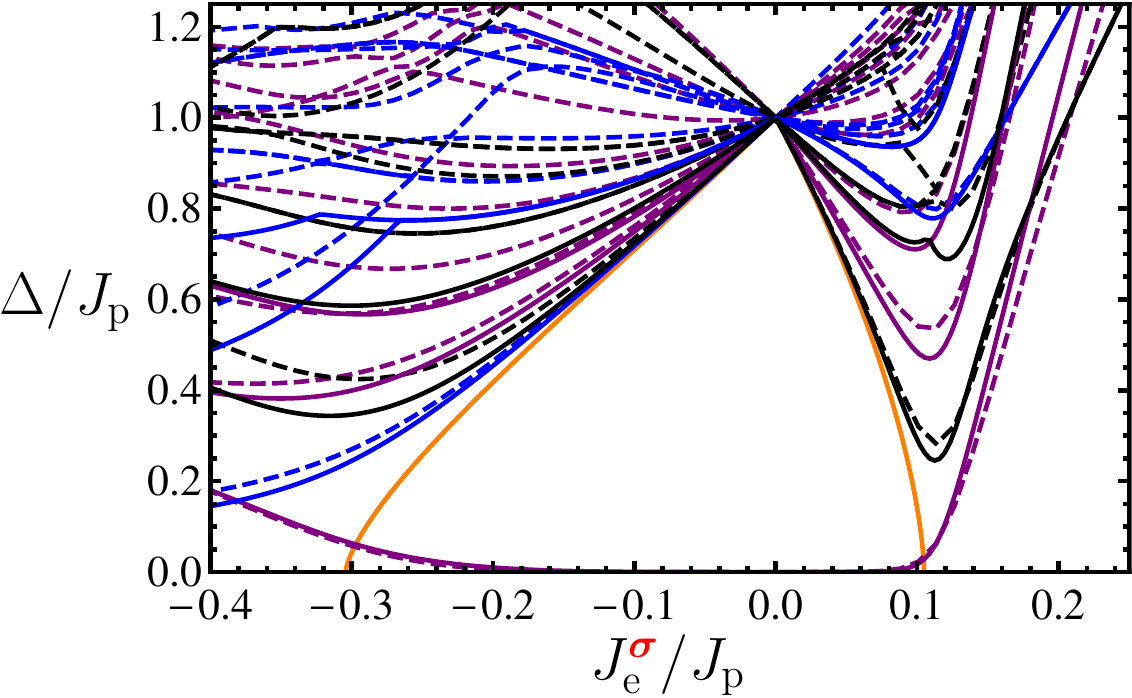}%
\\
\includegraphics[width=\columnwidth]{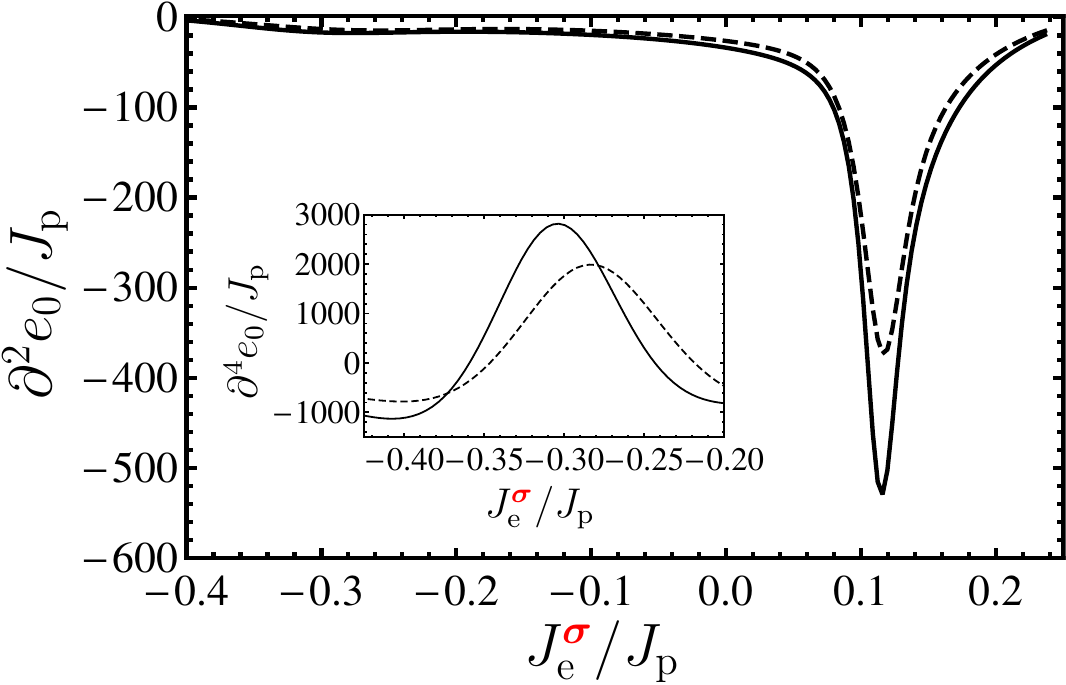}%
\caption{In the upper panel, we show here the low-energy spectrum of the effective model (\ref{eq:hameffpzt}) as a function of $\Jes$ (blue $\vec{k}=(\pm\frac{2\pi}{3},\mp\frac{2\pi}{3})$, black $\vec{k}=(0,0)$ even number of non-contractible strings, purple $\vec{k}=(0,0)$ odd number of non-contractible strings in at least one direction) for different system sizes ($N_{\rm p}=21$ plaquettes dashed, $N_{\rm p}=27$ plaquettes solid). Series expansion results are shown in orange. The spectrum reveals the topological trivial (unique ground state) for $\Jes\gg\Jp$, the $\mathbb{Z}_2$ topological phase by the degeneracy of the different topological sectors as well as the spontaneous symmetry broken (non-topological) phase signaled by the ground state degeneracy formed by states of different $\vec{k}$. In the lower panel, we show $\partial^2_{\Jes}e_0$, whose minima indicate the location of the phase transition. The inset shows the higher order derivative used to isolate the transition to the frustrated phase.}%
\label{fig:spectrum_pzt}%
\end{figure}
In Fig.~\ref{fig:spectrum_pzt}, we show the low-energy spectrum of Hamiltonian (\ref{eq:hameffpzt}) as a benchmark for our analytical and numerical findings for the transition out of the $\mathbb{Z}_2$-phase. For $\Jes\gg\Jp$, we find a topological trivial phase with a unique ground state. For $\left|\Jes\right|\ll\Jp$, we find the topological phase, whose presence is signaled by the degeneracy of the states in the different sectors defined by the parity of numbers of non-contractible loops around the torus in different directions. For $-\Jes\gg\Jp$, we find the translation symmetry broken phase signaled by the degeneracy (up to finite size effects) of levels of different $\vec{k}$ in the ground state. The reduction from the complete model (\ref{eq:fullham}) to the effective one (\ref{eq:hameffpzt}) allows for larger system sizes for the exact diagonalization as well as higher orders in the perturbation theory. The coefficients of the series expansions are given in App.~\ref{app:z2series}. The corresponding series for the dispersion fit the expressions for the low-energy gap given in Ref.~\onlinecite{he90} for the unfrustrated and Ref.~\onlinecite{powalski13} for the frustrated case. Our results of the location of the transition from the topological to the unfrustrated phase from series expansions ($\left.\Jes/\Jp\right|_c=0.103$) and from exact diagonalization ($\left.\Jes/\Jp\right|_c=0.115$) are consistent with the value given in the literature ($\left.\Jes/\Jp\right|_c=0.104$)\cite{bloete02}. The same holds for the transition from the topological to the frustrated phase, where we obtain the location of the phase transition ($\left.\Jes/\Jp\right|_c=-0.303$)\cite{isakov03} from series expansion as $\left.\Jes/\Jp\right|_c=-0.305$. In order to extract the value from the exact diagonalization results, it is instructive to consider higher order derivatives of the ground state energy in order to separate the phase transitions more clearly. This leads to a value $\left.\Jes/\Jp\right|_c=-0.304$. Note that this value is reached only for the largest system sizes, as finite size effects in the frustrated regime play a major role. Generally, one can attribute the larger uncertainty of exact diagonalization in this case to the fact that the single quasi-particle mode, whose condensation drives the phase transition, is not part of Hilbert space of the $\mathbb{Z}_2$ string-net model for periodic boundary conditions. Thus, in contrast to the non-Abelian original model (\ref{eq:fullham}), the low-energy mode signaling the phase transition stems from the two-particle continuum. As the free particle limit is reached only for larger system sizes, finite size effects are typically more significant in the Abelian models.

\subsection{Series expansions for the \texorpdfstring{$\mathbb{Z}_2$}{Z(2)} case}\label{app:z2series}
In this section, we give the low-energy spectrum of the Hamiltonian (\ref{eq:hameffpzt}) as a function of $J=\frac{\Jes}{\Jp}$ determined by series expansions. We give the ground state energy per plaquette $e_0$ as well as the individual hopping elements $t_{\vec{\boldsymbol{n}}}$ leading to the dispersion relation.

The ground-state energy per plaquette $e_0$ reads up to order $11$ in units of $\Jp$:
\begin{align}
\frac{e_0}{\Jp}=&-\frac{1}{2}-\frac{3\,J^2}{4}-\frac{3\,J^3}{2}-\frac{87\,J^4}{16}-\frac{99\,J^5}{4}-\frac{2139\,J^6}{16}\nonumber\\&
-\frac{6315\,J^7}{8}-\frac{1280037\,J^8}{256}-\frac{4263501\,J^9}{128}\nonumber\\&
-\frac{118233091\,J^{10}}{512}-\frac{40611961873\,J^{11}}{24576}.
\label{eq:e0z2pzt}
\end{align}
This coincide with the series given in Ref.~\onlinecite{he90} after rescaling the Hamiltonian accordingly.

For the dispersion of the low-energy excitations, we have $\omega(\vec{k})=\sum\limits_{\vec{\boldsymbol{n}}} e^{{\rm i} \vec{k}\vec{\boldsymbol{n}}} t_{\vec{\boldsymbol{n}}}$, where the hopping elements $t_{\vec{\boldsymbol{n}}}$ are given in units of $\Jp$:
\begin{align}
\frac{t_{\vec{0}}}{\Jp}=&\frac{1}{2}+\frac{3}{2}\,{J}^{2}+6\,{J}^{3}+{\frac {207\,{J}^{4}}{8}}+{\frac {303\,{J}^{5}}{2}}+963\,{J}^{6}\nonumber\\&
+{\frac {26697\,{J}^{7}}{4}}+{\frac {6136203\,{J}^{8}}{128}}+{\frac {45978297\,{J}^{9}}{128}}\nonumber\\&
+{\frac {1411200267\,{J}^{10}}{512}}+{\frac {266121366979\,{J}^{11}}{12288}}\\ 
 \frac{t_{ \vec{\boldsymbol{n}}_2 }}{\Jp}=&-\frac{J}{2}-\frac{{J}^{2}}{2}+\frac{{J}^{3}}{4}+\frac{9\,{J}^{4}}{4}+{\frac {149\,{J}^{5}}{8}}+{\frac {231\,{J}^{6}}{2}}\nonumber\\&
 +{\frac {53291\,{J}^{7}}{64}}+{\frac {91539\,{J}^{8}}{16}}+{\frac {21552013\,{J}^{9}}{512}}\nonumber\\&
 +{\frac {3809770171\,{J}^{10}}{12288}}+{\frac {58039640957\,{J}^{11}}{24576}}\\ 
 \frac{t_{2 \vec{\boldsymbol{n}}_2 }}{\Jp}=&-\frac{{J}^{2}}{4}-\frac{3\,{J}^{3}}{2}-{\frac {41\,{J}^{4}}{8}}-{\frac {73\,{J}^{5}}{4}}-{\frac {2223\,{J}^{6}}{32}}\nonumber\\&
 -{\frac {8003\,{J}^{7}}{32}}-{\frac {46195\,{J}^{8}}{64}}+{\frac {829807\,{J}^{9}}{1024}}\nonumber\\&
 +{\frac {481385009\,{J}^{10}}{12288}}+{\frac {9409868561\,{J}^{11}}{18432}}\\ 
 \frac{t_{ 3 \vec{\boldsymbol{n}}_2 }}{\Jp}=&-\frac{{J}^{3}}{4}-{\frac {15\,{J}^{4}}{4}}-{\frac {381\,{J}^{5}}{16}}-{\frac {543\,{J}^{6}}{4}}\nonumber\\&
 -{\frac {99105\,{J}^{7}}{128}}-{\frac {580905\,{J}^{8}}{128}}-{\frac {55803529\,{J}^{9}}{2048}}\nonumber\\&
 -{\frac {1026968231\,{J}^{10}}{6144}}-{\frac {38227654615\,{J}^{11}}{36864}}\\ 
 \frac{t_{ 4 \vec{\boldsymbol{n}}_2 }}{\Jp}=&-{\frac {5\,{J}^{4}}{16}}-{\frac {35\,{J}^{5}}{4}}-{\frac {347\,{J}^{6}}{4}}-{\frac {42729\,{J}^{7}}{64}}\nonumber\\&
 -{\frac {1216213\,{J}^{8}}{256}}-{\frac {16937505\,{J}^{9}}{512}}\nonumber\\&
 -{\frac {1413691361\,{J}^{10}}{6144}}-{\frac {118743077371\,{J}^{11}}{73728}}\\ 
 \frac{t_{ 5 \vec{\boldsymbol{n}}_2 }}{\Jp}=&-{\frac {7\,{J}^{5}}{16}}-{\frac {315\,{J}^{6}}{16}}-{\frac {9037\,{J}^{7}}{32}}-{\frac {716327\,{J}^{8}}{256}}\nonumber\\&
 -{\frac {49229795\,{J}^{9}}{2048}}-{\frac {396348875\,{J}^{10}}{2048}}\nonumber\\&
 -{\frac {223040672585\,{J}^{11}}{147456}}\\ 
 \frac{t_{ 6 \vec{\boldsymbol{n}}_2 }}{\Jp}=&-{\frac {21\,{J}^{6}}{32}}-{\frac {693\,{J}^{7}}{16}}-{\frac {54639\,{J}^{8}}{64}}-{\frac {10876353\,{J}^{9}}{1024}}\nonumber\\&
 -{\frac {13864707\,{J}^{10}}{128}}-{\frac {255268593\,{J}^{11}}{256}}\\ 
 \frac{t_{ 7 \vec{\boldsymbol{n}}_2 }}{\Jp}=&-{\frac {33\,{J}^{7}}{32}}-{\frac {3003\,{J}^{8}}{32}}-{\frac {625647\,{J}^{9}}{256}}\nonumber\\&
 -{\frac {153638895\,{J}^{10}}{4096}}-{\frac {14714945105\,{J}^{11}}{32768}}\\ 
 \frac{t_{ 8 \vec{\boldsymbol{n}}_2 }}{\Jp}=&-{\frac {429\,{J}^{8}}{256}}-{\frac {6435\,{J}^{9}}{32}}-{\frac {1717001\,{J}^{10}}{256}}\nonumber\\&
 -{\frac {2050184705\,{J}^{11}}{16384}}\\ 
 \frac{t_{ 9 \vec{\boldsymbol{n}}_2 }}{\Jp}=&-{\frac {715\,{J}^{9}}{256}}-{\frac {109395\,{J}^{10}}{256}}-{\frac {9112389\,{J}^{11}}{512}}\\ 
 \frac{t_{ 10 \vec{\boldsymbol{n}}_2 }}{\Jp}=&-{\frac {2431\,{J}^{10}}{512}}-{\frac {230945\,{J}^{11}}{256}}\\ 
 \frac{t_{ 11 \vec{\boldsymbol{n}}_2 }}{\Jp}=&-{\frac {4199\,{J}^{11}}{512}}\\ 
 \frac{t_{ \vec{\boldsymbol{n}}_1 + \vec{\boldsymbol{n}}_2 }}{\Jp}=&-\frac{{J}^{2}}{2}-\frac{3\,{J}^{3}}{2}-4\,{J}^{4}-{\frac {47\,{J}^{5}}{4}}-{\frac {259\,{J}^{6}}{8}}\nonumber\\&
 -{\frac {737\,{J}^{7}}{32}}+{\frac {88575\,{J}^{8}}{128}}+{\frac {10288015\,{J}^{9}}{1024}}\nonumber\\&
 +{\frac {308119475\,{J}^{10}}{3072}}+{\frac {136697766643\,{J}^{11}}{147456}}\\ 
 \frac{t_{ \vec{\boldsymbol{n}}_1 + 2 \vec{\boldsymbol{n}}_2 }}{\Jp}=&-\frac{3\,{J}^{3}}{4}-5\,{J}^{4}-{\frac {51\,{J}^{5}}{2}}-{\frac {1033\,{J}^{6}}{8}}\nonumber\\&
 -{\frac {86911\,{J}^{7}}{128}}-{\frac {237909\,{J}^{8}}{64}}-{\frac {21190753\,{J}^{9}}{1024}}\nonumber\\&
 -{\frac {1432677109\,{J}^{10}}{12288}}-{\frac {31678495615\,{J}^{11}}{49152}}\\ 
 \frac{t_{ \vec{\boldsymbol{n}}_1 + 3 \vec{\boldsymbol{n}}_2 }}{\Jp}=&-\frac{5\,{J}^{4}}{4}-{\frac {245\,{J}^{5}}{16}}-{\frac {1835\,{J}^{6}}{16}}-{\frac {3075\,{J}^{7}}{4}}\nonumber\\&
 -{\frac {160697\,{J}^{8}}{32}}-{\frac {67463475\,{J}^{9}}{2048}}\nonumber\\&-{\frac {896422249\,{J}^{10}}{4096}}-{\frac {434747248931\,{J}^{11}}{294912}}\\ 
 \frac{t_{ \vec{\boldsymbol{n}}_1 + 4 \vec{\boldsymbol{n}}_2 }}{\Jp}=&-{\frac {35\,{J}^{5}}{16}}-{\frac {693\,{J}^{6}}{16}}-{\frac {14231\,{J}^{7}}{32}}\nonumber\\&
 -{\frac {958651\,{J}^{8}}{256}}-{\frac {59524081\,{J}^{9}}{2048}}\nonumber\\&
 -{\frac {1747975\,{J}^{10}}{8}}-{\frac {239463246341\,{J}^{11}}{147456}}\\ 
 \frac{t_{ \vec{\boldsymbol{n}}_1 + 5 \vec{\boldsymbol{n}}_2 }}{\Jp}=&-{\frac {63\,{J}^{6}}{16}}-{\frac {231\,{J}^{7}}{2}}-{\frac {6279\,{J}^{8}}{4}}-{\frac {8284299\,{J}^{9}}{512}}\nonumber\\&
 -{\frac {300462381\,{J}^{10}}{2048}}-{\frac {20429078349\,{J}^{11}}{16384}}\\ 
 \frac{t_{ \vec{\boldsymbol{n}}_1 + 6 \vec{\boldsymbol{n}}_2 }}{\Jp}=&-{\frac {231\,{J}^{7}}{32}}-{\frac {4719\,{J}^{8}}{16}}-{\frac {165363\,{J}^{9}}{32}}\nonumber\\&
 -{\frac {262953129\,{J}^{10}}{4096}}-{\frac {22060824855\,{J}^{11}}{32768}}\\ 
 \frac{t_{ \vec{\boldsymbol{n}}_1 + 7 \vec{\boldsymbol{n}}_2 }}{\Jp}=&-{\frac {429\,{J}^{8}}{32}}-{\frac {186615\,{J}^{9}}{256}}-{\frac {4121403\,{J}^{10}}{256}}\nonumber\\&
 -{\frac {488124205\,{J}^{11}}{2048}}\\ 
 \frac{t_{ \vec{\boldsymbol{n}}_1 + 8 \vec{\boldsymbol{n}}_2 }}{\Jp}=&-{\frac {6435\,{J}^{9}}{256}}-{\frac {449735\,{J}^{10}}{256}}-{\frac {24552385\,{J}^{11}}{512}}\\ 
 \frac{t_{ \vec{\boldsymbol{n}}_1 + 9 \vec{\boldsymbol{n}}_2 }}{\Jp}=&-{\frac {12155\,{J}^{10}}{256}}-{\frac {1062347\,{J}^{11}}{256}}\\ 
 \frac{t_{ \vec{\boldsymbol{n}}_1 + 10 \vec{\boldsymbol{n}}_2 }}{\Jp}=&-{\frac {46189\,{J}^{11}}{512}}\\ 
 \frac{t_{2 \vec{\boldsymbol{n}}_1 + 2 \vec{\boldsymbol{n}}_2 }}{\Jp}=&-{\frac {15\,{J}^{4}}{8}}-{\frac {35\,{J}^{5}}{2}}-{\frac {1965\,{J}^{6}}{16}}-{\frac {25231\,{J}^{7}}{32}}\nonumber\\&
 -{\frac {80249\,{J}^{8}}{16}}-{\frac {16511537\,{J}^{9}}{512}}-{\frac {40479223\,{J}^{10}}{192}}\nonumber\\&
 -{\frac {12901176995\,{J}^{11}}{9216}}\\ 
 \frac{t_{2 \vec{\boldsymbol{n}}_1 + 3 \vec{\boldsymbol{n}}_2 }}{\Jp}=&-{\frac {35\,{J}^{5}}{8}}-{\frac {945\,{J}^{6}}{16}}-{\frac {17143\,{J}^{7}}{32}}-{\frac {537139\,{J}^{8}}{128}}\nonumber\\&
 -{\frac {7969807\,{J}^{9}}{256}}-{\frac {928510405\,{J}^{10}}{4096}}\nonumber\\&
 -{\frac {485047715845\,{J}^{11}}{294912}}\\ 
 \frac{t_{2 \vec{\boldsymbol{n}}_1 + 4 \vec{\boldsymbol{n}}_2 }}{\Jp}=&-{\frac {315\,{J}^{6}}{32}}-{\frac {3003\,{J}^{7}}{16}}-{\frac {68523\,{J}^{8}}{32}}-{\frac {20517711\,{J}^{9}}{1024}}\nonumber\\&
 -{\frac {5462787\,{J}^{10}}{32}}-{\frac {1422572163\,{J}^{11}}{1024}}\\ 
 \frac{t_{2 \vec{\boldsymbol{n}}_1 + 5 \vec{\boldsymbol{n}}_2 }}{\Jp}=&-{\frac {693\,{J}^{7}}{32}}-{\frac {9009\,{J}^{8}}{16}}-{\frac {509553\,{J}^{9}}{64}}\nonumber\\&
 -{\frac {359701515\,{J}^{10}}{4096}}-{\frac {27921342885\,{J}^{11}}{32768}}\\ 
 \frac{t_{2 \vec{\boldsymbol{n}}_1 + 6 \vec{\boldsymbol{n}}_2 }}{\Jp}=&-{\frac {3003\,{J}^{8}}{64}}-{\frac {6435\,{J}^{9}}{4}}-{\frac {14263821\,{J}^{10}}{512}}\nonumber\\&
 -{\frac {1470047415\,{J}^{11}}{4096}}\\ 
 \frac{t_{2 \vec{\boldsymbol{n}}_1 + 7 \vec{\boldsymbol{n}}_2 }}{\Jp}=&-{\frac {6435\,{J}^{9}}{64}}-{\frac {1130415\,{J}^{10}}{256}}-{\frac {5925205\,{J}^{11}}{64}}\\ 
 \frac{t_{2 \vec{\boldsymbol{n}}_1 + 8 \vec{\boldsymbol{n}}_2 }}{\Jp}=&-{\frac {109395\,{J}^{10}}{512}}-{\frac {3002285\,{J}^{11}}{256}}\\ 
 \frac{t_{2 \vec{\boldsymbol{n}}_1 + 9 \vec{\boldsymbol{n}}_2 }}{\Jp}=&-{\frac {230945\,{J}^{11}}{512}}\\ 
 \frac{t_{3 \vec{\boldsymbol{n}}_1 + 3 \vec{\boldsymbol{n}}_2 }}{\Jp}=&-{\frac {105\,{J}^{6}}{8}}-{\frac {3465\,{J}^{7}}{16}}-{\frac {18837\,{J}^{8}}{8}}\nonumber\\&
 -{\frac {5470485\,{J}^{9}}{256}}-{\frac {365718357\,{J}^{10}}{2048}}\nonumber\\&
 -{\frac {23482123047\,{J}^{11}}{16384}}\\ 
 \frac{t_{3 \vec{\boldsymbol{n}}_1 + 4 \vec{\boldsymbol{n}}_2 }}{\Jp}=&-{\frac {1155\,{J}^{7}}{32}}-{\frac {3003\,{J}^{8}}{4}}-{\frac {1241625\,{J}^{9}}{128}}\nonumber\\&
 -{\frac {414943293\,{J}^{10}}{4096}}-{\frac {31076174115\,{J}^{11}}{32768}}\\ 
 \frac{t_{3 \vec{\boldsymbol{n}}_1 + 5 \vec{\boldsymbol{n}}_2 }}{\Jp}=&-{\frac {3003\,{J}^{8}}{32}}-{\frac {315315\,{J}^{9}}{128}}-{\frac {4816383\,{J}^{10}}{128}}\nonumber\\&
 -{\frac {920970315\,{J}^{11}}{2048}}\\ 
 \frac{t_{3 \vec{\boldsymbol{n}}_1 + 6 \vec{\boldsymbol{n}}_2 }}{\Jp}=&-{\frac {15015\,{J}^{9}}{64}}-{\frac {984555\,{J}^{10}}{128}}-{\frac {70804305\,{J}^{11}}{512}}\\ 
 \frac{t_{3 \vec{\boldsymbol{n}}_1 + 7 \vec{\boldsymbol{n}}_2 }}{\Jp}=&-{\frac {36465\,{J}^{10}}{64}}-{\frac {11778195\,{J}^{11}}{512}}\\ 
 \frac{t_{3 \vec{\boldsymbol{n}}_1 + 8 \vec{\boldsymbol{n}}_2 }}{\Jp}=&-{\frac {692835\,{J}^{11}}{512}}\\ 
 \frac{t_{4 \vec{\boldsymbol{n}}_1 + 4 \vec{\boldsymbol{n}}_2 }}{\Jp}=&-{\frac {15015\,{J}^{8}}{128}}-{\frac {45045\,{J}^{9}}{16}}-{\frac {5300295\,{J}^{10}}{128}}\nonumber\\&
 -{\frac {3958082595\,{J}^{11}}{8192}}\\ 
 \frac{t_{4 \vec{\boldsymbol{n}}_1 + 5 \vec{\boldsymbol{n}}_2 }}{\Jp}=&-{\frac {45045\,{J}^{9}}{128}}-{\frac {1276275\,{J}^{10}}{128}}-{\frac {42814629\,{J}^{11}}{256}}\\ 
 \frac{t_{4 \vec{\boldsymbol{n}}_1 + 6 \vec{\boldsymbol{n}}_2 }}{\Jp}=&-{\frac {255255\,{J}^{10}}{256}}-{\frac {4295577\,{J}^{11}}{128}}\\ 
 \frac{t_{4 \vec{\boldsymbol{n}}_1 + 7 \vec{\boldsymbol{n}}_2 }}{\Jp}=&-{\frac {692835\,{J}^{11}}{256}}\\ 
 \frac{t_{5 \vec{\boldsymbol{n}}_1 + 5 \vec{\boldsymbol{n}}_2 }}{\Jp}=&-{\frac {153153\,{J}^{10}}{128}}-{\frac {4849845\,{J}^{11}}{128}}\\ 
 \frac{t_{5 \vec{\boldsymbol{n}}_1 + 6 \vec{\boldsymbol{n}}_2 }}{\Jp}=&-{\frac {969969\,{J}^{11}}{256}}
 \label{eq:hoppingz2pzt}
\end{align}
The remaining hopping elements are related by lattice symmetries of the underlying triangular lattice and can be obtained via the relations
\begin{align}
t_{\vec{r}}=&t_{-\vec{r}},\\
t_{n_2\vec{\boldsymbol{n}}_1+n_1\vec{\boldsymbol{n}}_2}=&t_{n_1\vec{\boldsymbol{n}}_1+n_2\vec{\boldsymbol{n}}_2},\\
t_{-n_2\vec{\boldsymbol{n}}_1+(n_1+n_2)\vec{\boldsymbol{n}}_2}=&t_{n_1\vec{\boldsymbol{n}}_1+n_2\vec{\boldsymbol{n}}_2},\\
t_{(-n_1-n_2)\vec{\boldsymbol{n}}_1+n_1\vec{\boldsymbol{n}}_2}=&t_{n_1\vec{\boldsymbol{n}}_1+n_2\vec{\boldsymbol{n}}_2}.
\label{eq:symrelhoppings}
\end{align}

For positive $J$, the minimum of the dispersion is obtained for $\vec{k}=\vec{0}$ and reads
\begin{align}
\frac{\Delta^+}{\Jp}=&\frac{1}{2}-3\,J-6\,J^2-21\,J^3-126\,J^4-\frac{3153\,J^5}{4}\nonumber\\&
-\frac{44379\,J^6}{8}-\frac{2570661\,J^7}{64}-\frac{9821055\,J^8}{32}\nonumber\\&
-\frac{1222762161\,J^9}{512}-\frac{39126191841\,J^{10}}{2048}\nonumber\\&
-\frac{7619833519319\,J^{11}}{49152}.
\label{eq:dispz2pos}
\end{align}
This series coincides with the series given in Ref.~\onlinecite{he90} after rescaling the Hamiltonian accordingly (up to an obvious typo in Ref.~\onlinecite{he90}).

For negative $J$, the minimum of the dispersion is obtained for $\vec{k}=(\pm\frac{2\pi}{3},\mp\frac{2\pi}{3})$ and reads
\begin{align}
\frac{\Delta^-}{\Jp}=&\frac{1}{2}+\frac{3\,J}{2}+\frac{3\,{J}^{2}}{4}+{\frac{15\,{J}^{3}}{4}}+{\frac{243\,{J}^{4}}{16}}+{\frac{1671\,{J}^{5}}{16}}\nonumber\\&
+{\frac{22275\,{J}^{6}}{32}}+{\frac{162855\,{J}^{7}}{32}}+{\frac{9700617\,{J}^{8}}{256}}\nonumber\\&
+{\frac{595490847\,{J}^{9}}{2048}}+{\frac{9308111103\,{J}^{10}}{4096}}\nonumber\\&
+{\frac{1777064899901\,{J}^{11}}{98304}}.
\label{eq:dispz2neg}
\end{align}
This series coincides with the series given in Ref.~\onlinecite{powalski13} after rescaling the Hamiltonian accordingly.

\section{Finite size scaling for the transition out of the \texorpdfstring{$\mathbb{Z}_2^{\prime}$}{Z'(2)} topological phase}\label{app:fss_z2prime}
Here, we present the finite size scaling for the phase transition out of the $\mathbb{Z}_2^{\prime}$-phase as determined for the effective model (\ref{eq:hameffc}).

\begin{figure}[htp]%
\includegraphics[width=\columnwidth]{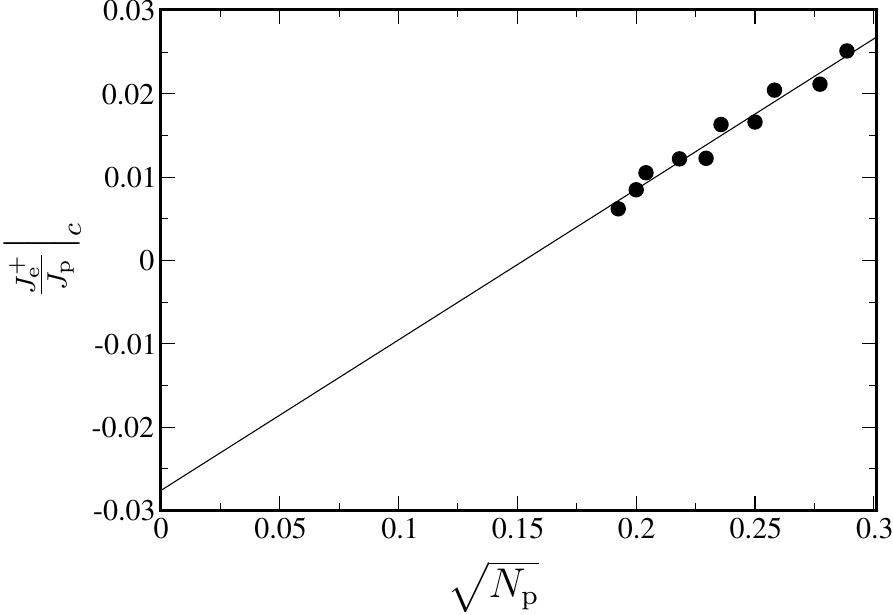}%
\caption{Finite size scaling for the transition from the topological to the trivial phase. From the minima of $\partial^2e_0$, we obtain the points, which extrapolate to the limit $\left.\frac{J_{\rm}^+}{\Jp}\right|_c=-0.0276$.}%
\label{fig:fss2triv}%
\end{figure}

In Fig.~\ref{fig:fss2triv}, we show the locations of the transition to the trivial phase obtained for several system sizes by exact diagonalization by considering the minimum of $\partial^2_{J_{\rm e}^+} e_0$ (shown in Fig.~\ref{fig:cornerspectrum}). Extrapolating the location in $\frac{1}{\sqrt{N_{\rm p}}}$, we obtain the location of the thermodynamic to be $\left.\frac{J_{\rm}^+}{\Jp}\right|_c=-0.0276$.

A similar analysis of the transition to the columnar phase ${\rm{col}}_{\mone}$ shows a much stronger system-size dependence than for the unfrustrated case, which leads us to conclude that the transition locations determined for the full model (\ref{eq:fullham}) on systems up to $N_{\rm p}=12$ plaquettes may deviate from the ones in the thermodynamic limit significantly. Extrapolating the results for the effective model (\ref{eq:hameffc}) to the thermodynamic limit, we obtain the location of the transition to be at $\left.\frac{J_{\rm}^+}{\Jp}\right|_c=-0.275$.
\begin{figure}[htp]%
\includegraphics[width=\columnwidth]{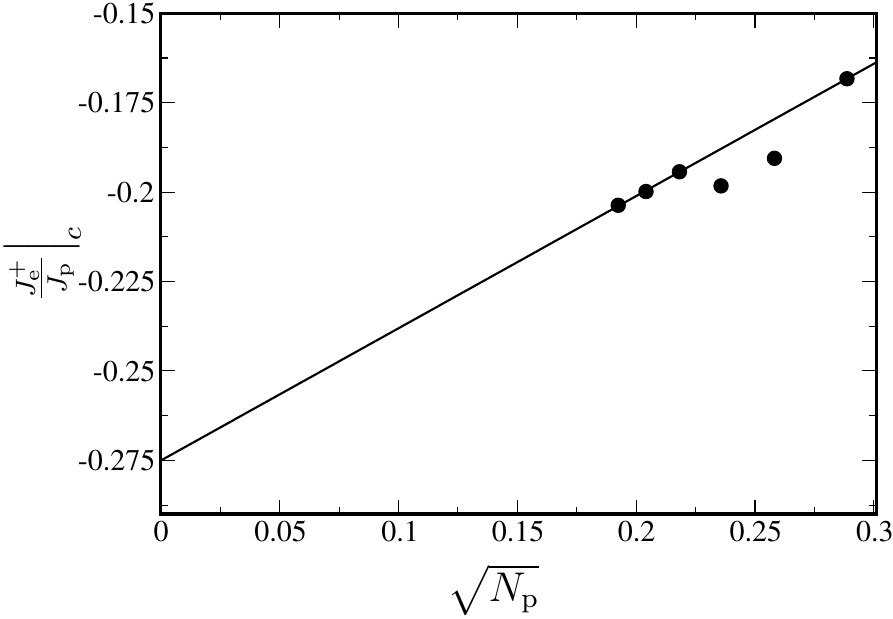}%
\caption{Finite size scaling for the transition from the topological to the columnar phase. From the minima of $\partial^2e_0$, we obtain the points, which extrapolate to the limit $\left.\frac{J_{\rm}^+}{\Jp}\right|_c=-0.275$.}%
\label{fig:fss2col}%
\end{figure}

%

\end{document}